\documentstyle{article}
\makeatletter
\def\be{%
\@ifnextchar[%   if the char after \be is [ 
{\def\ee{\end{equation}}\begin{equation}\l@b}% 
{\def\ee{\]}\[}%
}
\def\l@b[#1]{\label{#1}}
\def\eq#1{\hbox{\rm(\ref{#1})}}
\def\Eq#1{\hbox{Eq.\rm(\ref{#1})}}

\def\Eqs#1-#2{\hbox{Eqs.\rm\eq{#1}--\eq{#2}}}
\def\eqs#1-#2{\eq{#1}--\eq{#2}}
\@addtoreset{equation}{section}
\def\PART#1{\newpage\part{\rm #1}}
\def\SECTION{\section}
\def\sect#1{sec\-tion~\hbox{\rm\ref{#1}}}
\def\Sect#1{Section~\hbox{\rm\ref{#1}}}
\def\sects#1-#2{sections {\rm\ref{#1}--\ref{#2}}}
\def\Sects#1-#2{Sections {\rm\ref{#1}--\ref{#2}}}
\def\SUBSECTION{\subsection}
\def\ssect#1{{\rm\S\ref{#1}}}
\def\Ssect#1{Subsec\-tion~{\rm\S\ref{#1}}}
\def\ssects#1-#2{{\rm\S\ref{#1}--\S\ref{#2}}}
\def\Ssects#1-#2{\Sub\-sec\-tions {\rm\S\ref{#1}--\S\ref{#2}}}

\def\@sect#1#2#3#4#5#6[#7]#8{\ifnum #2>\c@secnumdepth
     \def\@svsec{}\else
     \refstepcounter{#1}\edef\@svsec%
{\ifnum #2=2{\S\kern-.43em\S\kern.20em}\fi%  
\csname the#1\endcsname%
.%    -- this is the dot after the (sub)section number.
     \hskip 0.8em }\fi
     \@tempskipa #5\relax
      \ifdim \@tempskipa>\z@
        \begingroup #6\relax
          \@hangfrom{\hskip #3\relax\@svsec}{\interlinepenalty \@M #8\par}%
        \endgroup
       \csname #1mark\endcsname{#7}\addcontentsline
         {toc}{#1}{\ifnum #2>\c@secnumdepth \else
                      \protect\numberline{%
\ifnum #2=2\S\fi%
                      \csname the#1\endcsname}\fi
                    #7}\else
        \def\@svsechd{#6\hskip #3\@svsec #8\csname #1mark\endcsname
                      {#7}\addcontentsline
                           {toc}{#1}{\ifnum #2>\c@secnumdepth \else
                             \protect\numberline{%
\ifnum #2=2\S\fi%
                             \csname the#1\endcsname}\fi
                       #7}}\fi
     \@xsect{#5}}
\def\asop{$As$-operation}
\def\asex{$As$-expansion}
\def\r*op{$R^{*}$-operation}
\def\rop {$R$-operation}
\def\trop{$\tilde R$-operation}
\def\textbox#1{{\mathchoice%
{\mbox{#1}}%
{\mbox{#1}}%
{\mbox{{\scriptsize#1}}}%
{\mbox{{\tiny#1}}}%
}}

\def \R  {{\bf R}}
\def \r  {{\bf r}}
\def \tR {{\bf\tilde R}}
\def \tr {{\bf\tilde r}}
\def \T  {{\bf T}}
\def \t  {{\bf t}}
\def \As {{\bf As}}
\def \as {{\bf as}}

\def \Tk  { \T_\kappa  }

\def \cB  {{ \cal B  }}
\def \cC  {{ \cal C  }}
\def \cD  {{ \cal D  }}
\def \cF  {{ \cal F  }}
\def \cG  {{ \cal G  }}
\def \cL  {{ \cal L  }}
\def \cM  {{ \cal M  }}
\def \cO  {{ \cal O  }}
\def \cP  {{ \cal P  }}

\def \cS  {{ \cal S  }}
\def \a   { \alpha          }
\def \b   { \beta           }
\def \G   { \Gamma          }
\def \GG  {{ G\backslash\G  }}
\def \Gg  { G\backslash\g   }
\def \g   { \gamma          }
\def \k   { \kappa          }
\def \d   { \delta          }
\def \D   { \Delta          }
\def \e   { \epsilon        }
\def \l   { \lambda         }
\def \L   { \Lambda         }

\def \o   { \omega          }
\def \O   { \Omega          }
\def \vf  { \varphi         }
\def \0{{\vphantom{0}}}
\def \.{\raise0.3ex\hbox{$\scriptscriptstyle\circ$}} 
\let\Cspace=C
\def\sup{\mathop{{\rm sup}}}
\def\inf{\mathop{{\rm inf}}}
\def\min{\mathop{{\rm min}}}
\def\max{\mathop{{\rm max}}}
\def\rad{\mathop{{\rm rad}}}

\def\supp{\mathop{{\rm supp}}}
\def\dim{\mathop{{\rm dim}}}
\def\const{\mathord{{\rm const}}}
\def\bydef{\ \mathrel{{\displaystyle\mathop{=}^{\rm def}}}\ }

\def\equals{\mathop{\mathstrut=\mathstrut}}
\let\@cap\cap
\let\@cup\cup
\def\cap{\mathop{\@cap}}
\def\cup{\mathop{\@cup}}
\def\crcaccent#1{\vbox{\ialign{##\crcr
\hfil${\scriptscriptstyle \circ}$\,\crcr
\noalign{\kern2pt\nointerlineskip}
$\hfil\displaystyle{#1}\hfil$\crcr}}}
\def\scriptcrcaccent#1{\vbox{\ialign{##\crcr
\hfil${\scriptscriptstyle \circ}$\crcr
\noalign{\kern1pt\nointerlineskip}
$\hfil\scriptstyle{#1}\hfil$\crcr}}}

\makeatother

\begin{document}

\thispagestyle{empty}
\vbox to1in{\vfill}         % VERTICAL SKIP
\hfill \hbox{PSU-TH-108/92}
\begin{center}
%  #[ title :
%
{\large\bf
    TECHNIQUES OF DISTRIBUTIONS\\
[2mm]
    IN PERTURBATIVE QUANTUM FIELD THEORY\\
[4mm]
    (I) {\bf Euclidean asymptotic operation\\
         for products of singular functions}\\
[8mm]}
%
%  #] title :
%  #[ authors :
%
{\bf A.~N.~Kuznetsov, F.~V.~Tkachov and V.~V.~Vlasov} \\ [5mm] 
Institute for Nuclear Research \\
of the Russian Academy of Sciences, \\ 
Moscow 117312, Russia
%  #] authors :
\end{center}
%  #[ abstract :
%
{
\centerline {\em Abstract}
\vskip2mm
\noindent
We present a systematic description of the mathematical techniques for
studying multiloop Feynman diagrams which constitutes a full-fledged
and inherently more powerful alternative to the BPHZ theory.  
The new techniques
emerged as a formalization of the reasoning behind a recent series of
record multiloop calculations in perturbative quantum field theory. It
is based on a systematic use of the ideas and notions of the
distribution theory.  We identify the problem of asymptotic expansion
of products of singular functions in the sense of distributions as a
key problem of the theory of asymptotic expansions of multiloop
Feynman diagrams. Its complete solution for the case of Euclidean
Feynman diagrams (the so-called {\it Euclidean asymptotic operation for products of singular functions}) is explicitly constructed
and studied.
}
%
%  #] abstract : 
\setcounter{footnote}{0}
\newpage

%  #] heading :

% r1inro.tex

\thispagestyle{myheadings}\markright{}

\SECTION {Introduction.}
\label {intro}

\SUBSECTION {Asymptotic expansions of Feynman diagrams in applications.}
\label {fdappl}

%\PARAGRAPH {Role of pQFT.}

The stupendous accelerator projects of the present-day high-energy
physics (see e.g.\ \cite{rubbia}) have transformed perturbative
quantum field theory (pQFT) into a virtual
branch of engineering science---the complexity of particle scattering
events being such that obtaining physically meaningful information
from experimental data is impossible without extensive and systematic
calculations of radiative corrections, background processes etc.

%\PARAGRAPH {Complexity of Feynman diagrams: 
%number of parameters {\it vs} number of loops.} 

The objects of such calculations are Feynman diagrams which are,
essentially, multiple integrals depending on the dimensional
parameters of the process under consideration (momenta and masses of
both observable and virtual particles). The number of parameters is
determined by the specific problem and experimental setup while the
number of integrations, by the number of ``loops'' in the
corresponding diagram which increases with the order of perturbation
theory.  The larger the number of parameters of a diagram, the harder
it is to calculate---even numerically, owing to the highly singular
nature of Feynman diagrams, cancellations of divergent terms due to UV
renormalization, cancellations characteristic of the gauge models of
particle interactions etc.

%\PARAGRAPH {Examples of calculations.}

If one deals with a problem without an intrinsic momentum scale (the
case of no dimensional parameters) one can systematically perform 3-
and 4-loop calculations, and with luck go as far as the 5 loop
approximation in models in 4 space-time dimensions%
\footnote{
cf.\ the 5-loop analytical calculation of renormalization group
functions in a scalar model \cite{5loops} with applications to the
phase transition theory \cite{crit}. In gauge models such calculations
are theoretically always possible through 4 loops \cite{fvt81},
\cite{mincer} and the 4-loop QED $\b$-function was calculated in
\cite{QED4loops}.  The algebraic complexity of non-abelian gauge
models has so far prevented one from going beyond the ground-breaking
3-loop result of \cite{QCD3loops}.  To this class also belong
calculations of 3-loop corrections to various sum rules for the deeply
inelastic lepton-hadron scattering \cite{DIS3loops}.  Still another
example of analytical 3-loop and numerical 4-loop calculations is the
anomalous magnetic moment of the electron in QED---see e.g.\
\cite{g-2}.
\relax}. 
On the other hand, if there are more than two parameters in the
problem, only tree-level and 1-loop calculations are possible (see
e.g.\ \cite{hoo-vel-1loop}, \cite{old-ver}), while already 2-loop
calculations are, as a rule, unfeasible by straightforward methods.

%\PARAGRAPH {Calculational simplifications via asymptotic expansions.}

One can see that a reduction of the number of parameters can result in
dramatic calculational simplifications. On the other hand, the only
way to achieve such a reduction is via an asymptotic expansion with
respect to some of the parameters when they are considered as being
much less or much larger than the others. Even if one deals with a
problem involving several parameters, one can obtain valuable
information by first studying the behaviour of the corresponding
amplitude near boundaries of the phase space.%
\footnote{
Note e.g.\ that determining singularities of an amplitude for some
momentum going on mass-shell is a special case of an asymptotic
expansion near a singular point in the phase space of the problem.
\relax}

From a more theoretical point of view, considering amplitudes and
Green functions (represented as perturbative collections of Feynman
diagrams) in various asymptotic regimes allows one to simplify
theoretical description of the underlying physics and separate
dependences that can be calculated within perturbative framework from
strictly non-perturbative contributions. The prime example here is the
famous parton model \cite{parton} which, with appropriate
modifications (see e.g.\ the review papers \cite{rad:rev},
\cite{factoriz}), can be regarded as a summary description of the
``leading-twist'' terms of expansions of the corresponding amplitudes
for a special class of asymptotic regimes (high-momentum-transfer,
short-distance and/or light-cone etc.\ regimes) within perturbative
Quantum Chromodynamics (QCD).

Given that even UV renormalization can be regarded from the technical
point of view as subtraction of some asymptotic terms from the bare
integrand%
\footnote{
Different representations of the \rop\ exhibiting this aspect of UV
renormalization were presented in \cite{dslavnov-as} and
\cite{gms}; see also \cite{rmp2}.
\relax},
one cannot fail to conclude that {\it the technique of asymptotic
expansions of Feynman diagrams with respect to their
parameters---masses and momenta---is a central one in the
computational arsenal of applied QFT}.

However, despite the efforts of many theoreticians over many years and
the obvious importance of the asymptotic expansion problem within
pQFT, its complete and explicit solution remains a major theoretical
challenge.

\SUBSECTION {Theory of asymptotic expansions in pQFT. State of the art.}
\label{asinappQFT}

Asymptotic expansions of Feynman diagrams with respect to momenta and
masses in applied QFT were studied by many authors, the existing
literature is enormous, and compiling a complete list of references on
the subject is well beyond the scope of the present paper.  Therefore,
we will limit our discussion to general trends and the most important
contributions.  Suffice it to mention that ``the tricky subject of
asymptotic behaviour''%
\footnote{
p.67 in \cite{chew}.
\relax}
(usually, in the so-called Regge, or large-$s$, limit) was extensively
(and inconclusively) investigated in the context of analytical
properties of scattering amplitudes in the 50's and 60's (see e.g.\
\cite{chew}). 

In applied QFT, one deals with quantum-mechanical amplitudes of
scattering processes---or, more generally, with Green functions of
elementary or composite operators which depend on momentum and mass
parameters.  Such amplitudes and Green functions are represented as
infinite sums over a perturbative hierarchy of Feynman diagrams, and
the initial problem falls into two parts.

First, one has to expand individual Feynman diagrams with respect to
external parameters---masses and momenta.  This is the analytical part
of the problem:

Power-and-log bounds for the leading behaviour of multiloop diagrams
in a large class of high-energy limits were established by S.~Weinberg
{}\cite{wein60}. On the other hand, in the more practical context of
the studies of the Regge limit, the so-called techniques of the
leading logarithm approximation was developed to a considerable extent
by the Leningrad school (see e.g.\ \cite{lead-logs}; see also
\cite{sudakov}).  Another approach \cite{efr-ginz}, \cite{rad:rev}
used Mellin transforms of multiloop diagrams in parametric
representations%
\footnote{
Formal aspects of such techniques were studied by the French school
\cite{french}.
\relax}. 
Variants of the leading logarithm approximation techniques were used
in the context of the so-called QCD factorization theorems
\cite{factoriz}. All that research was oriented towards investigation
of specific physical problems and obtaining concrete formulae for
description of physical phenomena while formal rigour, naturally,
remained a secondary issue.

An important class of results of a more formal character concerned the
analytical nature of the asymptotic expansions. Thus, D.~Slavnov
\cite{dslav73} proved that for a wide class of asymptotic regimes
expansions of individual multiloop diagrams run in powers and
logarithms of the expansion parameter. That result was
extended to other asymptotic regimes using various techniques in
\cite{french}--\cite{kosins-masl}. Expansions that involve simplest
functions of the expansion parameter (powers and logarithms) are
exactly what is ultimately needed in applications%
\footnote{
A detailed discussion of the requirements that should be satisfied for
an expansion to be phenomenologically useful, can be found in \cite{fvt91}.
\relax},
and even such a limited information on the analytical form of expansions
can be useful%
\footnote{
cf.\ an early calculational algorithm for coefficient functions of
operator product expansions \cite{gluing:82}.
\relax}. 
But attempts to obtain explicit closed expressions for expansions of
Feynman diagrams in powers and logarithms of the expansion parameter
for general asymptotic regimes have so far failed---except for the
case of the so-called {\it Euclidean regimes} where solution has been
first obtained \cite{fvt82}--\cite{IV} using novel methods which it is
the purpose of the present paper to formally describe. 

The main difficulty with asymptotic expansions of multiloop diagrams
is that formal Taylor expansions of the integrand result in
non-integrable singularities. This indicates that the integrals depend
on the expansion parameter non-analytically.  There are, as has become
clear \cite{paradigm}, \cite{fvt91}, two alternative ways to approach
the problem based on two diametrically opposite points of view on the
nature of the expansion problem.  The papers
\cite{wein60}--\cite{kosins-masl} followed, with variations, the old
idea of splitting integration space---whether in momentum or
parametric representations---into subregions in such a way as to allow
one to extract the non-analytic (usually logarithmic) contribution
from each subregion, eventually, by explicit integration.  However,
complexity of multiloop diagrams exacerbated by ultraviolet
renormalization made obtaining convenient closed expressions for
coefficients of such expansions unfeasible except for the leading
terms \cite{lead-logs}, \cite{rad:rev}, \cite{factoriz}.  The
alternative approach is based on the observation \cite{fvt82} that the
expansion problem in pQFT is of the distribution-theoretic nature (see
below \sects{CoordinateRepresentation}-{MomentumRepresentation}), and
makes a systematic use of the recursive structures inherent in
perturbative collections of Feynman diagrams, which allows one to
efficiently organize the reasoning and arrive at the results in a
deterministic fashion (see below \ssect{MethodOfAsop}).

\SUBSECTION {Combinatorial aspect and the BPHZ paradigm.}
\label{combinat}

The second, the combinatorial part of the perturbative expansion problem
addresses the issue of the global form of expansions of the entire
infinite collection of Feynman diagrams contributing to the amplitude
under consideration. Here one should take into account that Feynman
diagrams form a recursively organized set since they are typically
obtained by, say, iterating the Schwinger-Dyson equations. 

Among the first to realize that asymptotic expansions of
non-perturbatively defined objects (e.g.\ Green functions) should have
a non-perturbative form (i.e.\ be representable in terms of other
Green functions etc.) was K.~Wilson \cite{wilson:69} who introduced
the notion of operator product expansion (OPE) for the simplest kind
of asymptotic regimes (``short distance'' regimes that are equivalent
to Euclidean large momentum regimes).  It was also realized that the
short-distance OPE is closely related to other expansion problems
e.g.\ the decoupling of heavy particles in the low-energy effective
Lagrangian \cite{decoupl}, the problem of low-energy effective
Lagrangians of weak interactions \cite{effL}; in fact, the
short-distance OPE and related problems constitute a subclass of
Euclidean regimes studied in the present paper.

A different type of results in this direction is due to
L.~Lipatov \cite{lipa} who described the Regge behaviour in the
leading logarithm approximation using what is now known as Lipatov
equation.  Further examples are the so-called factorization theorems
for inclusive hadronic processes (see \cite{rad:rev}, \cite{factoriz}
and refs.\ therein) and the results on the leading asymptotic
behaviour of exclusive hadronic processes \cite{exclus}.  Again, as in
the case of individual diagrams, the form of the expansions beyond the
leading term remains unknown except for the case of short-distance
regime (Wilson's OPE) and its generalization---Euclidean regimes.

The first accurate proof of this type of results is due to
W.~Zimmermann \cite{zim70}, who used what became known as the
Bogoliubov-Parasiuk-Hepp-Zimmermann (BPHZ) techniques (see e.g.\
\cite{bog-shir}, \cite{hepp69}, \cite{zav79}) and demonstrated that it
is indeed possible to construct an OPE for a class of short-distance
asymptotic regimes.

The main achievements of \cite{zim70} were, first, the demonstration
of how the expansion in a ``global" OPE form is combinatorially
restored from terms corresponding to individual Feynman diagrams;
second, the required smallness of the remainder of the expansion was
proved in presence of UV renormalization.  However, unlike the
expansions obtained in \cite{dslav73}, the coefficient functions of
the OPE of \cite{zim70} were not pure powers and logarithms of the
expansion parameter (i.e.\ short distance or large momentum transfer)
but also contained a non-trivial dependence on masses of the particles
(apparently because masses were used as regulators for infrared
divergences). Therefore, although the results of \cite{zim70} are
firmly established theorems, they fell short of providing an adequate
basis for applications (for a more detailed discussion see
\cite{fvt91}): first, the expressions for coefficient functions were
unmanageable from purely calculational point of view%
\footnote{
This explains why the first large-scale calculations of OPE beyond
the tree level \cite{bardeen:78} were performed by ``brute force": the
coefficient functions of an OPE were found by straightforward
calculation of asymptotics of the relevant non-expanded amplitudes and
then by explicit verification of the fact that the asymptotics have
the form which agrees with the OPE ansatz.  The more sophisticated
methods of \cite{alg:83} were discovered \cite{fvt:ope83} outside the
BPHZ framework---using the ideas described in \cite{fvt91} and
systematically discussed in the present paper.
\relax};
second, infinite expansions could not be obtained in models with
massless particles, e.g.\ QCD; third, the logarithms of masses
contained in coefficient functions are---as became clear later
\cite{fvt:mass}---non-perturbative within QCD in the sense that they 
cannot be reliably evaluated within perturbation theory using
asymptotic freedom. All the above drawbacks have the same origin: the
lack of the property known as {\it perfect factorization\/}
\cite{fvt:ope83} (for a detailed discussion see \cite{fvt91})
which at the level of individual diagrams is equivalent to
strictly power-and-log form of the expansions.

Lastly, although the BPHZ techniques can be useful as an instrument of
verification of the results found by other methods%
\footnote{
Recall that existence of OPE was guessed \cite{wilson:69} using the
heuristics of coordinate representation; the more general results of
\cite{fvt82}-\cite{IV} on Euclidean asymptotic expansions were
discussed within the BPHZ framework in \cite{gorishny}--\cite{ch:mpi}.
\relax},
its heuristic potential turned out to be inadequate: it proved to be
of little help in finding {\it new\/} results.  Indeed, both the
formula of the Bogoliubov $R$-operation \cite{bog-shir} and the OPE
were discovered using heuristics that are foreign to the BPHZ method,
while the attempts to obtain full-fledged OPE-like results (i.e.\ to
construct an appropriate generalization of the Zav'yalov-Zimmermann
forest formula \cite{zav79}, \cite{zim70}) for a wider class of
asymptotic regimes (non-Euclidean regimes; cf.\ e.g.\ the Regge limit
mentioned above) have so far largely failed.  Given the central role
of the expansion problem in pQFT, the importance of finding better
ways to deal with multiloop Feynman diagrams than those suggested by
the BPHZ paradigm, is indisputable.

\SUBSECTION {Theory of asymptotic operation.}
\label{MethodOfAsop}

One such potentially more powerful alternative emerged
\cite{fvt82}--\cite{IV} as a formalization of the reasoning behind the
algorithms used in the mentioned series of multiloop calculations
\cite{5loops}--\cite{QED4loops}, \cite{DIS3loops}.
It provided a fully satisfactory solution of the expansion problem for
the class of Euclidean asymptotic regimes, including efficient
calculational formulae for coefficient functions of OPE \cite{alg:83}
and for renormalization group calculations
\cite{r*}, \cite{kuz-fvt-3}.  
The derivation in \cite{fvt82}--\cite{IV} employed a novel mathematical
techniques based on the concept of asymptotic expansions in the sense
of distributions.  Ideologically, the abstract extension principle
\cite{fvt82} provided a deterministic recipe for concrete construction
of such expansions.  Technically, a systematic use of the techniques
of the distribution theory allowed one to make efficient use of the
recursion structures in the hierarchies of Feynman diagrams.
Algorithmically, the key technical instrument here was the so-called
{\it asymptotic operation} ($As$-operation) for products of singular
functions%
\footnote{
The term \asop\ was first applied \cite{piv-fvt84} to a formula for
expansions of integrated Feynman diagrams.  The two flavours of
\asop\ are closely related: the \asop\ for integrals is an integrated
version of the \asop\ for products.
\relax}. 
Combinatorially, uniqueness of \asop\ \cite{fvt91} ensures that
expansions of individual diagrams inherit structural properties of the
non-expanded integrals, so that global factorization/exponentiation,
gauge properties etc.\ can be studied in a rather straightforward
fashion. As has become clear \cite{paradigm}, the techniques based on
the \asop\ offers a comprehensive, flexible and powerful alternative
to the conventional methods.

The aim of the present paper is to present a systematic and
consolidated description of the mathematical part of the theory of
asymptotic operation as developed in \cite{fvt-vvv}--\cite{IV}.  Its applications to
the theory of multiloop diagrams will be discussed in a companion
publication \cite{rmp2} (see also \cite{gms}--\cite{IV}).

From the point of view of mathematics, the theory of \asop\ is a theory of
asymptotic expansions of products of singular functions with respect
to one (or more) parameter in the sense of distributions. We believe
it deserves a separate discussion for several reasons.

{\bf First,} the level of formal rigour attained in the BPHZ
theory---widely known, widely acknowledged as being completely
``rigorous'', and the most successful by far among the mathematically
oriented approaches to the asymptotic expansion problem in
pQFT---establishes a de facto standard to which any alternative
techniques must conform.  Moreover, taking into account the overall
complexity of the problem and the fact that the BPHZ theory evolved
over many years in many papers%
\footnote{
cf.\ e.g.\ \cite{zav79} and refs.\ therein.
\relax},
it is clear that mathematics of multiloop diagrams is more than just a
technical appendix to perturbative QFT: the theory of \asop, arguably,
plays the same role in pQFT as the theory of ordinary differential
equations does in classical mechanics.

{\bf Second,} 
the original derivation of
\cite{fvt82}--\cite{kuz-fvt-3} made a heavy use of the techniques of
dimensional regularization \cite{dreg} since it was aimed at obtaining
practical formulae in a shortest way and the dimensional
regularization provides the most convenient framework for practical
calculations. However, the use of dimensional regularization is by no
means essential for the theory of
\asop, and it is important to understand the latter in a 
regularization independent way. Further interest to this point is
added by potential importance of supersymmetric models of elementary
particles at high energies \cite{rubbia}---and
supersymmetry is incompatible with dimensional regularization%
\footnote{
Moreover, there are indications \cite{coll-ellis} that dimensional
regularization is not an appropriate tool for studying
non-Euclidean asymptotic regimes.
\relax}.
This makes it necessary to work out the theory of \asop\ in a
regularization independent manner and in as mathematically general
form as possible, in order to accomodate various asymptotic regimes and
prepare ground for the non-Euclidean generalizations.

{\bf Third,} the theory of \asop\ adds a new flavour to the theory of
distributions. To our knowledge, there exists no other
systematic study of asymptotic expansions in the sense of
distributions%
\footnote{
although special cases of such expansions did occur both in theory and
applications---explicitly or implicitly---see e.g.\ eq.(46) in
\cite{ellis}, as well as \cite{pohl:82}. 
Which only proves our point that a full-fledged theory of such
expansions is both a theoretical and calculational necessity.
\relax}.
Moreover, the extension principle which constitutes a basis for such
studies (see below \sect{ExtensionPrinciple}) is interesting in that
it considers the problem of a distinctly applied nature (asymptotic
expansions) at a very general level of abstract functional analysis.
Also, asymptotic expansions of multiloop integrals provide an
excellent example of a problem that can be formulated entirely within
classical integral calculus, but whose complete solution can only be
meaningfully discussed in terms of the distribution theory.  On the
other hand it is quite interesting that an essentially linear theory
of distibutions can be so effectively applied to a study of non-linear
objects like products of singular functions. All this, we believe,
makes the theory of \asop\ interesting from a purely mathematical
point of view.

{\bf Fourth,} 
obtaining asymptotic expansions in pQFT for general
non-Euclidean asymptotic regimes in a perfectly factorized form (in the
sense of \cite{fvt91}) remains a major theoretical
chal\-len\-ge---attempts to derive them using the BPHZ techniques have so
far been unsuccessful%
\footnote{
The so-called light-cone OPE derived in \cite{anik-zav} suffers from
the usual drawbacks of the BPHZ-type results---most notably, the lack
of perfect factorization. Moreover, it is not immediately useful in
applications since the somewhat simplified light-cone asymptotic
regime in coordinate representation studied in \cite{anik-zav} is not
directly connected to phenomenological problems although some of the
mathematical structures found in \cite{anik-zav} (namely, light-cone
operators) have a general significance. Still, there is a long way to
go before a satisfactory solution to the non-Euclidean asymptotic
expansion problem will be found, and there are all indications that the
BPHZ approach is not the optimal one for that.
\relax}.
On the other hand, the deterministic recipe of the extension principle
and a highly flexible techniques of the distribution theory (cf.\
\sect{As-expansion} below) offer an intriguing opportunity to 
attack the non-Euclidean asymptotic expansion problem in a rather
general context \cite{workinprogress}.

{\bf Lastly,} unlike existence theorems, calculation-intensive
applications require one to be in a complete command of the details of the singularity structure of
expressions one has to deal with. The systematic formalism
we develop should be helpful in this respect, too.

\SUBSECTION {Plan of the paper.}
\label{plan}

The paper consists of sections grouped into 6 parts and subdivided
into subsections (marked with \S). Each part and section starts with a
preamble with information on its content and further comments.

Part I (\sects{CoordinateRepresentation}-{MomentumRepresentation})
contains an informal discussion of why the use of the distribution
theory is dictated by the very nature of the problems of perturbative
QFT.  First in \sect{CoordinateRepresentation} we consider the
(Euclidean) theory of UV renormalization in coordinate representation
to illustrate the dilemma one faces when dealing with multiloop
diagrams, as well as two ways to resolve it---one that follows 
Bogoliubov's seminal paper \cite{bog52} on the origin of UV
divergences and treats the problem from the point of view of
functional analysis as a problem of extension of functionals, and the
other one typical of the BPHZ theory.  In
\sect{MomentumRepresentation} we consider the asymptotic expansion
problem in pQFT and demonstrate its distribution-theoretic character,
which makes it to a considerable extent similar to the theory of \rop.
We also exhibit numerous recursive patterns within the expansion
problem. Our discussion provides, hopefully, sufficient
motivation for the formalized mathematical constructions of
subsequent sections.

In part II (\sects{Graphs}-{CompleteSubgraphs}) we develop a formalism
for description of products of singular functions (``universum of
graphs'') which allows one to make efficient use of the ubiquitous
recursive patterns of the theories of $R$- and \asop s. The crucial
notion of complete singular subgraphs ($s$-subgraphs) is introduced
using universal analytical language. When applied to specific problems
involving Feynman diagrams, it transforms into various types of
``subgraphs'' encountered in pQFT---the UV subgraphs in the original
Bogoliubov's definition \cite{bog-shir}, and those in the standard
BPHZ definition (used e.g.\ in the formulation of the MS scheme
\cite{MS}; see also \cite{zav79}), as well as various diagrammatic 
kinds of IR subgraphs in the context of the theory of \asop. 
The central issue addressed in this part of the paper is
geometrical classification of singularities of products of singular
functions and the corresponding classification of subproducts.

If part II mostly deals with geometry of singularities, part III
(\sects{Decompositions}-{Subtractions}) studies their analytical
structure and develops the corresponding analytical tools.  First in
\sect{Decompositions} various decompositions of unit are introduced
which are used to reduce the study of singularities localized on
manifolds to singularities localized at isolated points.  Then in
\sect{IsolatedSing} very useful formalisms of $\L$-functions and
$\cS$-inequalities are developed for description of singularities from
the analytical point of view.  The techniques of $\cS$-inequalities
is important because it exhibits the essential similarity of proofs
between the cases of distributions and ordinary functions%
\footnote{
Recall in this respect that every distribution is 
a derivative of a continuous function. It would be interesting to 
clarify this connection.
\relax}.
In \sect{Subtractions} we study the so-called subtraction operators to
deal with isolated singularities. Three variants of subtraction
operators (generic, with oversubtractions, and ``special'' ones that
preserve scaling properties to a maximal degree) are considered in the
context of the problem of extension of functionals.

Part IV (\sects{R-operation}-{Variations}) considers a general
definition and structure of $R$-like operations (\sect{R-operation}).
Then in \sect{GenericR} a generic (Bogoliubov-Parasiuk) \rop\ is
defined and a very compact proof of the corresponding existence
theorem is presented (compactness and an algebraically explicit
character of our proof are a spin-off of the stringent requirements
imposed on the formalism by the more demanding theory of asymptotic
expansions and \asop). In \sect{SpecialR} we consider the so-called
{\it special\/ \rop} (\trop, in formulae denoted as $\tR$) that will
play an important role in a subsequent construction of the \asop. It
differs from the generic one in that it preserves scaling properties
of the products to which it is applied. Finally, in \sect{Variations}
we discuss the structure of
\rop s in terms of finite and infinite counterterms.

In the ideologically central part V
(\sects{As-expansion}-{ExtensionPrinciple}) we first consider
definitions of asymptotic expansions in functional spaces
(\sect{As-expansion}) and study their general properties. The
considerable flexibility of the techniques is only partly utilized in
the theory of Euclidean asymptotic expansions---its potential is fully
realized in the non-Euclidean theory \cite{workinprogress}.
\Sect{As-operation} introduces the notion of \asop\ for products of singular 
functions and exhibits its structure. \Sect{ExtensionPrinciple}
considers the problem of its existence at a very abstract level (the
extension principle) which allows one to complete the definition of
the \asop.

Part VI (\sects{ExpansionObject}-{MainProof}) is devoted to proving
concrete inequalities that guarantee correctness of the construction
of the \asop. In \sect{ExpansionObject} a precise description of the
products to be expanded is given, \sect{FormalExpansion} studies
singularities of formal Taylor expansions, in \sect{Existence} our
results on the existence, explicit formulae, and properties of the
\asop\ are summarized, while \sect{MainProof} contains the
technical part of the proofs.

In the companion text \cite{rmp2} we apply the results of the
mathematical theory of \asop\ to problems of a more immediate
physical interest.  Fisrt, following \cite{gms}, we will consider the
problem of UV renormalization from the point of view of asymptotic
behaviour of integrands in momentum representation---it turns out that the
\rop\ can be represented in an explicitly convergent form as a
subtraction from the integrand of its UV asymptotics. Such a
representation is closely connected to the MS scheme \cite{MS}.
In particular, we will see that the non-trivial coefficients
$\tilde{E}_a$ of the \asop\ (cf.\ eqs.\eqs{e2/9.2}-{e2/9.5} below) are
exactly UV renormalized subdiagrams of the diagram whose integrand is
being expanded.  Second, we will show (following \cite{IV}) how the
problem of asymptotic expansions of renormalized diagrams is reduced
to a study of double \asex s, and present the corresponding
generalizations of our theorems. We will see that our formalism
immediately results in convenient, compact and explicit formulae for
expansions of renormalized diagrams, so that the combinatorial part of
the derivation of factorization/exponentiation/OPE etc.\ for the
entire perturbation series, becomes a rather simple excercise.

{\hbox{\ }}

\newpage
\noindent{\it Acknowledgements.}

One of us (F.~T.) 
thanks J.~C.~Collins and H.~Grotch for their hospitality at the Penn
State University where part of this work was done, E.~Kazes for his
lively interest in the subject and bringing the work by Epstein and Glaser \cite{EpGlas} to our attention, the CTEQ collaboration for
financial support, and H.~Epstein for a discussion.  
We thank D.~A.~Slavnov for a discussion of our work and advice. 
This work was supported in part by the Russian Foundation for Basic Research under grant 95-02-05794.

\newpage\thispagestyle{myheadings}\markright{}
\PART {Feynman Diagrams and Distributions: Examples and Motivations.}

In the following two sections we will consider simple
examples---first, from the theory of {UV} renormalization in
coordinate representation (\sect{CoordinateRepresentation}) and then
from the theory of asymptotic expansions of multiloop diagrams
(\sect{MomentumRepresentation}). Our aim will be to provide
motivations for the general formalism that we are going to develop, as
well as to convince the reader that the nature of multiloop Feynman
diagrams can be completely understood only from the point of view of
the distribution theory. In both cases we will demonstrate the central
dilemma of the theory (recursive structure of the problems {\it vs}.\
singular nature of the participating objects) and how it is resolved
within the conventional framework and in the new formalism that relies
on the distribution theory in order to make full use of the recursive
structures.

\SECTION{Feynman diagrams in coordinate representation.}
\label{CoordinateRepresentation}

Bogoliubov's coordinate-representation arguments that led him to the
discovery of the \rop\ \cite{bog52} were the first example of
distribution-theoretic reasoning in the theory of Feynman diagrams.%
\footnote{
Apparently, the distribution theory proper (in the final form due to
L.~Schwartz \cite{sch52}; see also \cite{gel-shi}, \cite{hormander}
and the more elementary texts \cite{sch-methodes} and \cite{vsvlad})
was unknown to Bogoliubov in 1952 when his first key paper on the
theory of renormalization in coordinate representation appeared
\cite{bog52}.  Bogoliubov was, however, aware of an earlier version of
the theory of ``generalized functions'' due to L.~Sobolev
\cite{sob35}, which provided enough insight into the nature of the
problem to allow him to write down the \rop\ with correct
subtractions. But, from technical point of view, Sobolev's theory was
geared to a special class of problems in the theory of differential
equations and lacked several basic notions (localization of
distributions and localized test functions) which make the
distribution theory so powerful.  Therefore, for the purposes of
rigourous justification, Bogoliubov and Parasiuk chose to switch to a
technique based on parametric representations that later evolved into
what is known as BPHZ theory.  Another reason for such a choice was
that using a parametric representation allowed one to avoid having to
deal with the light-cone singularities of propagators.  Such
singularities are harmless in the context of theory of UV
renormalization, but a convenient techniques to deal with them was
only introduced by H\"ormander and Sato in the 60's
(see \cite{reed-sim}).
\relax}
Such a reasoning does not fit into the paradigm of the BPHZ
theory \cite{hepp69}, \cite{zim70}, \cite{zav79}
which for a long time dominated ``rigorous'' theory of multiloop
diagrams.  But it is such reasoning that exhibits the essential
features of the \rop\ most clearly and is of great heuristic and
didactic value.  Therefore, we find it instructive to consider the
theory of UV renormalization in Euclidean coordinate representation
as a useful example to illustrate our techniques.

It should be stressed that the distribution-theoretic interpretation
of the \rop\ in coordinate representation is as old as the
Bogoliubov's paper \cite{bog52} and has since been discussed not
once.%
\footnote{
See e.g.\ Chapter 5 in \cite{bog-shir} and the remarks in
\cite{reed-sim}.
\relax}
From such a point of view our results offer only technical
improvements (although the compactness of our proof of the localized
version of the BPH theorem---see 
\ssects{GenericR.Proof.Plan}-{GenericR.Proof.PowerCounting}---seems 
to be hard to beat) and we by no means wish to develop a complete
theory of \rop\ in coordinate representation which should cover many
topics including renormalization in models with massless particles and
IR singularities---the coordinate space approach is not the
most appropriate one for that anyway, since concrete physical problems are
normally formulated in momentum representation. The main emphasis of
our work is on the theory of asymptotic expansions of Feynman
diagrams in
momentum representation which requires more subtle techniques than what would have been sufficient should we limit ourselves to the analysis of UV
renormalization in coordinate representation. 

Another remark is that Bogoliubov's distribution-theoretic construction of perturbative QFT that served as a starting point for the research that led to the theory of asymptotic operation, was discussed from mathematical point by Epstein and Glaser \cite{EpGlas}. However, the emphasis of is quite different from our work: the discussion of ref.\ \cite{EpGlas} focuses on chronological products of operators, accurate treatment of the causality condition, and construction of perturbation series for the $S$-matrix operator. Our aim, on the contrary, is to clarify the universal ``microscopic" analytical mechanism of the $R$-operation that is independent of any details specific to operators, and to understand it in a form useful for the problem of asymptotic expansions of Feynman diagrams.

\SUBSECTION{UV renormalization in coordinate representation.}
\label{Coordinate.UV-renormalization}

The coefficient function of a Euclidean%
\footnote{
We limit our discussion to the Euclidean case to avoid unnecessary
complications due to light-cone singularities. It would be more
appropriate to consider the latter in the context of studying the
asymptotic expansion problem for the most general non-Euclidean
asymptotic regimes to which a separate publication will be devoted.
\relax}
Feynman diagram $G$ in coordinate representation can be described as
follows (for simplicity we here consider only scalar models and
interactions without derivatives). Let $v$ enumerate the vertices of
the diagram, then $x_{v}$ is the corresponding coordinate which runs
over $D$-dimensional Euclidean space (the coordinate for one of the
vertices is usually set to $0$). The coefficient function is a product
of factors of the form $\D_{F}(x_{v_1}-x_{v_2})$ where $\D_{F}$ is the
standard Feynman propagator: $\D_{F}(x) \sim (x^{2})^{(2-D)/2}$ as
$x\to 0$ for $D > 2$, or $\log x^{2}$ for $D = 2$. If the theory to
which corresponds the diagram under consideration, involves non-scalar
particles and/or interactions with derivatives then the functions
$\D_{F}$ are replaced by various derivatives of the scalar propagator,
so that the singularity indices of different factors become different.
One normally considers cases $D = 2,3,4$ or $6$.  Occasionally, it is
convenient to consider toy models of such products in $D = 1$.

To obtain physical amplitudes, in the final respect, one has to
evaluate Fourier transforms of such products with respect to some
$x_{v}$ and perform integrations in infinite limits over the other.
The theory of UV renormalization in coordinate representation
considers the problem of local integrability of such
products.

To this end, one studies singularities of such products due to
singularities of individual factors. It is not difficult to present a
rather typical example where the overlapping of singularities of
individual factors generates non-integrable singularities for the
product as a whole.  Let $x$ and $y$ be 4-dimensional
Euclidean variables. The example is:
\be[Eq.Example1]
   {1\over x^{2}} {1\over (x-y)^{2}} \left({1\over y^{2}}\right)^2 .
\ee
The corresponding Feynman diagram is shown in Fig.~1a.  Each line
of Fig.~1a corresponds to one factor in \eq{Eq.Example1} which
depends on the difference of arguments associated with the vertices
connected by the line.  

The geometry of singularities of such products can be best studied if
one considers $x$ and $y$ as components of an aggregate 8-dimensional
variable $(x,y)$. The pattern of singularities is schematically shown
in Fig.~1b. It is easy to see that the singularities localized at the
plane $y = 0$ and the point $x = y = 0$, are non-integrable by power
counting.

The crucial observation (essentially due to Bogoliubov
\cite{bog52}) is that, as far as local properties of
\eq{Eq.Example1} are concerned, fundamental physical restrictions such
as unitarity and causality only require that the coefficient functions
of Feynman diagrams coincide with expressions like \eq{Eq.Example1}
away from the singularities, and that there is no physical requirement
that would forbid one to modify such expressions at their singular
points so as to make them locally integrable with smooth test
functions.  In modern language, for all physical purposes it is
sufficient to define coefficient functions of diagrams as
distributions coinciding with expressions \eq{Eq.Example1} at
non-singular points. Note that the expression \eq{Eq.Example1} only
describes a distribution that is well-defined on a subspace of test
functions that have zeros of sufficiently high order at the points of
singularities of \eq{Eq.Example1}.  Bogoliubov \cite{bog52} explicitly
phrased the problem in terms of extension of functionals
(distributions, in modern language) corresponding to \eq{Eq.Example1}
from the subspace of such test functions to the entire space of
arbitrary test functions.  At a practical level, the solution amounts
to adding counterterms localized at the singular points of
\eq{Eq.Example1}---a procedure known as the Bogoliubov \rop\
\cite{bog-shir}.

\SUBSECTION {Recursive aspect of the problem.}
\label{Coordinate.Recursive}

The axioms that should be satisfied by the distributions to be
constructed from the formal singular expressions \eq{Eq.Example1}, are
described in \cite{bog-shir}. A more formal version for the Euclidean
case, which we will have in view, was presented in \cite{lang-liesn}.
One of the axioms---the microcausality condition---exhibits an
important aspect of the theory of the \rop, namely, that one deals not with
a single diagram but with a hierarchy of such diagrams, and that the
\rop s on different members of the hierarchy should be correlated. 
Microcausality defines the \rop\ up to the so-called subtraction
operators, and has an explicitly recursive form. In the case of our
simple example \eq{Eq.Example1} it looks as follows.

First of all, besides the diagram \eq{Eq.Example1} itself, one has to
define the \rop\ on the diagram of Fig.~2a whose analytical
expression is
\be[Eq.Example1.Subgraph]
     \left({1\over y^{2}} \right)^2 .
\ee
It corresponds to a subdiagram of the diagram of Fig.~1a, and the \rop\
on \eq{Eq.Example1.Subgraph} can be written in a simple form:
\be
     \R\. \left({1\over y^{2}} \right)^2
   = \r\. \left({1\over y^{2}} \right)^2 ,
\ee
where $\r$ is defined on any distribution $F(y)$ with a logarithmic
singularity at $y=0$ as follows:
\be[Eq.Example1.Subgraph.r-Def]
     \int d^{4}y\, [\r\. F(y)] \,\vf (y)
     \bydef  \int d^{4}y\, F(y) \,
     \left[ \vf (y)-\Phi(\mu y)\vf (0) \right],
\ee
where $\Phi$ is a smooth cutoff function that is equal to $1$ in a
neighbourhood of $y=0$ and behaves at large $y$ so as to ensure
convergence of the integral (it can e.g.\ vanish for all sufficiently
large $y$; note also that integration around $y=0$ is in general
understood in the sense of principal value with smooth cutoffs;
precise definitions can be found in \sect{Subtractions}).  The fact
that the terms subtracted from the integrand on the r.h.s.\ of
\eq{Eq.Example1.Subgraph.r-Def} is proportional to $\vf (0)$, shows
that $\r$ modifies the distribution on which it acts only at $y=0$.

The latter fact can be demonstrated more explicitly: if one introduces
an appropriate regularization into $F(y)$ on the r.h.s.\ of
\eq{Eq.Example1.Subgraph.r-Def} so as to make the integrals of the two
terms exist separately, then the definition
\eq{Eq.Example1.Subgraph.r-Def} can be rewritten in a form with
``infinite counterterms" (cf. \ssect{Append.Regularization},
eq.\eq{eA.1}):
\be[rGZ]
     \r\. F(y) \sim  F(y) + Z \,\delta (y),
\ee
where $Z$ diverges as the regularization is removed. Note that, as a
general rule, $Z$ is not determined uniquely: one can always add to it
a finite constant---which corresponds to the arbitrariness in the exact
form of the cutoff $\Phi$.

The \rop s on the two expressions \eq{Eq.Example1} and
\eq{Eq.Example1.Subgraph} are related by the following Euclidean
version \cite{lang-liesn} of the microcausality condition: 
\be[Eq.Example1.Locality]
     \R\. \left\{
       {1\over x^{2}} {1\over (x-y)^{2}}
       \left(
           {1\over y^{2}}
       \right)^2
     \right\}
= {1\over x^{2}} {1\over (x-y)^{2}}
  \R\. \left\{
         \left(
            {1\over y^{2}}
         \right)^2
       \right\},
   \quad x\not=0 \hbox{\ or\ } y\not=0,
\ee
which says that away from the singular point $x=y=0$, the l.h.s.\
factorizes into renormalized subdiagram of Fig.~1 times the remaining
factors.  It is also implied that the \rop\ on the r.h.s.\ is
insensitive to the $x$-dependence.  The expression
\eq{Eq.Example1.Locality} can be taken as a definition of the \rop\ on
the l.h.s.\ everywhere except for the point $x=y=0$.  Such a
definition makes sense because the factors $x^{-2}(x-y)^{-2}$ are smooth
in small neighbourhoods around the points on the plane $y=0$ away from
$x=y=0$.

To complete the definition one should eliminate the remaining
singularity localized at an isolated point $x=y=0$.  This is important
because singularities at isolated points can be eliminated in a rather
elementary fashion using a straightforward modification of the
subtraction operators $\r$ introduced in
\eq{Eq.Example1.Subgraph.r-Def} (they will be studied in detail in
\sect{Subtractions} below) provided the expression to which a
subtraction operator is to be applied has a power-type singularity.
If the latter condition is true, then one can immediately write
\be[Eq.Example1.R-Def]
\R\. \left\{
       {1\over x^{2}} {1\over (x-y)^{2}}
       \left(
         {1\over y^{2}}
       \right)^2
     \right\}
=
\r\. \left\{
       {1\over x^{2}} {1\over (x-y)^{2}}
       \R\. \left\{
               \left(
                 {1\over y^{2}}
               \right)^2
            \right\}
      \right\},
\ee
with $\r$ defined analogously to \eq{Eq.Example1.Subgraph.r-Def}:
\be[Eq.Example1.Subgraph.r-Def2]
     \int d^4x\,d^{4}y\, [\r\. F(x,y)] \,\vf (x,y)
   = \int d^4x\,d^{4}y\, F(x,y) \,
     \left[ \vf (x,y)-\Phi(\mu x,\mu y)\vf (0,0) \right].
\ee

Thus one sees that the recursive nature of the \rop\ on the hierarchy
of Feynman diagrams allows one to recursively define the \rop\ in
terms of elementary subtraction operators $\r$.  It only remains to
justify the use of the operator $\r$ of the form
\eq{Eq.Example1.Subgraph.r-Def2} on the r.h.s.\ of
 \eq{Eq.Example1.R-Def}. 

\SUBSECTION {The BPH theorem from the point of view of
the distribution theory.}
\label{Coordinate.Distributions}

The key analytical point of the theorem of existence of the \rop\
(the Bogoliubov-Pa\-ra\-siuk-Hepp theorem \cite{bog-par},
\cite{hepp69}), \cite{bog-shir} is to determine whether the
singularity of the r.h.s.\ of
\eq{Eq.Example1.Locality} is of power-and-log type, and to determine
the corresponding power exponent explicitly, so that one could
determine the number of subtractions needed in an expression like
\eq{Eq.Example1.R-Def}.

From the point of view of the above analytical picture, the answer is
rather obvious. Indeed, a straightforward calculation shows that
\be
\r\.   \left(
         {1\over (\l y)^{2}}
       \right)^2
=
{1\over \l^{4}}
\r\. 
       \left(
        {1\over y^{2}}
       \right)^2
+ {\log  \l \over \l^{4}}\, \const \, \delta (y).
\ee
It follows immediately that the r.h.s.\ of \eq{Eq.Example1.Locality}
scales as $\l^{-4}\log \l$ under simultaneous scaling $x\to \l x$,
$y\to \l y$, plus terms $O(\l^{-4})$.  In other words, the singularity
at $x=y=0$ after subtraction of the singularity at $y=0$ is
modified only by a logarithmic correction as compared to the initial
expression and the naive power counting. It follows immediately that
the operator $\r$ on the r.h.s.\ of \eq{Eq.Example1.R-Def} should have
the form \eq{Eq.Example1.Subgraph.r-Def2}.

It is not difficult to understand that the same mechanism will work in
more complicated cases, so that the statement of the BPH theorem (at
least its critical part describing the structure and number of
subtractions) becomes entirely obvious. It is only a matter of a
careful choice of notations to exhibit the essentially trivial
power-counting nature of the BPH theorem in the most general case.

\SUBSECTION{The fundamental dilemma.}
\label{Coordinate.Dilemma}

Before we compare the above reasoning with the BPHZ approach, it is
worthwhile to dwell on the recursion described in
\ssect{Coordinate.Recursive}.

The relation \eq{Eq.Example1.Locality} exhibits the fundamental
dilemma of the theory of multiloop diagrams which emerges not only in
the theory of \rop\ in coordinate representation but also in the
general theory of asympotic expansions (\sect{MomentumRepresentation}
and \sect{As-operation}, eq.\eq{e2/2.7}). \Eq{Eq.Example1.Locality} has
an explicit and compact recursive structure which allows one to
construct the \rop\ in more complicated cases using solutions of
simpler ones. But one cannot benefit directly from such a recursion
unless one has at one's disposal a special mathematical techniques for
manipulating distributions in the recursive expressions like
\eq{Eq.Example1.Locality}---in particular, in order to perform the
power counting at $x,y\to 0$ taking into account the effect of
subtraction in the subdiagram. 

There are two ways to resolve the dilemma. The first is to follow the
nature of the problem and systematically use the reasoning similar to
the one presented in \ssect{Coordinate.Recursive}.  Transparency and
fundamental simplicity of such a reasoning makes it very appealing.
The catch, however, is that in order to be able to do so, one needs to
use the language and apparatus of the distribution theory
\cite{sch52}, \cite{gel-shi}, \cite{vsvlad}, \cite{hormander}.%
\footnote{Cf.\ the footnote at the beginning of 
\sect{CoordinateRepresentation}.
\relax}
The other alternative corresponds to the philosophy of the mainstream
BPHZ theory \cite{hepp69}, \cite{zim70}, \cite{zav79}. It consists in
a systematic reduction of the reasoning to the familiar techniques of
absolutely convergent integrals by expanding all the subtraction
operators according to their definitions
\eq{Eq.Example1.Subgraph.r-Def} and \eq{Eq.Example1.Subgraph.r-Def2}.

\SUBSECTION {The BPH theorem and the BPHZ theory.}
\label{Coordinate.BPH}

Within the BPHZ paradigm, one does not attempt to understand how the
formula for the \rop\ comes about. One simply attempts to prove
finiteness of the integrals obtained by expanding the operators $\r$.
For the expression \eq{Eq.Example1.R-Def} one would have:
%OBB
\be
     \int d^{4}x\, d^{4}y\,  e^{i(p\cdot x+q\cdot y)}
     \R\. \left\{
          {1\over x^{2}} {1\over (x-y)^{2}}
          \left(
             {1\over y^{2}}
          \right)^2
        \right\}
     \equiv
     \int d^{4}x\, d^{4}y\,
     \left(
       {1\over y^{2}}
     \right)^2,
\ee\be[Eq.Example1.MustBeFinite]
     \times\left\{
        \left( e^{i(px+qy)} - \Phi(\mu x,\mu y) \right) 
        {1\over x^{2}} {1\over (x-y)^{2}}
      - \Phi(\mu y) \left( e^{i(px   )} - \Phi(\mu x,    0) \right) 
        {1\over x^{2}} {1\over x    ^{2}}
     \right\}
\ee
where we have replaced the arbitrary test function $\vf (x,y)$ by a
typical exponent $e^{i(px+qy)}$, and expanded the operators $\r$ using
their definitions \eq{Eq.Example1.Subgraph.r-Def} and
\eq{Eq.Example1.Subgraph.r-Def2}. (One could also go to momentum
or parametric representation replacing the cutoffs $\Phi$ by
appropriate oscillating factors---that would further camouflage the
origin of the expression.) 

In order to prove convergence of \eq{Eq.Example1.MustBeFinite}, one
splits the integration region into the subregions:
$|y|<\frac{1}{2}|x|$ etc.\ (which are known as Hepp sectors
\cite{hepp69} in the context of parametric representations) and
constructs estimates for the integrand in each sector. We do not
attempt to present such estimates because they are quite cumbersome
even in our simple example.

The notorious complexities of the BPHZ theorem (the ``resolution of
singularities" using Hepp sectors and the forest formula needed to
handle the combinatorial entanglement of expressions
\eq{Eq.Example1.MustBeFinite}) are a logical consequence of such an
approach. The difficulties are further exacerbated by the use of
parametric representations which spoil the factorization properties of
the integrands, substantially contributing to complexity of proofs for
the models with non-scalar particles. Admittedly, parametric
representations are less sensitive to specifics of the UV
renormalization problem in its Minkowskian variant as compared to the
Euclidean case.  But the benefits achieved thereby are minor and
hardly justify the overall counterintuitive character of the approach
and the loss of the heuristic potential of the higher-level language
of the distribution theory and the defining recursion relations of the
\rop.

\SECTION{Feynman diagrams in momentum representation.}
\label{MomentumRepresentation}

Let us show that the same two characteristics---recursive structure
and distribution-theoretic nature---apply to the problem of asymptotic
expansions of multiloop diagrams in masses and momenta.  As
previously, we will consider only Euclidean diagrams, although the
reasoning remains essentially the same in the Minkowskian case.

\SUBSECTION{Distribution-theoretic nature of the expansion problem.}
\label{Momentum.Distribution}

Consider the following one-loop integral:
\be[Eq.Example2]
     \int d^{2}p\, {1\over (p^{2}+m^{2})} \, {1\over (p^{2}+M^{2})}.
\ee
(We take the integration momentum to be two dimensional in order to
avoid {UV} divergences, but this is inessential.) Consider expansion
of this integral at $m \ll M$.  Due to homogeneity of the integral
with respect to $m$ and $M$ one can choose to let $m\to 0$ with $M$
fixed. The problem one runs into is that the formal Taylor expansion
in $m$ of the integrand contains progressively more singular terms.%
\footnote{
If one were to expand in $M\to\infty$ with $m$ fixed, the problem would
be that the formal expansion of the integrand contains terms
non-integrable at $p\to\infty$.
\relax}
The problem can be resolved in two ways---which is similar to the two
ways to treat UV renormalization in coordinate representation
discussed in the previous section.

The traditional approach is to split the integration region into two
parts: a neighbourhood ${\cO}$ of the point $p=0$, and the complement
of ${\cO}$, and then to apply ad hoc tricks to explicitly
extract the non-analytical part of the expansion coming from the
integral over ${\cO}$. However, such a method quickly becomes far too
inefficient in more complicated cases of multiloop integrals.

The second approach is based on the following crucial observation.
Indeed, in applications there are many specific problems of this kind.
For example, one can consider a class of integrals with the same
$m$-dependent factor but with different second factors, e.g.:
\be
     \int d^{2}p\, {1\over (p^{2}+m^{2})}\, {1\over (p-Q)^{2}},
\ee
where $Q$ is an external momentum such that $Q^{2} \gg m^{2}$; another
example corresponds to a triangle diagram:
\be
\int d^{2}p\, {1\over (p^{2}+m^{2})} \,{1\over (p-Q)^{2}}\,
              {1\over ((p-Q')^{2}+M^{2})},
\ee
etc. 

Furthermore, the propagator $(p^{2}+m^{2})^{-1}$ can occur as a part
of multiloop integrals where, again, contributions from the
singularity at $p=0$ will have to be studied anyway, while the rest of
the integrand can vary greatly but ``almost always" is a smooth
function at $p=0$. It would be only reasonable to solve first the
simpler problem of expanding $(p^{2}+m^{2})^{-1}$ integrated against a
generic smooth function, and after that address the full problem with
the propagator embedded in a more complicated integrand.

Now, one cannot fail to notice the distinctly functional-analytic
character of this class of problems: the $m$-dependent factor plays
the role of a kernel of a linear functional, while the remaining
factors play the role of ``test functions". Moreover, the most
difficult part of the problem is how to handle the singularities of
the formal expansion, so that as a first step one can replace the
factors that are independent on $m$ (or whose dependence is trivial)
by an arbitrary smooth function with compact support.  Then one
arrives at the problem of expanding the propagator
$(p^{2}+m^{2})^{-1}$ (and more general products in more general cases)
in the sense of distributions \cite{fvt82}, \cite{fvt84},
\cite{fvt91}.  We postpone a precise formulation of the problem till
\sect{As-operation}.  Let us only note that the general solution is
given by the so-called {\it asymptotic operation\/} (\asop) for
products of singular functions (\cite{piv-fvt84}, \cite{fvt91} and
\sects{As-expansion}-{MainProof} below) which is similar in form to
Bogoliubov's \rop\ but possesses the important property of uniqueness.

\SUBSECTION {Construction 
of asymptotic expansions as extension of functionals.}
\label{Momentum.ExtensionPrinciple}

Consider the problem of expanding the propagator $(p^{2}+m^{2})^{-1}$
in $m\to 0$ in the sense of distributions. The obvious starting point
is the formal Taylor expansion:
\be[Eq.Example2.Subgraph]
     {1\over p^{2}+m^{2}}
     =
     {1\over p^{2}} - m^{2} {1\over p^{4}} + \cdots .
\ee
Since it contains progressively more singular terms, it cannot be
substituted directly into integrals such as \eq{Eq.Example2}, and
cannot be a correct expansion in the sense of distributions. However,
once one started regarding the problem from the viewpoint of
functional analysis, it is natural and appropriate to think in terms
of functionals and their domains of definition.  Then one notices that
the expansion \eq{Eq.Example2.Subgraph} is perfectly correct if one
restricts the space of all test functions to the subspace of test
functions that are equal to zero in a neighbourhood of the point
$p=0$.

Thus we see that the problem of constructing expansions in the sense
of distributions is essentially a variant of the generic problem of
constructing extensions of functionals possessing certain properties.
The novel feature in our case is only the condition which the
extensions should satisfy, namely, the approximation property. A
constructive solution at the level of generality of the Hahn-Banach
theorem was proved to exist (the so-called {\em extension
principle}---\cite{fvt82} and \sect{ExtensionPrinciple} below).  It
provides an important heuristic clue to the problem of
constructing asymptotic expansions of products of singular functions
in the sense of distributions.

In practical aspect, the modification that needs to be done to
\eq{Eq.Example2.Subgraph} according to the extension principle 
in order to obtain the desired solution consists in modifying the
r.h.s.\ of \eq{Eq.Example2.Subgraph} by adding to it counterterms
proportional to $\d$-functions localized at the point
$p=0$---similarly to what is done in the case of the \rop\ in
coordinate representation. Which is no wonder given the close
similarity of the two problems at an abstract level. One should keep
in mind, however, that there are important differences as well.  For
instance, while the construction of \rop\ allows considerable
arbitrariness, the \asop, as discussed in \sect{As-expansion}, is
unique. In other words, the finite constants that remained arbitrary
in the case of the \rop\ (recall eq.\eq{rGZ} and the remark
thereafter), are uniquely fixed by the approximation requirement in
the case of the \asop.

\SUBSECTION {UV renormalization via \asop.}
\label{Momentum.UV-renormalization}

Consider now an $l$-loop Feynman integral. Its momentum space
integrand $F(p_1\ldots p_{l})$ is a function of $l$ $D$-dimensional
integration momenta $p_{i}$, among other parameters.  In general, $F$
is a product of propagators of the form $(l^{2}+\cM ^{2})^{-1}$ where
$l$ is a linear combination of $p_{i}$ and some external momenta $Q$,
while $\cM $ is a real parameter.  It is important that a propagator
is a uniform function of $l$ and $\cM $. $F$ may also contain
polynomials of $p_{i}$ as overall factors. In the final respect, $F$
has to be integrated over all $p_{i}$ in infinite limits.  The {UV}
divergences are due to a slow decrease of $F$ in some or all
directions in the space of integration momenta.  Irrespective of what
is known about {UV} divergences from the point of view of coordinate
representation, for purely practical reasons it is highly interesting
to know their precise structure in momentum representation.  Following
\cite{gms}, consider the integral of $F$ with a smooth cutoff at
$p=(p_1\ldots p_{l}) \sim \L $:
\be
     \int d^{D\times l}p\, \Phi(p/\L)\,F(p,\cM ,Q) 
     = \L^{\o}
     \int d^{D\times l}p\, \Phi(p)\,F(p,\cM /\L,Q/\L) ,
\ee
where the smooth $\Phi(p)$ is $1$ around $p=0$ and $0$ around
$p=\infty$, and we have explicitly indicated the dependence of $F$ on
the sets of parameters $\cM $ and $Q$; the r.h.s.\ has been obtained
by rescaling $p\to \L p$ and making use of the uniformity properties
of the factors of which $F$ is built.

One can immediately see that studying the details of behaviour of the
l.h.s.\ at $\L \to \infty $ is a special case of the problem of
expansion of $F(p,\cM ,Q)$ at $\cM ,Q\to 0$ in the sense of
distributions. Moreover, it turns out (\cite{gms} and \cite{rmp2})
that subtraction from the integrand of exactly those terms of such an
expansion that are responsible for UV divergences, is equivalent to the
standard Bogoliubov $R$-operation.

\SUBSECTION{Recursion patterns.}
\label{Momentum.Recursion}

Consider a more complicated example of a two-loop self-energy diagram
in a scalar $\vf^{3}$-type model in four dimensions (cf.\ Fig.~3a):
\be
     \int d^{4}p_1\, d^{4}p_2\,
     \left\{
          {1\over (p^{2}_1+m^{2}_\0)(p^{2}_2+m^{2}_\0)
                  ((p_1-p_2)^{2}+m^{2})}
      \right\} 
\kern1cm\ee\be[Eq.Example3]\kern2cm
     \times
     {1\over ((p_1-Q)^{2}+M^{2})((p_2-Q)^{2}+M^{2})}.
\ee
We have separated the factors that become singular after a formal
expansion in $m$ (the product in braces) from those that stay smooth.

First of all, we see that if one studies expansion of \eq{Eq.Example3}
in $m\to 0$, one has to deal with the distribution in braces---the
remaining factors may vary, producing other diagrams with the same
product of $m$-depending factors.  The corresponding pattern of
singularities is visualized in Fig.~3b. Note that it is essentially
the same as in the example we considered in
\ssect{Coordinate.UV-renormalization} in connection with the \rop\ in
coordinate representation (cf.\ Fig.~1).

Denote the solution for the expansion problem in
the sense of distributions for the expression in braces in
\eq{Eq.Example3} as
\be[oiuy]
     \As\.\left\{
            {1\over (p^{2}_1+m_\0^{2})(p^{2}_2+m_\0^{2})
                    ((p_1-p_2)^{2}+m^{2})}
     \right\}.
\ee
Consider a test function $\vf (p_1,p_2)$ such that it is
non-vanishing in a region non-intersecting with the axes $p_2=0$
(where the singularities of the second ppropagator is localized) and
$p_1=p_2$ (corresponding to the singularities of the third
factor), as shown in Fig.~3b. Considered on such test functions,
the distribution in braces factorizes as
\be[effTest]
     {\vf (p_1,p_2)
         \over
     (p^{2}_1+m_\0^{2})(p^{2}_2+m_\0^{2})((p_1-p_2)^{2}+m_\0^{2})}
     = 
     {1 \over (p^{2}_1+m_\0^{2})} \, \tilde\vf (p_1,p_2,m),
\ee
where
\be
     \tilde\vf (p_1,p_2,m)
     =
     {\vf (p_1,p_2)
          \over
      (p^{2}_2+m_\0^{2})((p_1-p_2)^{2}+m^{2})}
\ee
is essentially a test function with a highly regular expansion in $m$,
while $p_2$ is a purely spectator parameter.  This means that a
correct expansion of \eq{effTest} can be obtained by applying the
\asop\ to its first factor and Taylor-expanding the second factor.
In terms of the initial problem, this means that on the subspace of
the test functions $\vf$ with the described properties, the asymptotic
operation $\As$ can be represented as
\be
     \As\.\left\{
       {1 \over (p^{2}_1+m_\0^{2})(p^{2}_2+m_\0^{2})((p_1-p_2)^{2}+m^{2})}
     \right\}
\kern5cm\ee\be[Eq.Example3.Locality]
     =
     \As\.\left\{
         {1 \over (p^{2}_1+m_\0^{2})}
     \right\}
     \times \T \.\left\{
         {1 \over (p^{2}_2+m_\0^{2})((p_1-p_2)^{2}+m^{2})}
     \right\},
\ee
where $\T$ denotes the Taylor expansion in $m$.

\hbox{\Eq{Eq.Example3.Locality}} is a special case of the general 
{\it locality condition\/} for the \asop\ (to be described in detail in
\ssect{As-op.Locality}) and has a typical recursive form similar to
what we had for the \rop\ (cf.\ \eq{Eq.Example1.Locality}).  Note,
however, that unlike the causality condition, the locality condition
\eq{Eq.Example3.Locality} has no direct physical interpretation.

Another type of recursion in the expansion problem emerges as follows.
As will be shown in \sect{As-operation}, construction of the \asop\
for the distribution in braces involves expressions of the form
\be[serapW]
     \int d^{4}p_1\, d^{4}p_2\,
     \cP(p_1,p_2)
     \left\{
        {1\over (p^{2}_1+m^{2})(p^{2}_2+m^{2})
                ((p_1-p_2)^{2}+m^{2})}
     \right\},
\ee
where $\cP$ is a polynomial. Such integrals diverge in the ultraviolet
limit (i.e.\ at $p_{i}\to \infty)$, and once again we are facing the
problem of studying {UV} behaviour in the form discussed in the
preceding subsection. For instance, $\cP=1$ corresponds to the
two-loop vacuum-energy diagram from the scalar $\vf^3$ theory shown in
Fig.~4a.

\SUBSECTION{Summary.}
\label{Momentum.Summary}

The above examples demonstrate the numerous recursive connections that
exist between various kinds of asymptotic expansion problems for
multiloop diagrams. It is extremely interesting to develop a general
formalism that could allow one to make efficient use of such
recursions. Such a formalism, however, has to be developed with great
attention to detail in order to achieve the degree of precision needed
to accomodate various problems---and in order for such formalism to be
of any use from the calculational point of view.

On the other hand, it should also be clear that all analytical
problems here essentially boil down to construction of asymptotic
expansions of products of singular functions in the sense of
distributions (\asop). Once an explicit expression for the \asop\ has
been constructed and its properties studied, applications to the
theory of Feynman diagrams is a rather straightforward excercise of
``algebraic'' nature \cite{IV}, \cite{gms}, \cite{rmp2}.

\newpage\thispagestyle{myheadings}\markright{}
\PART{Universum of Graphs.}

Papers devoted to ``rigorous" studies of multiloop Feynman diagrams
often start with a formal definition of what a Feynman diagram is,
exactly. We will not follow that tradition. Instead, we will develop
an abstract formalism to describe hierarchical families of products
of singular functions and the relationships between them of the type
encountered in such studies in various situations. Only those aspects
of multiloop diagrams that are relevant and important for analyzing
singularities will be reflected in the formalism. The formalism,
nevertheless, allows further refinements, should one wish to consider
special problems.

In a sense, the role of the definitions presented in the following
sections is analogous to studying the structure of parametric
representations often used in rigorous studies of multiloop diagrams.%
\footnote{
It is interesting to note that the use of a parametric representation
at this stage creates a psychological illusion that something more
meaningful is being done than just devising a system notations to
refer to and manipulate complex objects and overcoming problems of
iatrogenic character (e.g.\ the destruction of the fundamental
multiplicative structure of Feynman integrands that takes place after
transition to parametric representations).
\relax}

Our formalism is purely algebraic, almost no reference is made to
Feynman-diagrammatic images (except in examples). This is because,
again, the details of structure of the Feynman integrals that make
their graphical interpretation possible have no relevance to the
analytical problems we consider. Keeping such details out of the way
simplifies the formulae a great deal and, moreover, makes the
formalism applicable equally well to quite different problems.
Nevertheless, we use the terms ``graph/subgraph" to denote products of
singular functions and the associated structures for the sake of
readability (alternatives like ``product of singular functions" are
cumbersome, while contractions like PSF/sub-PSF, too criptic).

The crucial point is to ensure a smooth transition from a problem for
a graph to corresponding problems for its subgraphs, so as to avoid
inconsistencies. Another important point is to develop systematic
rules for ``abusing notations" so that the symbolism that one actually
uses in proofs were simple and helped one to focus directly on what is
really important without causing ambiguities or distracting attention
by irrelevant minutiae.

\SECTION{Graphs and their coefficient functions.}
\label{Graphs}%1

\SUBSECTION{Individual factors.}
\label{Graphs.Factors}%1.1

Let $g$ be a discrete variable; it will be used for enumerating factors in
products. For each $g$ define:

$P_g \bydef$ a finite dimensional linear space; its elements will be
denoted as $p_g$. It is convenient to fix a smooth norm in $P_g$
denoted as $|p_g|$; its specific form is inessential.

$F_g \bydef$ a function $P_g \to \Cspace$; its properties will be
fixed separately depending on the specific problem.

In general $F_g$ will take values in some vector space $\cB_g$
(\sect{NonScalarFactors}); $F_g$ may even be a formal series (as in
the problem of constructing asymptotic
expansions---\sect{As-operation}); however, such complications are
inessential for understanding the analytical aspects of the problems
which we will study, so that $F_g$ can be considered scalar-valued in
the first reading.

The point $p_g = 0$ can be {\it singular\/} for $F_g$ in some
sense (depending on the specific problem) in which case the factor
$F_g$ is called {\it singular\/}; otherwise it is a {\it regular}
factor.  This is the only essential property of $F_g$ that has to be
known at this stage.

\SUBSECTION{Examples.}
\label{Graphs.Example}%1.2

Descriptions of Feynman diagrams can be found in any textbook on
quantum field theory (see e.g.\ \cite{bog-shir}). We only note that we
consider the multiloop integrals without trivial kinematical factors
such as $\d$-functions expressing global momentum conservation. Such
factors play no role in the problems we consider.

     ($i$) Let $P_g$ be the space of $D$-dimensional Euclidean
momenta, and $F_g(p_g)$ the scalar propagator of the form $1/(p^{2}_g
+ m_g^{2})$.  If one considers integrability of products of such
propagators for fixed $m_g$'s then $p_g = 0$ is a regular point of
this function for $m_g\not=0$. But if expansions at $m_g\to 0$ are
studied, then the point $p_g = 0$ will have to be considered as
singular, and $F_g$ will also be considered as a singular factor,
essentially because the Taylor expansion of such $F_g$ in powers of
$m_g$ will contain terms non-integrable at $p_g = 0$ in all
sufficiently high orders in $m_g$.

     ($ii$) Let now $g$ correspond to a vertex with $v$ incident lines
in a Feynman diagram, and let $p_g = (p_1,\ldots p_{v-1})$ be the
collection of independent $D$-dimensional momenta flowing into the
vertex, so that $P_g$, the manifold of such $p_g$, is a space of
dimension $D\times (v-1)$.  Let $F_g(p_g)$ be the polynomial of
$p_g$ that corresponds to the vertex $g$ according to Feynman rules.
Without loss of generality one can take this polynomial to be
homogeneous.

This example motivates introduction of spaces $P_g$ of different
dimensions for different $g$. Note also that a polynomial factor may
be considered as either regular or singular, at wish, even within
the same problem. The choice of the point of view here is a matter of
convenience---the final results do not depend on this.

     ($iii$) If one wishes to consider the problem of UV
renormalization in coordinate representation then one has to deal with
(Euclidean) Feynman diagrams in the coordinate representation, and one
would normally make each factor to correspond to a line of the
diagram.  Then $P_g$ is $D$-dimensional coordinate space (whose
points are usually denoted as $x$, $y$ etc.)  while $F_g(x) \sim
\int d^{D}p\, \exp (ipx)/(p^2+m^2)$ is the Feynman propagator
in coordinate representation.  If the interaction vertices of the
diagram involve derivatives, it is easiest to include such derivatives
into the propagators on which they act (although more sophisticated
schemes along the lines of \sect{NonScalarFactors} are possible).

In this example all factors have to be considered as singular.  The
leading singularity at $x\to 0$ of the above function $F_g(p)$ is
$(x^2)^{(2-D)/2}$, $D > 2$, or $\log x^2$ for $D = 2$.

\SUBSECTION{Graphs.}
\label{Graphs.Graphs}%1.3

Let $G$ be a finite set of values of the label $g$. Consider the
following objects associated with it:

$P_G \bydef$ a finite dimensional affine space; its elements are
denoted as $p_G$ and the corresponding norm as $|p'_G-p''_G|$;
often a point, e.g.\ $p_G = 0$, is singled out in $P_G$ so that it
can be considered as a vector space;

$l_{g(G)}(p_G) \bydef$ an affine mapping $P_G \to P_g$ defined for
each $g \in G$. It can always be taken to be surjective.

     {\bf Convention concerning the subscript $(G)$.} Note the use of
the subscript in brackets $(G)$ in the notation $l_{g(G)}$ to indicate
that the argument $p_G$ of $l_g$ runs over the space $P_G$.  In
general, a presence of such subscript means that the object thus
marked is considered ``in the context of the space $P_G$". The exact
meaning of this phrase will be clear from the corresponding
definitions; as a typical example, the same mapping $l_g$ will induce
related mappings $l_{g(G)}$ on $P_G$ for a family of sets $G$.
Omitting the subscript $(G)$ everywhere may lead to confusion, while
keeping it everywhere will unnecessarily complicate formulae. We will
choose a median strategy: our formulae will normally have such a
structure that all factors in an expression will have to be considered
``in the context of" the same space, say, $P_G$ and should carry the
same subscript $(G)$. For this reason, we will be omitting such a
subscript in a formula everywhere except for a few occurences, and
write $p$ instead of $p_G$.  Whenever a confusion is likely, full
notation will be used.

For a set consisting of a single element $G = \{g\}$: \ $P_G \bydef
P_g$ \ and \ $l_{g(G)} \bydef 1$.

For the reasons explained above such sets $G$ are called {\it
graphs\/}; their coefficient functions are defined as
\be[1/1.1]
     F_G(p) \bydef \prod_{g\in G} F_g(l_{g(G)}(p)), 
     \qquad p \in P_G.
\ee
It is convenient to assume that all components of $P_G$ are
essential---which means that there are no translations in the space
$P_G$ that would leave $F_G$ invariant.

\SUBSECTION{Examples.}
\label{Graphs.MoreExample}%1.4

One example of such a product is provided by the integrand of
\eq{Eq.Example3} corresponding to Fig.~3a: $P_G$ is the space of
integration variables $p_1$ and $p_2$, the mappings $l_g$ are:
$l_1(p_1,p_2) = p_1$, $l_2(p_1,p_2) = p_2$, $l_{3}(p_1,p_2) =
p_1-p_2$, $l_{4}(p_1,p_2) = p_1-Q$ etc.  The reader will easily see
how other expressions (e.g.\ the integrand of \eq{serapW} or
\eq{Eq.Example1}) fit into the general scheme.

\SUBSECTION{``Structure" of graphs.}
\label{Graphs.Structure}

We will say that two graphs {\it have the same structure\/} if they
differ only by the analytical form of the functions $F_g$; the sets
$G$, the spaces $P_g$ and $P_G$, the mappings $l_g$ and the
property of a given factor to be singular or regular, should be the
same (or isomorphic) in both graphs.  We then can consider linear
combinations of graphs with the same structure and treat them
simultaneously. This is convenient, typically, if one has to consider
graphs obtained by differentiation of another graph with respect to a
parameter on which several factors of the initial graph depend---cf.\
\ssect{Complete.Co-subgraphs} and \ssect{As-op.Formal}.

\SUBSECTION{Simplified notations for subproducts.}
\label{Graphs.Simplified}

Consider a subset $\g \subset G$ (in what follows, we will be
considering subsets satisfying additional restrictions and call such
subsets {\it subgraphs}\/).  Define its coefficient function
$F^\0_{\g(G)}$ as a subproduct of \eq{1/1.1}:
\be[1/1.2]
     F^\0_{\g(G)}(p) 
     \bydef  \prod_{g\in \g} F_g(l_{g(G)}(p)),
     \qquad p \in  P_G.
\ee
For the empty set $\g = \emptyset$ it is convenient to define
\be[1/1.3]
     F_{\emptyset (G)}(p) \bydef 1, 
     \qquad P_{\emptyset} \bydef \{0\},
     \qquad l_{\emptyset (G)} \bydef 0.
\ee
For two subsets $\g' \cap \g'' = \emptyset $ the following is true
\be[1/1.4]
     F^\0_{\g'\cup\g''}(p) 
   = F^\0_{\g'}(p)\times F^\0_{\g ''}(p).
\ee
To make formulae shorter, we will use the same symbol to denote both
sets and the corresponding coefficient functions and write $p$ instead
of $p_G$:
\be[1/1.5]
     \g_{(G)}(p) \bydef  F^\0_{\g (G)}(p)
     \qquad {\rm for\ any\ \ }\g \subset G.
\ee
One special case of this convention is:
\be[1/1.6]
     g_{(G)}(p) \bydef  F_g(l_{g(G)}(p)).
\ee
Now \eq{1/1.1} takes the form:
\be[1/1.7]
     G(p) = \prod_{g\in G} g(p)
\ee
(This formula is the first example of the convention of omitting
the subscript $(G)$ introduced in \ssect{Graphs.Graphs})

We will also use the notation:
\be
     \Gg (p) = F_\Gg(p).
\ee
Under the above conventions, the following expressions are possible:
$G = \Gg \times \g$, or even $\emptyset (p) = 1$. We hope no
misunderstanding will occur because it is usually clear from the
context whether symbols like $G$, $\g$, $\emptyset$ denote functions
or sets.  For the same reason, we will use the term graph (subgraph)
to denote not only sets (subsets) but also their coefficient
functions.

To denote graphs and subgraphs we will systematically use symbols $G$,
$H$, $\G$, $\g$.

Note that to any graph $G$, one can always add fictitious elements $g$
such that $F_g \equiv 1$ with $P_g$ and $l_g$ chosen
arbitrarily.  The coefficient function of the graph is not affected
thereby. Whether fictitious elements are considered as singular or
regular is decided in each case by convenience.

\SUBSECTION{Graphs as distributions.}
\label{Graphs.asDistributions}%1.5

Graphs $G(p)$ will be regarded as distributions (linear functionals)
of the type
\be[1/1.8]
     \langle G, \vf_{(G)}\rangle
     = \int_{P_G} dp\, G(p)\, \vf_{(G)}(p),
\ee
where $\vf_{(G)}$ is our standard notation for test functions from
the space $\cD(P_G)$ \cite{sch52} (i.e.\ each $\vf_{(G)}$ is
smooth and has a compact support denoted as $\supp \vf_{(G)}$).

Note that the integral in \eq{1/1.8} can be ill-defined for some
$\vf$ owing to singularities of some of the factors in $G$. One of
the problems we are going to consider will be to determine the class
of functions $\vf $ for which the expression \eq{1/1.8} is
well-defined, and to study extensions of the functional $G$ onto the
entire space $\cD(P_G)$.

\SUBSECTION{$i$-graphs.}
\label{Graphs.iGraphs}

We will also have to consider formal integrals of the form
\be[i-graph]
     \int dp\, \cP(p) \,G(p),
\ee
where $\cP(p)$ is a polynomial of $p$.  We will call such objects
$i$-graphs (``integrated graphs") to distinguish them from the graphs
in the sense of the preceding subsections, i.e.\ from the graphs
considered as distributions over $p$.  Ordinary unrenormalized Feynman
diagrams are special cases of $i$-graphs.

$i$-graphs will naturally appear in our construction of the
\asop, so that with each subgraph $\g $ an $i$-graph
\be[i-subgraph]
     \int dp_\g\, \cP(p_\g) \g (p_\g)
\ee
is associated.

\SECTION{Non-scalar factors.}
\label{NonScalarFactors}

In \sect{Graphs} we required that all $F_g$ took scalar values and
$F_G$ were constructed using scalar multiplication.  However, Feynman
diagrams for non-scalar theories do not fit directly into such
framework. For instance the propagator of a spinor particle is
naturally considered as a function over $P_G$ taking values in the
Dirac algebra, and the corresponding products are formed using matrix
multiplication, trace evaluation and contractions over Lorentz
subscripts. Though one can always consider such products of non-scalar
factors component-wise, it would be advantageous to preserve a
covariant point of view, e.g.\ for the purposes of studying Lorentz
invariance, quantum anomalies, internal symmetries, supersymmetry etc.
The definitions and syntax conventions introduced below allow one to
consider diagrams for non-scalar theories with no apparent changes in
formulae as compared to the scalar case.  Note that in the mainstream
variant of the BPHZ theory which uses $\a$-parametric representation
\cite{zav79}, the analysis of diagrams with numerators is complicated
to a very great extent by the fact that such representations destroy
the multiplicative structure of the original integrand.

First (\ssect{NonScalar.Subscripts}) we discuss our conventions in
terms of {\em subscripts} and {\em contractions}.  Then
(\ssects{NonScalar.General}-{NonScalar.Syntax}) a more formal approach
is described in order to show that, in general, one is not restricted
to finite dimensional spaces and there is a possibility to generalize
our formalism to models with infinite number of component fields; an
interest in such models arose due to progress of superstring
models---see e.g. \cite{kras}.

\SUBSECTION{Subscripts and contractions.}
\label{NonScalar.Subscripts}%2.1

Assume that each $F_g$ carries {\em subscripts}---e.g.\ Lorentz,
internal group sub\-/su\-per\-scripts etc. Assume that the product
\eq{1/1.1} implies contractions of some pairs of such subscripts. Then
if the product \eq{1/1.2} contains a pair of subscripts that are
contracted in \eq{1/1.1} then, by definition, this contraction is
implied in \eq{1/1.2} as well.

Let now $\G'$ and $\G''$ be non-intersecting subproducts of $G$.
There may exist a contraction over such a pair of subscripts that one
of them belongs to $\G '$ and the other one to $\G ''$. (It is assumed
that $\G '$ and $\G ''$ are constructed according to the rules of the
preceding paragraph.) Then the multiplication in
\be[1/2.1]
     \G '\times \G ''
\ee
(cf.\ \eq{1/1.4}) involves contraction over this pair of subscripts.

We will construct new expressions from $G$ by replacing some
subproducts $\G $ by some expressions, say, $Z_\G$ where $Z_\G$
carries the same subscripts as $\G $ does (cf.\ eg.\ \eq{e1/6.8} and
\eq{e3/2.11}).  Then in products like
\be[1/2.2]
     \G'\times Z_{\G''},\qquad  Z_{\G'}\times \G''\qquad{\rm etc.,}
\ee
the same contractions as in \eq{1/2.1} are implied.

\SUBSECTION{General formalism.}
\label{NonScalar.General}%2.2

Let us present more general and formal definitions which avoid the use
of components of non-scalar objects.

    ($a$) For each factor $g$ let $\cB_g$ be a Banach space in which
the function $F_g$ takes values. Analogously, for any $H \subset G $
(including $H = G$) let $\cB_{H}$ be the Banach space in which $F_{H}$
takes value. (For $H = \emptyset$, $\cB_{\emptyset}$ is
the field of scalars).

    ($b$) For any decomposition of $H$ into non-intersecting subsets
$H_k$: \ $H = \cup_k H_k$, \ let \ $M^{H}_{\{H_k\}_k}$ \ be a linear
mapping $\otimes_k \cB_{H_k} \to \cB_{H}$. We require $M$ to be
bounded, i.e.\ for any $b_k \in \cB_{H_k}$ :
\be[1/2.3]
     \left| M^{H}_{\{H_k\}_k}
            (\mathop{\otimes}_k  b_k) 
     \right|
     \leq \const\cdot \prod_k |b_k|,
\ee
where all norms are evaluated in the corresponding spaces. (We require
$M$ to be bounded in order to be able to obtain estimates for $H(p)$
from estimates for factors.)

     ($c$) {\bf Structural transitivity.} Assume that a set of further
decompositions $H_k = \cup_{l} H_{kl}$ is given. Then the
composition of the mappings $M^{H_k}_{\{H_{kl}\}_{l}}$ and
$M^{H}_{\{H_k\}_k}$ should coincide with the mapping
$M^{H}_{\{H_{kl}\}_{kl}}$. In other words we require commutativity of
the following diagram:
\be[1/2.4]
\begin{array}{ccccc}
\left( \mathop{\otimes}_l \cB_{H_{1l}} \right)
&\otimes\; \ldots\; \otimes
&\left( \mathop{\otimes}_l \cB_{H_{kl}} \right)
&\equiv
&\left( \mathop{\otimes}_{k,l} \cB_{H_{kl}} \right) \\
\downarrow
&
&\downarrow
&
&\downarrow\\
\cB_{H_1}
&\otimes\; \ldots\; \otimes
&\cB_{H_k}
&\to
&\cB_{H}
\end{array}
\ee

\SUBSECTION{Syntactic conventions.}
\label{NonScalar.Syntax}%2.3

Let us use the idea known from the practice of programming languages
(cf.\ the notion of a polymorphic operation in \cite{[35]}). The idea
is that the exact interpretation of the symbol of an
operation---multiplication in our case---depends on the types of the
operands:

if $b_k$ are expressions with values in $\cB_{H_k}$, where all
$H_k$ do not intersect pairwise, then $\prod_k b_k$ is to be
understood as $M^{H}_{\{H_k\}_k}(\otimes_k b_k)$ with values in
$\cB_{H}$ where $H = \cup_k H_k$.

Note that the formal associativity of such multiplication is provided
by the structural transitivity. Formal commutativity holds due to the
fact that all the factors belong to different (albeit perhaps
isomorphic) spaces, so that their order is inessential.

One can see that this syntactic convention is equivalent to that
described in \ssect{NonScalar.Subscripts}.

\SUBSECTION{Conventions for test functions.}
\label{NonScalar.TestFunctions}%2.4

We assume that for distributions taking values in $\cB_{H}$ the
corresponding test functions take values in the dual space $\cB^{\ast
}_{H}$.  So the result of multiplication of a $\cB_{H}$-valued
distribution by a $\cB^{\ast}_{H}$-valued test function is a number.
As before, the usual multiplicative notation can be used here.

In some cases we will also need to multiply a $\cB_{\GG
}$-valued distribution (function) by a $\cB^{\ast}_G$-valued (test)
function (we are only interested in the case when $\G \subset G$).
There exists the only mapping $M^{\#}$: $\cB_\GG
\otimes
\cB^{\ast}_G \to
\cB^{\ast}_\G$
\ such that
\be[1/2.5]
   \langle
       M(b_\G \otimes  b_\GG),
       b^{\ast}_G
  \rangle
  =
  \langle
       b_\G,
       M^{\#}(b_\GG \otimes  b^{\ast}_G)
  \rangle
\ee
(this equation is simply the definition of $M^{\#}$; the expression
$\langle b,b^{\ast}\rangle$ denotes the value of $b^{\ast}$ on~$b$).
It is sufficient to extend the syntactic convention by allowing to
write down both sides of~\eq{1/2.5} as $b_\G \times b_\GG \times b^{\ast
}_G$; all possible interpretations of this expression are equivalent
due to the definition of $M^{\#}$ and the transitivity property.

For example, the expression $\langle G, \vf_{(G)} \rangle =
\langle \G, \GG \, \vf_{(G)} \rangle$ may be interpreted in two
equivalent ways: besides the straightforward interpretation of the
l.h.s., we may consider it as the result of integration of a $\cB_{\G
}$-valued distribution $\G $ with a $\cB^{\ast}_\G$-valued test
function $\tilde{\vf} = \GG \, \vf_{(G)}$.

\SECTION{Graphs and subgraphs.}
\label{Subgraphs}%3

Subproducts (subgraphs) in a graph may in turn be considered as graphs
in their own right. As was shown in 
\ssect{Coordinate.Recursive} and \ssect{Momentum.Recursion} (see also below
\ssect{R-op.Recursive} and  \ssect{As-op.Recursive}) the  problem
for a graph gives rise to analogous subproblems for the subgraphs.  In
this section we study the connection between the problem for a graph
and the corresponding subproblems for its subgraphs.

\SUBSECTION{Subgraphs.}
\label{Subgraphs.Subgraphs}%3.1

Consider a graph $G$. The coefficient function $G(p)$ is defined on
$P_G$. While analyzing $G(p)$, we will have to study its subproducts
\be
     \g_{(G)} (p) = \prod _{g\in\g\subset G} g(p).
\ee
Let $\Pi_{\g (G)}$ be the linear space of translations of $P_G$
which leave all $l_{g(G)}(p)$, $g \in \g $, and consequently
$\g_{(G)}(p)$, invariant. The factor space
\be[1/3.1]
     P_{\g (G)} \bydef  P_G / \Pi_{\g (G)}
\ee
is a natural domain of definition of $\g_{(G)}$.  Denote the canonical
projection $P_G \to P_{\g (G)}$ as $l_{\g (G)}$.

Set $P_\g \bydef P_{\g (G)}$ and for any $g \in \g $ define
$l_{g(\g)}$ from the equation:
\be[1/3.2]
     l_{g(G)} = l_{g(\g)}\. l_{\g (G)}.
\ee
As a result the subset $\g $ together with $P_\g$ and $l_{g(\g)}$
can in itself be considered as a true graph.

\SUBSECTION{Uniqueness of the construction of subgraphs.}
\label{Subgraphs.Uniqueness}%3.2

There are other ways to construct a graph for the same subset $\g$.
Indeed, define first a subgraph $H \subset G$ such that $H \supset
\g$, $P_{H} = P_G / \Pi_{H(G)}$.  Now a graph for the subset $\g$
can be defined as a subgraph in $H$.  Then $P'_\g = P_{H} / \Pi_{\g
(H)} = (P_G / \Pi_{H(G)}) / \Pi_{\g (H)}$.  However, $P_\g$ as
well as $P'_\g$ may be considered as sets of subspaces of $P_G$,
and for this reason $P_\g$ and $P'_\g$ are not only isomorphic
but identical. In virtue of this an identity holds:
\be[1/3.3]
     l_{\g (G)} = l_{\g (H)}\. l_{H(G)},
\ee
so that the functions $l_{g(\g)}$ arising in these two constructions
(cf.\ \eq{1/3.2}) are equal as well.

All this allows one to view $\g$ as a graph with its own space $P_\g$
and functions $l_{g(\g)}$ which are defined naturally, uniquely and
independently of the higher graphs in which $\g$ may be embedded.

\SUBSECTION{The universal graph.}
\label{Subgraphs.Universum}%3.3

A characteristic feature of the problems of quantum field theory is
that one usually has to deal with infinite hierarchies of Feynman
diagrams, so that any Feynman diagram is embedded into a more
complicated one.  And the problem, say, of constructing the
\rop\ for the former turns out to be a subproblem in, and
analogous to the problem for, the latter.  It may be helpful to
introduce a {\it universal graph\/} $\cG$ (with its own space
$P_{\cG}$ and functions $l_{g(\cG)}$ where the set $\cG$ is not
necessarily finite and $P_{\cG}$ not necessarily finite dimensional),
so that any graph $G$ under consideration would be a (finite) subgraph
in $\cG$. Then $P_G$ is $P_{G(\cG)}$ etc.

Introduction of the universal graph gives a certain completeness to
the entire construction (the ``universum of graphs").

\SUBSECTION{Example.}
\label{Subgraphs.Example}%3.4

Let us explain how the notions introduced above are used.  Given the
graphs $\g \subset \G \subset G$, let $f_\g$ be an arbitrary
function on $P_\g$.  Then the following functions
are defined on $P_\G$ and $P_G$, respectively:
\be[1/3.4]
     f_{\g (\G)} = f_\g \,\.\, l_{\g (\G)},\qquad
     f_{\g (G)}  = f_\g \,\.\, l_{\g (G)} ,
\ee
which are invariant under translations of their arguments by vectors
from $\Pi_{\g (\G)}$ and $\Pi_{\g (G)}$.  In this case a connection
follows from the property \eq{1/3.3}:
\be[1/3.5]
     f_{\g (G)} = f_{\g (\G)} \,\.\, l_{\G (G)} .
\ee
For instance, in order to study $F_G$ we will first have to study its
subproducts $\g_{(G)}$ and $\G_{(G)}$ while the latter two problems
are reduced to studying $\g$ and $\G$. But within the problem for $\G$
we will encounter $\g_{(\G)}$.  The above reasoning guarantees that we
will come to the same problem~$\g$ whether we start from $\g_{(\G)}$
or $\g_{(G)}$. Thus the connection between the problem $G$ and its
sub-problems $\G$ and $\g$ is correct.

Note that expressions like
$
     \langle \g, \vf_{(G)} \rangle
$
are interpreted as
$
     \langle \g_{(G)}, \vf_{(G)} \rangle,
$
etc.\ (the convention of \ssect{Graphs.Graphs}).

\SECTION{Completeness condition and subgraphs.}
\label{CompleteSubgraphs}%4

As should be clear from the discussion in
\sects{CoordinateRepresentation}-{MomentumRepresentation}
one of the key technical problems is that of describing the
singularities of graphs.  In order to do that one needs to distinguish
substructures within graphs which may correspond to quite different
types of diagrammatic objects: in the case of \rop\ in coordinate
representation they are subgraphs in the sense of Bogoliubov and
Parasiuk \cite{bog-par}, \cite{bog-shir}, in the case of asymptotic
expansion problem they are various types of infrared subgraphs
\cite{fvt91}, \cite{r*}.  However, the original analytical criterion of
distinguishing such substructures is simple and essentially the same
in all types of problems \cite{fvt91}: once it has been decided which
factors are to be considered as singular
(\ssects{Graphs.Factors}-{Graphs.Example}), then the substructures one
will have to deal with are just the subproducts containing only
singular factors and satisfying the completeness condition of
\ssect{Complete.s-subgraphs}.  The completeness condition simply means
that all factors that contribute to the singularity localized at a
given submanifold should be included into the corresponding subgraph.

In applications to Feynman diagrams, in order to use the completeness
condition it is sufficient to invoke only the primary notions of
Feynman diagrams such as momentum conservation in vertices.  Herein is
another advantage of our techniques over parametric representations:
the structure of the latter has to be described in terms of co-cycles,
trees and other derivative graph-theoretic notions. This necessitates
a translation of the completeness condition into such a language, which
can be rather tricky and non-transparent and offers no practical
advantages.%
\footnote{
and, of course, it does not make the definition of the corresponding
subgraphs more ``explicit'', as is sometimes claimed.
\relax}
We prefer to use the analytical formulation of the completeness
condition which, in any case, reflects the heart of the matter
directly.

After considering the geometry of singularities in
\ssect{Complete.Planes}, in \ssect{Complete.s-subgraphs} 
the definition of complete singular subgraphs ($s$-subgraphs) is
presented.  The natural ordering in the set of $s$-subgraphs is
exhibited in \ssect{Complete.Ordering}. In
\ssect{Complete.Factorization} the properties of factorizable graphs
are studied. The so-called co-subgraphs and the resulting double
hierarchy of graphs are discussed in \ssect{Complete.Co-subgraphs} and
\ssect{Complete.Hierarchy}, respectively. Finally, in
\ssect{Complete.Examples} some examples are presented.

\SUBSECTION{Singular planes.}
\label{Complete.Planes}%4.1

As was pointed out in \ssect{Graphs.Factors}, the functions $F_g(p)$,
$p
\in P_g$, may have singularities at $p = 0$.  If $g \in G$, then the
function $G(p)$ inherits those singularities.

Let $g$ be a singular factor in the graph $G$.  The {\it singular
plane\/} of $g$ in $G$ is the set
\be[1/4.1]\kern3.8mm
     \pi_{g(G)} \bydef \{ p \in  P_G \mid l_{g(G)}(p) = 0 \}.
\ee
The singular plane of a regular factor is empty by definition.

The singularities of individual factors overlap at the points of
intersection of the corresponding singular planes. So, let us define
the {\it singular plane\/} of $\g \subset G$ as:
\be
     \pi_{\g(G)} \bydef \cap_{g\in\g} \pi_{g(G)},
     \quad\hbox{if}\quad \g\ne \emptyset,
\ee\be[1/4.2]
     \pi_{\emptyset(G)} \bydef P_G.\kern3.2cm
\ee
If the plane $\pi_{\g (G)}$ is non-empty then it is isomorphic to
$\Pi_{\g (G)}$ (\ssect{Subgraphs.Subgraphs}), so that:
\be[1/4.3]
     P_{\g (G)} = P_G / \pi_{\g (G)} .
\ee
Then it is convenient to take as a ``representative" of $P_{\g (G)}$
in intermediate constructions a subspace $\pi^{\rm T}_{\g (G)}$ such
that
\be[1/4.4]
     \pi^{\rm T}_{\g (G)} \oplus \pi^\0_{\g (G)} = P_G .
\ee
As a rule, the final results do not depend on the choice of 
$\pi^{\rm T}_{\g(G)}$.

\SUBSECTION {Completeness condition and $s$-subgraphs.}
\label {Complete.s-subgraphs}%4.2

The set of all subgraphs $\g \subset G$ is divided into
non-intersecting classes by the following equivalence relation:
\be[1/4.5]
     \g' \sim  \g''
     \;\;\;\quad \Leftrightarrow \quad\;\;\;
     \pi_{\g '} = \pi_{\g ''} .
\ee
In each class there exists a unique $\g $ containing all other
subgraphs of the same class (it is sufficient to take $\g = \cup
\g'$ where the union is taken over all subgraphs of the class). We
will say that such $\g $ possesses the {\it completeness
property\/} or is {\it complete\/}.  Any such $\g $, provided
$\pi_{\g (G)} \not= \emptyset$, is called a {\it singular
subgraph\/} in $G$ (in short: $s$-{\it subgraph\/}). From the
condition $\pi_{\g (G)} \not= \emptyset$ it follows that $\g$
contains only singular factors.

Note that the empty subgraph is, by definition, an $s$-subgraph, which
is convenient in a number of cases.

The point in distinguishing $s$-subgraphs is that all factors from $G$
that are singular on $\pi_{\g (G)}$ for an $s$-subgraph $\g\subset G$
belong to $\g$ so that almost all points of $\pi_{\g (G)}$
are non-singular for the function $\Gg (p)$.  Conversely, if $\g'
\not=\g $ but $\pi_{\g'(G)} = \pi_{\g (G)}$ (in which case $\g'
\subset \g)$, then the function $\Gg'(p)$ contains factors that are
singular on the entire $\pi_{\g (G)}$.

$S[G]$ will denote the set of all $s$-subgraphs in $G$.

Note that completeness of a subgraph is {\it not\/} its internal
property.  For instance, for the above relation between $G$, $\g $ and
$\g'$, the subgraph $\g'$ which is not complete in $G$, will be
complete in $H = G\backslash (\g \backslash \g')$.

\SUBSECTION{Ordering in $S[G]$.}
\label{Complete.Ordering}%4.3

The set of $s$-subgraphs is ordered with respect to inclusion. If
$\g$, $\G \in S[G]$ and $\g \subset \G$, $\g \not=
\G $ then $\pi_{\g (G)} \supset \pi_{\G (G)}$. In this
case we will write $\g < \G$.

In $S[G]$ there exists at least one maximal element. The set of such
elements is denoted as $S_{\max}[G]$. There may exist more than one
maximal elements, then $G \notin S[G]$ and $\pi_{G(G)} = \emptyset $.
If the maximal element is unique then it is equal to $G$; conversely,
if $G \in S[G]$ then $G$ is its own and only maximal $s$-subgraph.

For the universal graph $\cG$ (\ssect{Subgraphs.Universum}), $S[\cG]$
will denote the set of all graphs $G$ such that $G \in S[G]$.  We may
always assume that for any $G \in S[\cG]$, {\ }$\pi_{G(G)} = \{0\}$.
Such $G$ will be called $s${\it-graphs\/}.

If $G \in S[\cG]$ then it is useful to define {\it submaximal\/}
$s$-subgraphs.  The notation $\G \triangleleft G$ means that for any
$\G' \in S[G]$, $\G' > \G$ implies that $\G' = G$.

The minimal $s$-subgraph is always unique---it is the empty
subgraph.  Sometimes it is useful to consider a set $S_{\min}[G]$
consisting of all $\G \triangleright \emptyset $.

Note that if $\G \in S[G]$ then $S[\G ] \subset S[G]$.

\SUBSECTION{Factorizable graphs.}
\label{Complete.Factorization}%4.4

Let $G = G_1 \cup G_2$, where $G_1 \cap G_2 = \emptyset $.
Let also $\Pi_{G_1} \oplus \Pi_{G_2} = P_G$.  Then in $P_G$ we
may choose coordinates $p = p_1 + p_2$, \ $p_1 \in P_{G_1} =
P_G / \Pi_{G_1}$, \ $p_2 \in P_{G_2} = P_G / \Pi_{G_2}$,
\ $P_1 \oplus P_2 = P_G$, so that the graph $G_{1(G)}(p)$ depends
only on $p_1$ and $G_{2(G)}(p)$ only on $p_2$:
\be[1/4.6]
     G(p) = G_1(p_1)\times G_2(p_2).
\ee
Then we will say that the graph $G$ is {\it factorized\/} into a
product of $G_1$ and $G_2$ and write $G = G_1 \times G_2$.

A graph may, of course, be factorizable into more than two components,
but the decomposition into non-factorizable subgraphs is unique.

$S_{0}[\cG]$ will denote the subset of $S[\cG]$  containing
only  non-factorizable graphs.

Let us mention the following elementary properties of subgraphs in a
factorized graph $G = G_1 \times G_2$ :

 ($a$) $S[G_{i}] \subset S[G],\qquad i = 1, 2$;

 ($b$) if $\g \in S[G]$ then $\g = \g_1 \times \g_2$ where
$\g_1 \in S[G_1]$, $\g_2 \in S[G_2]$ (note that here all $\g $
are considered as {\it graphs\/} for which the operation $\times $ is
defined; if they are considered as {\it subsets\/} in~$G$ then $\g =
\g_1 \cup \g_2$).  In this case
\be[1/4.7]
     \pi_{\g (G)} 
     =
     \pi_{\g_1(G_1)} \oplus \pi_{\g_2(G_2)} ;
\ee

 ($c$) if $\g \triangleleft G$ then either $\g = G_1 \times \g_2$
where $\g_2 \triangleleft G_2$ or $\g = \g_1 \times G_2$
where $\g_1 \triangleleft G_1$.

\SUBSECTION{Co-subgraphs.}
\label{Complete.Co-subgraphs}%?4.5.

Another important and natural construction that leads to new graphs
from a given one is as follows. Let $G$ be a graph, and $\G$ its
$s$-subgraph.  Consider the complement of $\G $ in $G$. A typical
construction (cf.\ \eq{e3/2.11}) consists in adding to $G$ expressions
obtained by replacing a subgraph $\G$ in $G$ by a (linear combination
of derivatives of the) $\d $-function localized at $\pi^\0_{\G (G)}$:
\be
     G(p^\0_G) \to  G(p^\0_G) +
     Z \d^\a(p^\0_\G) \times  \GG (p^\0_G).
\ee
One is then led to consider expressions of the form:
\be[Eq.Co-subgraph.Def0]
     [\GG ]_\G
     \equiv
     \left(
         {\partial \over \partial p^\0_\G}
     \right)_{p^\0_\G=0}^\a
     \GG (p^\0_\GG, p^\0_\G),
\ee
where we have introduced the variable $p^\0_\GG$ that is complementary
to $p^\0_\G$ in $p^\0_G$:
\be
     p^\0_G = (p^\0_\GG,p^\0_\G).
\ee
In constructions such as \eq{Eq.Co-subgraph.Def0} one essentially has
to deal with $\GG$ considered as a function of $p^\0_\GG$,
with $p^\0_\G$ treated as an external parameter.  (Note that in
general eq.\eq{Eq.Co-subgraph.Def0} is a linear combination of several
products, but they all have the same ``structure" in the sense of
\ssect{Graphs.Structure}, and it is convenient to consider them
simultaneously.)

In general, $\GG$ factorizes as follows:
\be[Eq.Co-subgraph.Def]
     [\GG ]_\G
     = \sum \prod_i  \xi_i(p_i,p^\0_\G),
\ee
where $p_i$ are independent components of $p^\0_\GG$:
\be
     p^\0_\GG = (p_1,..,p_i,..).
\ee
It is convenient to call the expression \eq{Eq.Co-subgraph.Def} (or
each of the $\xi_i$, if one wishes to work with non-factorizable
components) {\it co-subgraph\/} corresponding to subgraph $\G$.  We
will see (\ssect{SpecialR.VariationTheorem}) that, say, the structure
of variations of \rop\ on a graph can be explained in terms of the
\rop\ on its co-subgraphs.

\SUBSECTION{Double hierarchy of subgraphs.}
\label{Complete.Hierarchy}

Thus, there are two ways to generate new graphs from a given $G$:
($i$) taking complete singular subgraphs $\g < G$
(\ssect{Complete.s-subgraphs}); ($ii$) taking complements $[G
\backslash \g ]_\g$ projected onto the singular plane of $\g$
(\ssect{Complete.Co-subgraphs}).  Therefore, now one has a double
hierarchy of graphs: $G \to \g$, and $G \to [\Gg ]_\g$.  Since both
transitions lessen the dimension of the corresponding coordinate
space, no loops with respect to the above operation $\to $, which
generates new subgraphs in the universum, are possible. In other
words, if, in addition to the relation $\g < G$, one requires that
$[\Gg ]_\g < G$, then the relation $<$ will remain a partial ordering
in the universum.

Similarly to consistency of the construction of subgraphs (see
\ssect{Subgraphs.Uniqueness}), one has to examine consistency of
the construction of co-subgraphs.  Indeed, let $\g < \G $ be two
$s$-subgraphs of $G$. Consider $[\GG ]_\G$. It can also
be arrived at as follows:
\be[3/2.5]
     G \to  G'
     \bydef
     [\Gg ]_\g
     \to  [G'\backslash \G']_{\G '}
     \equiv  [\GG ]_\G,
\ee
where $\G' = [\Gg ]_\g$.  Equivalence of the two
constructions can be easily established by using reasoning similar to
\ssect{Subgraphs.Uniqueness}, if one uses the fact that, given the
above relations between $\g$, $\G$ and $G$, $\G'$ is automatically an
$s$-subgraph in $G'$.

It is worth noting that there may occur cases when a subgraph $\G$ in
a graph $G$ is automatically a co-subgraph. Then the same is true
about $\GG $, and $G$ factorizes into $\G$ and $\GG $.
Therefore, the operation of factorization of a graph (which also
yields new graphs) is essentially equivalent to taking subgraphs or
co-subgraphs.

\SUBSECTION{Equivalences between subgraphs.}
\label{Complete.Equivalences}

It often happens in applications that two subgraphs---or an
$s$-subgraph and a co-subgraph---have the same structure in the
sense of \ssect{Graphs.Structure}. This is a rather general
situation: one often has to introduce equivalences in the universum
between graphs of the same structure (e.g.\ {UV} subtractions should
be done in the same manner etc.). Such equivalences can only exist
between graphs whose spaces $P_G$ have equal number of dimensions
while the ordering connects only graphs with non-isomorphic $P_G$'s:
if \ $G > \g $ \ then \ $\dim P_G > \dim P_\g$.  Therefore, one can
introduce equivalences in the universum between $s$-subgraphs and
co-subgraphs etc.\ without generating inconsistencies.

\SUBSECTION{Examples.}
\label{Complete.Examples}

$\mathstrut$\hbox{\indent}($i$) In Fig.~1a (the analytical expression
\eq{Eq.Example1}) the lines $[0-x]$ and $[x-y]$ constitute
$s$-subgraphs, as does the pair of lines $[0-y]$ (Fig.2a). The
co-subgraph of the latter in the graph of Fig.~1a is shown in Fig.~2b.
The graphs Fig.~2a and Fig.~2b have the same structure and may be
equivalent in the universum (\ssect{Complete.Equivalences}).

 ($ii$) Consider the graph in Fig.~3a with the analytical expression
\eq{Eq.Example3} within the framework of the problem of expansion in
$m$ (cf.\ \ssect{Momentum.Recursion}).  Then, as prescribed by the
general theory of the \asop, the factors containing $m$ are to be
considered as singular (because they generate singularities if
expanded in $m$), and those containing $M$, as regular.  Then the two
lower lines (through which the momenta $p_1$ and $p_2$ flow)---as well
as any other pair of light lines---do not constitute a complete
subgraph because setting $p_1=p_2=0$ also nullifies the momentum
flowing through the vertical line, and the corresponding factor blows
up. However, if one replaces $m\to M$ in the vertical line, thus
making it non-singular, the two lower lines become an $s$-subgraph.

 ($iii$) In Fig.~4b neither one, nor any pair of lines constitute an
$s$-subgraph, but only all three of them---because at the singular
plane $x=0$ all three factors are singular simultaneously.  This
agrees with the fact that this diagram has only one UV subgraph
according to the definition of Bogoliubov and Parasiuk \cite{bog-par},
\cite{bog-shir}. On the other hand, consider the integrand of the
corresponding momentum space expression \eq{serapW} (Fig.~4a).  As
explained in \ssect{Momentum.UV-renormalization}, studying the {UV}
convergence of the integral \eq{serapW} is essentially equivalent to
studying expansion of the intergrand in $m$ in the sense of
distributions.  Then non-trivial $s$-subgraphs (i.e.\ other than the
entire graph) consist of single lines.  Consider e.g.\ the
$s$-subgraph $\G$ consisting of the line with momentum $p_1$.  The
corresponding co-subgraph $\GG$ consists of the other two lines; then
$p_\GG=p_2$.  Incidentally, the empty $s$-subgraph corresponds to the
co-subgraph consisting of all the three lines. One sees that when
dealing with {UV} divergences in momentum representation, the
structures equivalent to {UV} subgraphs in Zimmermann's definition
\cite{zim70} (which is used e.g.\ in the formulation of the standard
MS scheme of UV subtractions \cite{MS}) emerge as co-subgraphs in the
expansion problem for the momentum space integrand.  This illustrates
the difference between the two definitions of UV subgraphs which lead
to physically equivalent \rop s.

\SUBSECTION{Summary.}
\label{Complete.Summary}

Within the universum of graphs defined above we have three natural
interrelated operations on graphs that generate new subgraphs that
should be considered together with the original graph within the same
universum: taking $s$-subgraphs, taking co-subgraphs, and
factorization. The three operations are consistent in the sense that
if a subgraph can be generated from a graph using several different
compositions of the three operations, then the final result will
always be the same.  There may also be equivalences between members of
the universum. Such equivalences are consistent with the existing
ordering. The constructions on graphs that we will consider will be
recursive in the sense that a construction on a graph is defined in
terms of the corresponding operations on its subgraphs (and, perhaps,
co-subgraphs).  

This completes our study of the structure of the universum of graphs.
The formalism introduced above will allow us in the following sections
to concentrate on the study of analytical properties of products of
singular functions of the described type.

\newpage\thispagestyle{myheadings}\markright{}
\PART {Analytical Tools for Studying Singularities.}

There are three groups of tools for studying analytical structure of
singularities in graphs, which we will consider in turn: 

--- decompositions of unit which allow one to split the integration
region for a graph into subregions in such a way as to reduce the
study of the $R$- or \asop\ on the graph in each subregion to a study
of subgraphs (\sect{Decompositions});

--- inequalities for description of power-and-log behaviour of
distributions near singular points (\sect{IsolatedSing});

--- subtraction operators which allow one to eliminate singularities
localized at an isolated point, providing a realization of a
construction of extension of functionals similar to that of the
Hahn-Banach theorem on extension of functionals (\sect{Subtractions}).

\SECTION{Decompositions of unit isolating singularities of 
$s$-subgraphs.}
\label{Decompositions}

The technique of decompositions of unit allows one to study local
structure of distributions and then to use the    local results in
construction of globally defined distributions.  It belongs to most
important elements of the distribution theory \cite{sch52},
\cite{vsvlad}.  In \ssect{Decomp.Subregions} we discuss how
decompositions of unit are used in the framework of our methods. The
decomposition of unit isolating singular planes of maximal
$s$-subgraphs is introduced in
\ssect{Decomp.Maximal}.  In \ssect{Decomp.Submaximal} a decomposition
of unit is constructed which isolates singular planes of submaximal
subgraphs in the important case of $s$-graphs. \Ssect{Decomp.Hepp}
explains connection of our decompositions of unit and Hepp sectors.
In \ssect{Decomp.Radial} we introduce an analogue of radial variables
for use in situations involving distributions. Lastly, in
\ssect{Decomp.Subgraphs} we consider cutoff functions with
special properties reflecting factorization properties of graphs and
their sub- and co-subgraphs.

\SUBSECTION{Factoring out singularities in subregions.}
\label{Decomp.Subregions}%{ss5.1}

We will use decompositions of unit as follows. Let $\cO$ be a region
in $P_G$, and let $\G $ run over some set of $s$-subgraphs in $G$.
Assume that a family of functions $\eta^\0_\G\in C^\infty({\cO})$
forms a decomposition of unit:
\be[e1/5.1]
     \sum_\G\eta^\0_\G (p)=1,\qquad p\in \cO.
\ee
Assume also that all $\eta^\0_\G=0$ in neighbourhoods of singular
points of the corresponding products $\GG(p)$.  Then the study of
singularities of $G(P)$ in $\cO$ is reduced to a study of
singularities of the $s$-subgraphs $\G(p)$ via the following
representation:
\be[e1/5.2]
     G(p) = G(p) \, \sum_\G\eta^\0_\G (p)
     =  \sum_\G \, \left[\, \eta^\0_\G (p) \, \GG (p) \,\right] \, 
           \G (p),
\ee
where the expressions in square brackets have no singularities in
$\cO$.  Therefore, in each term on the r.h.s.\ of \eq{e1/5.2},
singularities are only generated by $\G$ which is effectively factored
out of the product.

Furthermore, if one wishes to define an operation {\bf Op} to act on
$G$ which would yield a distribution in $\cO$, it can be done in the
following way:
\be[e1/1.3]
     \hbox{\bf Op}\. G
     =
     \sum_\G [\eta^\0_\G \, \GG ] \, \hbox{\bf Op} \. \G.
\ee
The above relation reduces construction of ${\bf Op}$ on $G$ to its
construction on the subgraphs $\G$.  The conditions that make such
definitions possible are the causality condition in the case of the
\rop\ (\ssect{Coordinate.Recursive},
\ssect{R-op.Axioms} and \ssect{SpecialR.Example}), 
and the localization property in the case of the
\asop\ (\ssect{Momentum.Recursion} and \ssect{As-op.Locality}).

\SUBSECTION{Isolating singularities of maximal $s$-subgraphs.}
\label{Decomp.Maximal}%{ss5.2}

The first example is as follows: $\cO=P_G$ and $\G$ runs over the set
$S_{\max} [G]$ of all maximal $s$-subgraphs.  It is clear that all
$\pi^\0_\G$ are pairwise non-intersecting.

The desired decomposition $\{\eta^\0_\G\}$ can be
constructed e.g.\ in the following way. Let
\be
     A_\G = P_G \backslash \cup_{g\notin \G}\pi_g.
\ee
Then the set $\{A_\G\}$ covers $P_G$ and for any $\G' \not= \G$ one
has $\pi^\0_\G \cap A^\0_{\G'} = \emptyset $.  It follows
\cite{sch-analyse} that there exists a decomposition of unit
$\{\eta^\0_\G \}$ such that $\supp\eta^\0_\G \subset A^\0_\G$.  Then
for any $g\not=\G $
\be[e1/5.4]
     \supp \eta^\0_\G \cap  \pi_g = \emptyset,
\ee
and there exists a neighbourhood $\cO^\0_\G \supset \pi^\0_\G$
in which $\eta^\0_\G \equiv 1$.

\SUBSECTION{Isolating singularities of submaximal subgraphs.}
\label{Decomp.Submaximal}%{ss5.3}

The second important example: $G$ is an $s$-graph, $\cO=P_G
\backslash \{0\}$, and $\G $ runs over the set of submaximal
subgraphs, $\G \triangleleft G$.  Here all $\pi^\0_\G$ are
closed and pairwise non-intersecting in $\cO$, so that this case is
quite similar to the one studied in the preceding subsection.
Moreover, due to the scale invariance of all the relevant geometric
structures one can construct a decomposition of unit $\{\theta_{\G
}\}$ such that
\be[e1/5.5]
     \theta_\G (\l \, p)
     = \theta_\G(p), \qquad p\not=0, \ \ \l >0.
\ee
The idea of the construction is to consider a smooth sphere $|p|=1$
instead of $\cO$, and instead of the singular planes to consider their
intersections with the sphere. A decomposition of unit $\{\theta_{\G
}\}$ on the sphere is constructed following the scheme of the
preceding subsection and then the functions $\theta_\G$ are
extended onto the entire $\cO$ using \eq{e1/5.5}.

If $S[G]=\{G,\emptyset \}$ then the only submaximal $s$-subgraph in
$G$ is the empty subgraph $\emptyset $. In this case
$\theta_{\emptyset}(p) \equiv 1$.

Note an important geometric property of the supports  of  functions
$\theta_\G$ thus constructed:
\be[e1/5.6]
     \supp \theta_\G \cap \pi_g
     = \{0\} \quad\hbox{\ \ \ for\ any}\quad g\notin \G.
\ee
In this case the products $\theta_\G(p)\, \GG(p)$ have only one
singularity which is localized at the point $p=0$.  This is important
because a relation similar to \eq{e1/1.3}, namely:
\be[e1.3']
     \hbox{\bf Op} \. G \Big|_{p_G\not= 0}
     =
     \sum_{\G \triangleleft G} 
          \left[\, \theta_\G\, \GG \,\right]
          \, \hbox{\bf Op} \. \G,
\ee
defines $\hbox{\bf Op} \. G$ everywhere away from $p=0$.  The problem
is thus reduced to studying an isolated singularity, which represents
a radical simplification. Methods to deal with singularities localized
at an isolated point will be described in
\sects{IsolatedSing}-{Subtractions}.

\SUBSECTION{Decompositions of unit $\theta_\G$ and Hepp sectors.}
\label{Decomp.Hepp}

The relation \eq{e1.3'} reflects the philosophy of constructing
$R$-like operations on a graph inductively using the corresponding
constructions on its subgraphs as building blocks. We are then not
interested in the internal structure of the latter. In the BPHZ
philosophy, on the contrary, one ``resolves" the singularities of $G$
by, essentially, iterating relations \eq{e1.3'}.  One then
arrives at a sum of terms, each of which contains a typical factor
\be
     \theta_\G\,\theta_{\G'}\,\theta_{\G''} \ldots,
     \qquad 
     G \triangleright \G \triangleright \G' \triangleright \G'' \ldots ,
\ee
Such products describe subregions in the integration space which are
equivalent to what is known as Hepp sectors in the techniques of
$\a $-parametric representation \cite{zav79}.

\SUBSECTION{Radial variables and cutoff functions.}
\label{Decomp.Radial}%{ss8.5}

Specifics of distributions do not allow one to perform some standard
manipulations like introduction of radial variables and cutoffs in a
straightforward manner. Rather, one has to resort to (standard) tricks
like decompositions of unit into spherical layers.

Let us fix $\Phi(p)\in \cD(P)$ such that
\be[e1/8.12]
     \Phi(p) \equiv 1,\qquad |p| \leq \const.
\ee
Denote:
\be[e1/8.13]
     \phi(p) \bydef {d \over d\l} \Phi(p/\l )\Big|_{\l=1}.
\ee
Then
\be[e1/8.14]
     \int^{\infty}_{0}{d\l \over \l} \, \phi(p/\l )
     \equiv 1, \qquad p\not=0,
\ee
so that the functions
\be[e1/8.15]
     \phi_\l(p)\bydef \phi(p/\l )
\ee
realize a continuous decomposition of unit into ``spherical layers" in
$P\backslash \{0\}$. Then \eq{e1/8.14} can be used to reduce an
integral over the entire $P$ to integrals over spherical layers of
radii $O(\l)$:
\be
     \int dp\,F(p) = \int_0^\infty \frac{d\l}{\l} \int dp\,F(p)\phi(p/\l).
\ee

It is also convenient to define {\it cutoff functions}
as follows:
\be
     \Phi^\L_{\e}(p)
     \bydef
     \int^\L_{\e}{d\l \over \l}\, \phi_\l(p),\kern15mm
\ee
\be[e1/8.16]
     \Phi^\L(p) \bydef \Phi^\L_{0}(p) \equiv \Phi(p/\L ),\kern8mm
\ee
\be
     \Phi_{\e}(p) 
     \bydef     \Phi^\infty_{\e}(p) \equiv 1-\Phi(p/\e ).
\ee

\SUBSECTION{Cutoffs and subgraphs.}
\label{Decomp.Subgraphs}

In the case of factorizable graphs, it is convenient to choose the upper
cutoff functions $\Phi^\L$ to be also factorizable. In particular, if
a graph $G$ is factorized, $G=G_1\times G_2$, then the space
$P_G$ is decomposed into the direct sum $P_{G_1} \oplus P_{G_2}$.
In this case it is convenient to choose
\be[e1/8.17]
    \Phi(p) \bydef \Phi_1(p_1) \times \Phi_2(p_2),
\ee
where $p=p_1+p_2$ while $p_1\in P_{G_1}$, $p_2\in
P_{G_2}$.  Then for the upper cutoff function one obtains:
\be[e1/8.18]
     \Phi^\L(p) = \Phi^\L_1(p_1) \times \Phi^\L_2(p_2).
\kern1.8mm\ee
Factorization properties---or the lack thereof---of other functions
($\phi_\l$, $\Phi_\e$ and $\Phi^\L_\e$) are of no consequence.

Now consider a non-factorizable graph $G$.  Simple geometrical
considerations allow one to choose $\Phi_G$ so that for each $\G \in
S[G]$ its projection on $p^\0_\G=0$ has a factorized form that
corresponds to the factorization of the co-subgraph $\GG$ into a
product of $\xi_i$ (cf.\ \eq{Eq.Co-subgraph.Def}):
\be[H.Factorized]
     \Phi_G(p^\0_G) \Big|_{p^\0_\G=0}
     = \prod_i \Phi_{\xi_i} (p^\0_{\xi_i}).
\ee
Moreover, one can choose $\Phi_G$ so that
\be[H."Flat"]
     D^{\b}_\G \Phi_G(p^\0_G) \Big|_{p^\0_\G=0} = 0.
\kern22.4mm\ee
Strictly speaking, the latter condition depends on the choice of the
variables $p^\0_\G$, but this is of no consequence except
that the variables $p_\g$ and $p^\0_\G$
for any pair $\g < \G$ should be chosen consistently.  Indeed,
choosing $p^\0_\G$ implies fixing a subspace $\pi^{\rm
T}_\G$ that is transverse to $\pi^\0_\G$.  Then the relation
\eq{H."Flat"} says that $\Phi_G$ does not change in the directions
parallel to $\pi^{\rm T}_\G$ in small neighbourhoods of $\pi_{\G
}$.  On the other hand, $\pi_\g \supset \pi^\0_\G$.  Then
for consistency of the set of requirements \eq{H.Factorized} and
\eq{H."Flat"} it is sufficient to choose $\pi^{\rm T}_\g \subset
\pi^{\rm T}_\G$.  The ordering among subgraphs ensures that such
conditions can be satisfied for all pairs of subgraphs simultaneously.

It is easiest to understand \eq{H.Factorized} and \eq{H."Flat"} by way
of visualization: one can e.g.\ consider $G$ such that the space
$P_G$ is three-dimensional: $p^\0_G=(x,y,z)$ and the structure of $G$
is, say, $G(x,y,z) = \D (x)\D (y)\D (x-z)\D
(y-z)\D (z)$. In this example a non-trivial factorization of
co-subgraph takes place for the $s$-subgraph consisting of the last
factor; its singular plane is described by $z=0$.

As a rule, our reasoning will not depend on the specific choice of the
cutoffs within the restrictions specified above.

\SECTION{Techniques for description of isolated singularities.}
\label{IsolatedSing}%6

First in \ssect{Isolated.Lambda} our concern is how to describe the
leading power asymptotic behaviour without being distracted by
irrelevant ``soft" (i.e.\ logarithmic etc.) factors. To this end we
develop a simple but extremely useful formalism of the so-called
$\L$-functions which is similar to the familiar techniques of the
Bachmann-Landau symbols $O$ and $o$ \cite{Oo}.  Then after
introducing some notations in
\ssects{Isolated.Deltas}-{Isolated.Seminorms2}, in
\ssect{Isolated.d-Inequalities} we present the techniques of so-called
{\it $\cS$-inequalities} for description of singularities of
distributions. Practically all our proofs will be expressed in the
language of the $\cS$-inequalities.

Throughout this section $P$ is a normed space of $N$ dimensions. (As a
rule $P=P_G$ for a non-factorizable $s$-graph.) We assume that a
coordinate system is fixed in {\it P}.

\SUBSECTION{$\L$-functions.}
\label{Isolated.Lambda}%{ss6.1}

To study singularities of a function means, essentially, to evaluate
its asymptotics at the corresponding points%
\footnote{
One can see that the problem of constructing \rop\ is closely
connected with the problem of constructing \asop\ which will be
considered later.  Indeed, one can prove the theorems about \rop\ and
\asop\ simultaneously. The resulting complicated inductive pattern is
such that the proof of \rop\ for a graph makes use of the \asop\ on
its subgraphs.  To simplify logistics, however, we opt for a somewhat
longer but more transparent separate treatment.
\relax}.  
In applications one normally deals with power asymptotic behaviour
modified by soft multipliers which are powers of logarithms.  The
essential information for constructing distributions etc.\ is the
exponent of the power part of the asymptotics which can be evaluated
by power counting.  On the other hand, to determine the exact form of
the soft corrections in the most general case would be both difficult
and superfluous. The formalism of $\L $-functions introduced here
shields one from the irrelevant details of the structure of the soft
part of the asymptotics and allows one to concentrate on the power
behaviour.

A non-negative locally integrable function $\L (t)$, $0<t<\infty $, is
called {\it double-sided} $\L $-{\it function} if its growth rate at
$t\rightarrow +0$ (t$\rightarrow +\infty$) does not exceed that of
any negative (positive) power of $t$; more precisely, for any $\a
>0$,
\be[e1/8.8]
     \L (t) = o(t^{-\a}), \qquad t\rightarrow +0,\kern2.3mm
\ee
\be[e1/8.9]
     \L (t)=o(t^{+\a}), \qquad t\rightarrow +\infty.
\ee
Examples  of $\L $-functions  are:  a  constant,  a   polynomial   of
$|\log t|$, $|\log|\log t||^{1/2}$  etc.

The use of $\L $-functions is very similar to that of the widely used
symbols $O$ and $o$ \cite{Oo}. First, one may write
\be[e1/8.10]
     f(t)=\L (t)\kern4mm
\ee
in order to indicate that a specific function $f(t)$ belongs to the
class of $\L $-functions. Second, $\L $-functions can be conveniently
used in inequalities like
\be[e1/8.11]
     |f(t)|\leq t^N\, \L (t),
\ee
which means that $f(t)$ possesses power singularities at $t\rightarrow
+0$ and $t\rightarrow +\infty $ of order not higher than $t^N$,
perhaps modified by soft factors.  Indices distinguishing different
$\L$-functions are usually omitted as well as the words ``there exists
a $\L$-function such that $\ldots $" in descriptions of such
inequalities.

A major advantage of the formalism of $\L $-functions is that it
simplifies evaluation of estimates like \eq{e1/8.11} because the class
of $\L $-functions is closed with respect to a number of operations:

 ($a$) \hskip2em
$\displaystyle
\a \, \L (t)+\b \, \L (t)
=\L (t), \qquad \a, \b \geq 0$;

 ($b$) \hskip2em
$\displaystyle
\L (t) \times \L (t)=\L (t)$;

 ($c$) \hskip2em
$\displaystyle
[\L (t)]^\a=\L (t),\qquad \a \geq 0$;

 ($d$) \hskip2em
$\displaystyle
\L (\a \, t)=\L (t),\qquad \a >0$;

 ($e$) \hskip2em
$\displaystyle
\int^{t}_{0}
{dt'\over t'} \, {t'}^\a
\, \L (t')
=
t^\a\, \L (t), \qquad \a >0$;

 ($f$) \hskip2em
$\displaystyle
\int^\infty_{t} {dt'\over t'}
\,
{t'}^{-\a}
\, \L (t')
= t^{-\a} \, \L (t), \qquad\a >0$;

 ($g$) \hskip2em
$\displaystyle
\left|
\int^{M}_{t}{dt'\over t'} \, \L (t')
\right|
= \L (t), \qquad M>0$;

 ($h$) \hskip2em
$\displaystyle\mathop{
\left\{
     \mathop{\vcenter{\hbox{$\sup$}\hbox{$\inf$}}}
\right\}}_{\a t \leq t' \leq \b \, t}
{t'}^\g \, \L (t')
=
t^{\g} \, \L (t),
\qquad \b \geq \a \geq 0$;

 ($i$) \hskip2em
$\displaystyle
\sup_{k=k_1 \ldots k_n} \L_k(t)=\L (t)$;

 ($j$) \hskip2em $1=\L(t)$.

The double-sided $\L $-functions  are  primarily  useful  in  the
study  of products of functions that possess certain scaling properties;
such functions emerge in the problem of asymptotic expansions. In
applications to the theory of UV renormalization in coordinate
representation which we  consider  as  an example  of  application  of
our  techniques,  one  is  only  interested  in asymptotics at
$t\rightarrow +0$. Then  it  is  convenient  to  consider  {\it
single-sided} $\L $-functions, i.e.\  possessing only the  property
\eq{e1/8.8}  but  not  necessarily \eq{e1/8.9}. For single-sided
$\L$-functions all the above properties are preserved except for ($f$)
and ($j$) instead of which the following takes place:

 ($f'$) \hskip2em
$\displaystyle
\left|
\int^{M}_{t} {dt'\over t'} \, {t'}^{-\a} \, \L (t')
\right|
= t^{-\a}\, \L (t), \qquad \a \geq 0, \quad M>0$,

\noindent
where, as in ($g$), $M$ is an arbitrary constant; 

 ($j'$) \hskip2em
$\displaystyle t^\a=\L (t), \qquad \a \geq 0$.

We will use $\L$-functions in description of properties of singular
functions for which our constructions are valid (cf. \eq{e2/7.2}); in the
definition of the seminorms $\cS$ (\ssect{Isolated.Seminorms2}); in
studying properties of subtraction operators (\sect{Subtractions}).
In the main proofs $\L$-functions will be mostly hidden within $\cS$'s
except for the case of asymptotic expansions where one has to deal
with several dependences and $\L$-functions reappear explicitly.

\SUBSECTION{$\d^\a$  and $\cP^\a$.}
\label{Isolated.Deltas}%{ss8.2}

Following standard practice, define a multi-index $\a $ as a
finite sequence of integer non-negative numbers $(\a_1\ldots
\a_N)$. Standard functions on multiindices are: the absolute
value $|\a|=\sum \a_i$; the factorial $\a!=\prod \a_i!$; the exponential
$p^\a=\prod p_i^{\a_i}$ where $p$ is an $N$-dimensional vector; the
partial derivative $D^\a\vf(p) = \vf^{(\a)}(p)$.

More generally, let $\a$ be a discrete label. Define the following
objects:

$\cP^\a(p)\bydef$ a complete set of homogeneous polynomials of $p\in
P$. The index of homogeneity is denoted as $|\a|$.

$\d^\a(p)\kern1mm\bydef$  a set of derivatives of the $\d$-function
which is dual to $\cP^\a$:
\be[e1/9.7]
     \langle \d^\a ,\cP^{\b}\rangle
     =
     \d^{\a \b} \qquad \hbox{(the Kronecker symbol)}.
\ee
Using the above notations, the Taylor expansion at $p=0$ can be
rewritten as follows:
\be[e1/9.8]
     \T^{\o}\. \vf
     = \sum_{|\a|\leq\o}
     \cP^\a \,
     \langle \d^\a,\vf \rangle.
\ee
It may be necessary (cf.\ \sect{NonScalarFactors} and, e.g.,
\eq{e3/2.11}) to assume that $\d^\a$ takes values in the space
$\cB_G$, then $\cP^\a$ takes values in the dual space $\cB^{\ast}_G$.

We do not specify the exact form of $\cP^\a$ and $\d^\a$
because it may be convenient in applications to choose these objects
to satisfy additional requirements, e.g.\ Euclidean covariance etc.
Still, if summations over full sets are involved (as e.g.\ on the
r.h.s.\ of \eq{e1/9.8}) then the results are often insensitive to the
specific choice and one can take e.g.\
\be[P&Delta.Def]
     \cP^\a(p) = p^\a,    
     \quad\hbox{and}\quad
     \d^\a(p) = \frac{(-)^\a D^\a}{\a!}\d(p),
\ee
with $\a $ being the usual multiindex as described in the
beginning of this subsection.

It will often be necessary to consider $\d^\a$'s and
$\cP^\a$'s on different spaces, e.g.\ on the spaces $P_\G$ for
graphs $\G$ from our universum.  In such cases it is convenient to use
subscripts as $\d^\a_\G$ and $\cP^\a_\G$.

\SUBSECTION{Seminorms $\|\cdot\|$.}
\label{Isolated.Seminorms1}%{ss6.2}

For any integer $k\geq 0$ and subset $A\subset P$ define:
\be[e1/8.5]
     \|\vf\|^k_{p\in A}
     \bydef
     \sup_{p\in A, |\a|=k}
     |D^\a\vf (p)|.
\ee
If the set $A$ is not specified, $A=P$ is assumed.  If $\vf$ takes
values in a normed space $\cB$ then $|\cdot|$ in \eq{e1/8.5} is the
$\cB$-norm.

We will make use of the following elementary proposition: Assume that
$\vf \in C^\infty(P)$ has zero of order $\o$ at $p=0$ and let
$K$ be a convex set containing the zero point. Then for any
non-negative integer $\o' \leq \o $
\be[e1/8.7]
     \|\vf\|^{\o'}_{p\in K}
     \leq
     \const\,\|\vf\|^{\o}_{p\in K} \, (\rad K)^{\o -\o'},
\ee
where the radius of a set is defined as:
\be[e1/8.2]
     \rad\, A \bydef \sup_{p\in A}|p|.
\ee
The result \eq{e1/8.7}  easily follows from the Taylor theorem.

\SUBSECTION{Seminorms $\cS$.}
\label{Isolated.Seminorms2}%{ss6.3}

On the right hand sides of inequalities describing singularities of
distributions, the following aggregates appear regularly:
\be[Seminorm.Def]
     \cS_{p\in A}[\vf ,d]
     \,\bydef\,
     \sum_{k\geq 0}\L_k(d) \, d^k\|\vf\|^k_{p\in A},
\ee
where $d$ is a constant (normally, $d\geq \rad\,\supp\vf$),
$\L_k$ are $\L$-functions whose exact form is inessential, and
summation runs over integer $k$ not exceeding a certain finite value
which is also inessential.  The use of the seminorms $\cS$ is,
to an extent, similar to the use of $\L $-functions (see the following
subsection).

The seminorms $\cS$, along with the usual properties of seminorms,
possess the following simple properties:

($a$)
if $A\subset A'$ then $\cS_{p\in A}[\vf ,d]
\leq \cS_{p\in A'}[\vf ,d]$.

($b$) {\it Scaling}:
if $t>0$  then $\cS_{p\in A}[\vf (t\, p),d]
=\cS_{p\in tA}[\vf (p),d/t]$.

($c$) {\it Factorization on products}:
for any $\vf_1$ and $\vf_2$
\be
     \cS_{p\in A}[\vf_1 \times \vf_2,d]
     \leq
     \cS_{p\in A}[\vf_1,d] \times \cS_{p\in A}[\vf_2,d].
\ee
This property is especially useful in formalized versions of power
counting.

($d$) We will also use other quite obvious relations e.g.\
$\a\cS[\vf,d]+\b\cS[\vf,d]\leq\cS[\vf,d]$ and
$\L(d)\cS[\vf,d]\leq\cS[\vf,d]$ etc.  The point here is that the
implicit upper limits of summation as well as $\L$-functions in
different occurences of $\cS$ (cf.\ the definition \eq{Seminorm.Def})
can be different, whence the inequality signs.

\SUBSECTION{$\cS$-inequalities.}
\label{Isolated.d-Inequalities}%1/8.6{ss6.4}

Let us show how the above notions are used to describe the singularity
of a distribution at an isolated point.  The pathos of the technique
of $\cS$-inequalities which we are now going to introduce is to stress
{\it the essential similarity of proofs between the cases of
distributions and ordinary functions}. A very important consequence of
such similarity is that ``rigorous" proofs become a rather
straightforward exercise once a result has been established by naive
power counting.

Given a distribution $\cF$ defined on $P$ without the zero point,
\be[e1/8.19]
     \cF\in \cD'(P\backslash \{0\}),
\ee
we would like to formalize the statement that $\cF$ behaves at
$p\rightarrow 0$ as
\be[e1/8.20]
     \cF(p) \sim p^{-\o -\dim P}, \qquad p\rightarrow 0.
\ee
The index $\o$ is defined so as to serve as an indicator of when an
integral of $\cF$ around $p=0$ exists in some natural sense (cf.\
\sect{Subtractions}): it is expected to exist for $\o<0$.

For ordinary function, the formalized version of \eq{e1/8.20} reads:
\be[e1/8.20.1]
     \cF(p) \leq |p|^{-\o -\dim P}\,\L(p).
\ee
But distributions that we are going to deal with will, in general,
possess singularities localized on manifolds passing through $p=0$, so
that their values are ill-defined along certain $p\not=0$ and
inequalities \eq{e1/8.20.1} become meaningless. Therefore, consider
$\vf \in \cD(P\backslash \{0\})$ and $\vf_\l(p)\equiv\vf
(p/\l )$.  Then the following estimate can be taken as a replacement
of \eq{e1/8.20.1} in the case of distributions:
\be[e1/8.21]
     |\langle \cF,\vf_\l \rangle| \leq \l^{-\o} \, \L (\l ).
\ee
In this inequality, however, the information about $\vf $ on the
r.h.s.\ is hidden within the $\L $-function. For our purposes, it will
be necessary to make the dependence on $\vf$ more explicit (which
is necessary in order to deal with the test functions defined as
products). On the other hand standard inequalities of the distribution
theory which have the form
\be[e1/8.22]
     |\langle \cF,\vf \rangle|
     \leq \sum^N_{k=0} C_k \,\|\vf\|^k
\ee
contain no information on the nature of singularity of $\cF(p)$ at
$p\rightarrow 0$.

Convenient estimates that combine advantages of inequalities
\eq{e1/8.21} and \eq{e1/8.22} make use of the seminorms $\cS$
introduced in the preceeding subsection:
\be[e1/8.23]
     |\langle \cF,\eta_\l \, \vf \rangle| 
     \leq \l^{-\o} \cS [\vf ,\l ].
\ee
Here $\vf$ may belong to $\cD(P)$ while $\eta_\l(p) \equiv \eta
(p/\l )\in \cD(P\backslash \{0\})$.  (In particular, one can take
$\eta=\phi$, the spherical layer function defined in
\ssect{Decomp.Radial}.) Note that the explicit form of $\cS$ here 
depends on $\eta_\l$. 

It is convenient to call the index $\o$ on the r.h.s.\ of
\eq{e1/8.23} {\it effective divergence index at $p=0$} of the
distribution $\cF\in \cD'(P\backslash \{0\})$. ``Effective'' because
it can, in general, depend on how the singularities of $\cF$ at
$p\not=0$ are treated (\ssect{Subtraction.Oversubtraction}).

Another convenient variant of \eq{e1/8.23} (for the case of
distributions defined on the entire $P$: $\cF\in\cD'(P)$) exhibits
dependence on the radius of support of $\vf
\in \cD(P_G)$ instead of the artificial parameter $\l$:
\be[e1/8.24]
     |\langle \cF,\vf \rangle| \leq d^{-\o}\cS[\vf ,d],
\ee
where
\be[e1/8.25]
     d = \rad \supp \vf.
\ee
The notation \eq{e1/8.25} together with the name ``$\cS$-inequality"
for expressions like \eq{e1/8.23} and \eq{e1/8.24} will be standard
throughout this work.

Taking into account that $\rad\supp\vf_\l= O(\l)$ and using
properties of $\L $-functions, one can easily restore \eq{e1/8.21}
from \eq{e1/8.24}.

A useful fact is that the inequality \eq{e1/8.24} (with the same
$\cS$) holds for $\vf'$ such that
\be
     d'=\rad\supp \vf' \leq d.
\ee
This is because one can always add $\e \, \d \vf$ to such
$\vf'$ provided
\be
     \rad \supp (\vf' + \e \, \d \vf )= d.
\ee
Then the l.h.s.\ of the inequality for $\vf'$ analogous to
\eq{e1/8.24}  tends  to $\langle\cF,\vf' \rangle$
as $\e \rightarrow 0$ while the r.h.s.\ becomes the same as in
\eq{e1/8.24}.

In the following section further examples of the use of
$\cS$-inequalities can be found.

\SECTION{Subtraction operators.}
\label{Subtractions}%{ss9}

As indicated in \sect{Decompositions}, decompositions of unit and
natural recursions allow one to reduce the general problem of defining
$R$-like operations to a much simpler case of isolated singularity.
In this section we study the so-called {\it subtraction operators}
which form the basis for solving this simpler problem of transforming
a distribution $\cF(p)\in \cD'(P\backslash \{0\})$ (i.e.\ a
distribution specified on the space $P$ without one point) into a
distribution that is well-defined over entire~$P$, $\r \. \cF(p) \in
\cD'(P)$. The construction corresponding to our subtraction operators
is well-known in the theory of distributions---it is the details of
the construction and notations that we are interested in.

The class of functionals $\cF \in \cD'(P\backslash \{0\})$ for which
the subtraction operators are defined is specified in
\ssect{Subtraction.Problem}.  In \ssect{Subtraction.Extension} we
demonstrate that such $\cF$ have a natural extension onto a subspace
in $\cD(P)$ of finite codimension. In \ssect{Subtraction.r} we define
``generic subtraction operators" $\r$. An $\cS$-inequality for the
extended functional $\r\. \cF$ is proved in
\ssect{Subtraction.d-Inequalities}.  In order to clarify the effect of
subtractions on the singularity of resulting distributions, in
\ssect{Subtraction.Oversubtraction} we consider the case of the 
so-called oversubtractions.  Then in
\ssect{Subtraction.UniformProblem} we turn to the extension problem
for distributions possessing uniformity properties, and the
corresponding ``special" subtraction operators are studied in
\ssect{Subtraction.r-tilde}.  Their characteristic property is
exhibited in the last \ssect{Subtraction.Minimality}. Note that the
construction of special subtraction operators is essentially similar
to that of the standard one-dimensional distributions $x_{\pm}^{-n}$
(see e.g.  \cite{gel-shi}).

\SUBSECTION{Extension problem for singularity localized at an
isolated point.}
\label{Subtraction.Problem}%{ss9.1}

Consider a class of functionals $\cF\in \cD'(P\backslash \{0\})$ such
that for any $\eta (p)\in \cD(P\backslash \{0\})$ and $\vf (p)\in
\cD(P)$ an estimate of the form \eq{e1/8.23} holds:
\be[e1/9.1]
     |\langle \cF, \eta_\l \, \vf \rangle|
     \leq \l^{-\o} \cS_{p\in \supp \eta_\l} [\vf ,\l ],
\ee
where $\eta_\l(p) \equiv \eta (p/\l )$; the $\L $-functions in the
definition of $\cS$ are assumed to be single-sided (the case of
double-sided $\L $-functions is considered in
\ssect{Subtraction.UniformProblem}).  The number $\o$ is 
referred to as {\it effective divergence index\/} of the
distribution $\cF(p)$ at $p=0$ (cf.
\ssect{Subtraction.d-Inequalities}).

Note that in the above inequality $\eta$ is considered as fixed, and
the exact form of the seminorm on the r.h.s.\ depends on it.

We wish to construct an operator $\r$ which extends the functional
$\cF\in \cD'(P\backslash \{0\})$ to the entire $\cD(P)$ (i.e.\ $\r\.
\cF\in \cD'(P))$ without changing it on the test functions from
$\cD(P\backslash \{0\})$.

Formally speaking, this is an extension problem in the sense that one
has to extend the functional $\cF$ defined on the subspace of test
functions which are identically zero around $p=0$, $\cD(P\backslash
\{0\})$, onto the space of all test functions on $P$, $\cD(P)$.  This
is similar to the Hahn-Banach problem except that there is no single
seminorm that bounds the functional to be extended (cf. the fact that
$\l$ in \eq{e1/9.1} takes arbitrary valuesand that such a bound
contains a non-trivial information on singularity of $\cF$).  Another,
less formal (and, probably, heuristically more useful) interpretation
is to say that the domain of definition of $\cF(p)$---namely,
$P\backslash
\{0\}$---is to be extended to include the point $p=0$ without changing
$\cF(p)$ at $p\not=0$.

The first point of view (in terms of functional spaces) is better
suited for proving existence of a solution, while the second point of
view helps one to understand the fact that the extension essentially
consists in subtracting (or removing) the singularity of $\cF$ with
the so-called counterterms that must be localized at $p=0$.

\SUBSECTION{Extension to a subspace of finite codimension.}
\label{Subtraction.Extension}%{ss9.2}

Suppose first that the divergence index is non-negative $(\o \geq 0)$.
Denote as $\cD_{\o+1}(P)$ the closed subspace of $\cD(P)$ formed by
functions possessing zero of order $\o +1$ at $p=0$, i.e.\ turning
to zero at this point together with all their derivatives of order
$\leq \o $.  Let us demonstrate that there exists a natural extension
of $\cF$ from $\cD(P\backslash \{0\})$ onto $\cD_{\o +1}(P)$ which is
defined as follows:
\be[e1/9.2]
     \langle \cF \ast \vf \rangle 
     \bydef
     \lim_{\e \rightarrow 0}\;
     \langle
          \cF,\Phi_\e\, \vf
     \rangle, \qquad \vf \in \cD_{\o +1}(P),
\ee
where $\Phi_\e$ is a cutoff defined in \ssect{Decomp.Radial}.
``Natural" means here that the limit in \eq{e1/9.2} does not depend on
the special choice of $\Phi$. (The symbol $\ast$ usually denotes
convolution which we never use so that confusion is excluded.)

Consider the expression under the limit in \eq{e1/9.2}.  Using the
integral representation \eq{e1/8.16} for $\Phi_\e$, represent it
as:
\be[e1/9.3]
     \int^{d\, \const}_\e
     {d\l \over \l} \,
     \langle \cF,\phi_\l\, \vf \rangle,
\ee
where the constant depends only on the choice of $\Phi_\l$. The
integrand in \eq{e1/9.3} can be estimated using \eq{e1/9.1} with $\eta
(p)=\phi(p)$.  Since $\vf $ has zero of order $\o +1$ at $p=0$, we may
use the property ($e$) of $\L$-functions (\ssect{Isolated.Lambda}) and
estimate the seminorm $\|\cdot\|^k$ by $\|\vf\|^{\o +1} \,
\const \, \l^{\o +1-k}$ for $k\leq \o $, leaving $\|\vf\|^k$
unaffected for $k\geq \o +1$. Then \eq{e1/9.1} is transformed into
\be[e1/9.4]
     |\langle
          \cF,\phi_\l\, \vf
      \rangle|
      \leq \l^{-\o} \, \sum_{k\geq \o +1}\|
      \vf\|^k \, \l^k \, \L(\l ).
\ee
Since summation here  runs  from $k=\o +1$,  it  follows,  first,
that  the integral \eq{e1/9.3} exists in the limit $\e \rightarrow 0$
and,  consequently,  the  limit  in \eq{e1/9.2} exists as well. Second, making
use of the properties of $\L$-functions one obtains the following
estimate from \eq{e1/9.3} and \eq{e1/9.4}:
\be[e1/9.5]
     |\langle \cF\ast \vf \rangle|
     \leq d^{-\o}\, \sum_{k\geq \o +1}\|\vf\|^k
     \, d^k \, \L (d).
\ee
Independence of the limit \eq{e1/9.2} of the choice of $\Phi$ is verified
as follows. Let $\Phi'$ be another cutoff. Then
\be[e1/9.6]
     \langle\cF,\Phi_\e\, \vf \rangle
     -
     \langle\cF,\Phi'_\e\, \vf \rangle
     = \langle \cF,\eta_\e\, \vf \rangle,
\ee
where $\eta_\e(p)=(\Phi-\Phi')(p/\e )$.  Using the estimate
\eq{e1/9.1} once again and repeating the above reasoning, one obtains
an estimate of the form \eq{e1/9.4} with $\l $ replaced by $\e
$. It follows from the estimate that in the limit $\e
\rightarrow 0$ the difference \eq{e1/9.6} vanishes as required.

For $\o <0$ a similar reasoning (without using the property ($e$),
\ssect{Isolated.Lambda}) proves existence of
$\langle\cF\ast \vf\rangle$ for any $\vf \in \cD(P)$ and its
independence of the shape of the cutoff, as well as an estimate
differing from \eq{e1/9.5} only in that the summation starts at $k=0$
instead of $\o +1$.

\SUBSECTION{Generic subtraction operator $\protect\r$.}
\label{Subtraction.r}%{ss9.4}

For the case $\o <0$ we simply require that the functional $\langle
\r\. \cF , \vf \rangle$ coincide with $\langle \cF \ast \vf
\rangle$ on the entire $\cD(P)$.

If $\o \geq 0$, we may at most require that $\r\. \cF$ coincide with
$\cF\ast $ on the subspace $\cD_{\o+1}(P)$.  This requirement emerges
naturally in most of the interesting problems where a need for the
construction \eq{e1/9.2} arises.  A weaker version of this requirement
is discussed in \ssect{Subtraction.Oversubtraction}.  

The codimension of the space $\cD_{\o+1}(P)$ in $\cD(P)$ is finite so
it is sufficient to define $\r\. \cF$ on a finite number of vectors
transverse to $\cD_{\o +1}(P)$.  For instance the vectors $\Phi^{\mu}
\, \cP^\a\in \cD(P)$, $|\a|\leq \o$ (where $\Phi^\mu$ is the cutoff from
\ssect{Decomp.Radial} are transverse to $\cD_{\o+1}(P)$ and linearly
independent. Let
\be[e1/9.9]
     \langle
         \r \. \cF , \Phi^{\mu}\, \cP^\a
     \rangle
     = z_\a, \qquad|\a|\leq \o.
\ee
It is easy to verify that a functional satisfying all the above
conditions is uniquely represented in the form:
\be[e1/9.10]
     \r \. \cF=\r_{0}\. \cF+\D,
\ee
where 
\be[e1/9.11]
    \langle \r_{0}\. \cF,\vf \rangle
    \bydef
    \langle \cF  * \r^{+}_{0}\.\, \vf \rangle,\kern11mm
\ee
\be[e1/9.12]
     \r^{+}_{0} \. \vf \bydef
     (1-\Phi^{\mu} \, \T^{\o})\vf
\ee
and
\be[e1/9.13]
     \D
     \bydef
     \sum_{|\a|\leq \o} z_\a \, \d^\a.\kern1mm
\ee
Note that $\r^{+}_{0}$ is a projector $\cD(P) \rightarrow \cD_{\o
+1}(P)$.  All the arbitrariness of the above construction for $\r\.
\cF$, in particular the arbitrariness in the choice of the cutoff
function $\Phi^{\mu}$ or the cutoff parameter $\mu $, is reduced to the
arbitrariness in the choice of the constants $z_\a$.

As usual, $\D $ will be referred to as {\it finite counterterm} or,
more generally, {\it finite renormalization}.

Although the arbitrariness in the construction of $\r\. \cF$ is
adequately reflected in $\D $, in practice it is important to make a
convenient choice of $\r_{0}$ from the very beginning. In our case the
identity
\be[e1/9.14]
     \langle \r_{0}\. \cF, \Phi^{\mu}\, \cP^\a\rangle = 0,
     \qquad|\a|\leq \o,
\ee
holds for any $\cF$.  Below we will encounter the case when $\r_{0}$
satisfies a different version of this condition
(\ssects{Subtraction.UniformProblem}-{Subtraction.Minimality}).

\SUBSECTION {$\cS$-inequality for $\protect\r\.\cF$.}
\label{Subtraction.d-Inequalities}%{ss9.5}

Let us prove the following estimate for $\r\. \cF$:
\be[e1/9.23]
     |\langle {\bf r\. \cF},\vf \rangle|
     \leq   d^{-\o}\, \cS[\vf ,d],
\ee
which holds for any $\vf \in \cD(P)$ such that $\rad\supp\vf
\leq d$.

Rearrange the definition of $\r\. \cF$ \eq{e1/9.10}--\eq{e1/9.13} as
follows:
\be[e1/9.15]
     \langle \r \. \cF ,\vf \rangle = D + M + C ,
\kern3cm\ee
where 
\be
     D  \equiv  \langle \cF\ast (1-\Phi^{d}\, \T^{\o})\vf \rangle,
\kern17mm\ee
\be
     M \equiv \sum_{|\a|\leq \o}
     \langle \d^\a, \vf \rangle
     \, \int^{\mu}_{d} {d\l \over \l}
     \, \langle \cF, \phi_\l \, \cP^\a \rangle,
\ee
\be
     C \equiv \sum_{|\a|\leq \o}
     z_\a \, \langle\d^\a,\vf \rangle.
\kern26mm\ee

We will estimate each term in \eq{e1/9.15} separately.  Taking into
account the fact that
\be
     \rad\supp [(1-\Phi^{d}\, \T^{\o}) \vf] \leq d\, \const,
\ee
one obtains the following estimate from \eq{e1/9.5}:
\be[e1/9.16]
     |D|  \leq  d^{-\o}\,\sum_{k\geq \o +1} 
          \| (1-\Phi^{d}\, \T^{\o}) \vf\|^k 
          \, d^{k} \, \L (d).
\ee
The seminorms in \eq{e1/9.16} can be estimated using the definition
\eq{e1/9.8} as follows:
\be[e1/9.17]
     \|(1-\Phi^{d}\, \T^{\o})\vf\|^k
     \leq  \|\vf\|^k
        +  \sum_{|\a|\leq \o}
               |\langle\d^\a,\vf \rangle|
                \cdot \|\Phi^{d} \, \cP^\a\|^k.
\ee
Using elementary properties of the seminorms $\|\cdot\|^k$, the
definition of the cutoff $\Phi^{d}$ \eq{e1/8.16} and the homogeneity
of the polynomials $\cP^\a$ (\ssect{Isolated.Deltas}), one obtains:
\be[e1/9.18]
     \|\Phi^{d}\, \cP^\a\|^k
     = d^{|\a|-k}\, \const.
\ee
Finally, using the obvious estimate:
\be[e1/9.19]
     |\langle \d^\a,\vf \rangle|
     \leq
     \| \vf \|^{|\a|}\, \const,
\ee
one gets from inequalities \eqs{e1/9.16}-{e1/9.17}:
\be[e1/9.20]
     |D|\leq d^{-\o}\cS[\vf ,d].
\ee

The term $C$ is estimated as
\be[e1/9.21]
     |C|  \leq  \sum^{\o}_{k=0} \const \, \|\vf\|^k
          \leq  d^{-\o} \, \cS[\vf ,d],
\ee
which is obtained using the property $(j')$ of single-sided
$\L$-functions (\ssect{Isolated.Lambda}) and the definition
\eq{Seminorm.Def}.

To estimate $M$, apply to the integrand the estimate \eq{e1/9.1} and
the equality obtained from \eq{e1/9.18} by replacing $\Phi\rightarrow
\phi$, $d\rightarrow \l $. One obtains:
\be[e1/9.22]
    |M|  \leq   \sum^{\o}_{k=0}
    \|\vf\|^k \cdot \left|
        \int^{\mu}_{d} {d\l \over \l} \, \l^{k-\o}
        \, \L (\l )
    \right|
    \leq d^{-\o}\cS[\vf ,d].
\ee
We have used the property ($f'$) of single-sided $\L$-functions,
\ssect{Isolated.Lambda}.

Adding the estimates for $D, C$ and $M$, one arrives at
\eq{e1/9.23}.

\SUBSECTION{Remarks.}

$\quad$($i$) Note that even if in the initial estimate \eq{e1/9.1} for
$\cF$ the $\L $-functions are replaced by constants, one can not get
rid of the $\L $-functions in \eq{e1/9.23} because they possess a
logarithmic singularity at $d\rightarrow 0$, which is clear from the
integrals in
\eq{e1/9.22}.

($ii$) The exponent $\o $ in the estimate \eq{e1/9.23} can be conveniently
called, as in \eq{e1/9.1}, {\it divergence index at $p=0$} of
$\r\. \cF$.  Then the inequality \eq{e1/9.23} can be rephrased in the
following way: 

{\narrower\it
the operator\/ $\r$ defined in {\rm \eqs{e1/9.10}-{e1/9.13}}
does not change the divergence index of the distribution it acts on.
}

\noindent
This proposition is crucial for the BPH theorem in the theory of
$R$-operation (see below \ssect{GenericR.Theorem}), as it eventually
implies that the number of subtractions determined by simple power
counting prior to any subtractions is indeed sufficient to ensure
convergence of renormalized diagrams.  

($iii$) There will be an
operator $\r$ defined for each $s$-graph $G$ from the universum of
graphs of \sects{Graphs}-{CompleteSubgraphs}.  Such operators (i.e.\
the form of the cutoffs used in the definition, the finite
renormalizations $z_\a$ as well as the index $\o$) can be chosen
independently for each $G$.  The dependence on $G$ will be indicated
with subscripts in brackets: $\r^\0_{(G)}$, $\Phi^{\mu}_{(G)}$ etc.
Note that the conventions of
\ssect{Graphs.Graphs} remain valid here, so that in order to simplify
formulae, such a subscript may be omitted if it can be restored from
the context.

\SUBSECTION{Oversubtractions.}
\label{Subtraction.Oversubtraction}%{ss12.1}

In \ssect{Subtraction.r} we required that the functional $\r\. \cF$
coincide with $\cF$ on $\cD_{\o +1}(P)$, the subspace of $\cD(P)$
consisting of functions possessing zero of order $\o+1$ at $p=0$. It
is instructive to see what happens if this requirement is weakened
as it may sometimes happen in applications---cf.~\cite{zim70}.  It
should be stressed, however, that the theory of asymptotic
expansions in our interpretation, unlike \cite{zim70}, never requires
the use of oversubtractions, and we consider this case only because it
provides further insight into how the subtractions work.

Consider the following general case.  The class of functionals $\cF$
from $\cD'(P\backslash \{0\})$ satisfying the estimate \eq{e1/9.1}
forms a linear space which we denote as $L_{\o}$.  For $\O >\o$ the
inclusion $L_{\o}\subset L_{\O}$ takes place.  Consider two
functionals: $\cF_1\in L_{\O}\backslash L_{\o}$ and $\cF_2\in L_{\o
}$.  Then $\cF_1+\cF_2\in L_{\O}\backslash L_{\o}$.  Define an
operator $\r^{\O}_{0}$ by \eq{e1/9.11} and \eq{e1/9.12} with $\o$
replaced by $\O$. Then the following identity is true:
\be[e1/12.1]
     \r^{\O}_{0} \. (\cF_1+\cF_2)
     = \r^{\O}_{0}\. \cF_1 + \r^{\O}_{0}\. \cF_2,
\ee
where the operator $\r_{0}$ acting on $\cF_2$ on the r.h.s.\ is
defined with oversubtractions. As a result $\r^{\O}_{0}\. \cF_2$
coincides with $\cF_2$ not on $\cD_{\o +1}(P)$ (where $\cF_2$ is
defined) but only on its subspace $\cD_{\O +1}(P)$.

Consider a subtraction operator $\r^{\O}$ acting on $L_{\O}$
(it  is  convenient  to indicate the dependence of $\r$ on $\O $
explicitly):
\be[e1/12.2]
     \r^{\O}  \bydef (1-\Phi^{\mu}\, \T^{\O})^{+}+\D_G.
\ee
It is convenient to define $\r^{\O} \. \cF\bydef \cF\ast $ \ for $\O
<0$ (cf.\ \ssect{Subtraction.r}).

Consider a functional $\cF$ from $L_{\o}$, $\o \leq \O $.  Then the
result of application of $\r^{\O}$ to $\cF$ can be represented as (we
consider only the non-trivial case $\O \geq 0$):
\be[e1/12.3]
     \langle \r^{\O} \. \cF,\vf \rangle
     = \langle \r^{\o}\. \cF,\vf \rangle
     + \sum_{\o \leq|\a|\leq \O}
       \bar{z}_\a
      \,
      \langle \d^\a,\vf \rangle,
\ee
where $\bar{z}_\a =  
       ( z_\a 
       - \langle
             \cF, \Phi^{\mu}\, \cP^\a
         \rangle )
$.
The operator $\r^{\o}$ is well-defined and continuous on
$L_{\o}$.

So, the action of $\r^{\O}$ with oversubtractions is equivalent to
introducing an additional finite renormalization (``counterterms") of
order higher than $\o $, the effective divergence index of $\cF$.

Using \eq{e1/12.3}, one derives the following
estimate for $\r^{\O}\. \cF$, which is analogous to \eq{e1/9.23}:
\be[e1/12.6]
     |\langle  \r^{\O}\. \cF, \vf  \rangle|
     \leq
     d^{-\O}\cS[\vf ,d],
\ee
where $\vf \in \cD(P)$, $d=\rad\supp\vf$.  ($\cS$ here differs
from that in \eq{e1/9.23}, of course.)

This estimate has the form which is typical of functionals with the
divergence index $\O $, though initially the divergence index of $\cF$
was $\o<\O $. So, a natural way to describe it is to say that 

{\narrower\it
oversubtractions increase the effective divergence index by the number
of oversubtractions}.

\SUBSECTION{Extension problem for homogeneous distributions.}
\label{Subtraction.UniformProblem}%2/4.1-2/4.2

The variation of the extension problem that we are going to consider
next emerges naturally in the context of constructing the {\it
As}-operation.  Essentially, the functionals $\cF\in \cD'(P\backslash
\{0\})$ considered here have certain homogeneity properties so that
they exhibit the same power behaviour both at $p\rightarrow 0$ and
$p\rightarrow \infty $. Then the natural additional requirement on the
subtraction operators is that they should preserve that power
behaviour of $\cF$.

Consider a class of functionals $\cF\in \cD'(P\backslash \{0\})$ with
the same power behaviour both at $p\rightarrow 0$ {\it and\/} at
$p\rightarrow \infty $.  To be more precise, we assume that the
inequality \eq{e2/9.1} is still true but the $\L$-functions of which
$\cS$ is built are double-sided.  (Recall that the estimate
\eq{e2/9.1} carried no information on the behaviour of $\cF(p)$ at
$p\rightarrow \infty $ because the behaviour of single-sided $\L
$-functions at the infinity is undefined.)

Let us describe the natural extension of the domain of definition of
$\cF$.  Define the operation $\ast $ by analogy with \eq{e1/9.2}:
\be[e2/4.2]
     \langle \cF \ast f\rangle
     \bydef
     \mathop {\lim _{\e \rightarrow 0}} _{\L \rightarrow \infty}
     \langle  \cF, \Phi^\L_\e \, f  \rangle
\ee
where $f(p)\in C^\infty(P)$ and $\Phi^\L_\e(p)$ is a
cutoff function (cf.\ \ssect{Decomp.Radial}).  

From now on the symbol $\ast $ will denote the presence of limits
$\e \rightarrow 0$ and/or $\L \rightarrow \infty $.  It will
always be clear from the context which of these limits are implied.
For instance, in \eq{e2/4.16} below the functional $\tilde\r \. \cF$
is well-defined for any $\vf \in \cD(P)$ so that only the limit
$\L \rightarrow \infty $ needs to be introduced via $\ast $ there.

The expression \eq{e2/4.2} is defined correctly if the limits $\e
\rightarrow 0$ and $\L \rightarrow \infty $ exist and do not depend on
the choice of the cutoff function $\Phi$.  In that case the two limits
commute.

The results of \ssect{Subtraction.Extension} imply that the limit
$\e \rightarrow 0$ is defined correctly for all functions
$f(p)\in C^\infty(P)$ possessing zero of order $\o +1$ at $p=0$.
Moreover, since the reasoning of \ssect{Subtraction.Extension} can be
repeated under the assumption that all $\L$-functions are double-sided
(which is possible due to similarity of the algebra of single-sided
and double-sided $\L $-functions---\ssect{Isolated.Lambda}), then
for $f=\vf (p)\in \cD_{\o +1}(P)$ one immediately obtains the
estimate \eq{e2/9.5} with double-sided $\L $-functions.

Consider the limit $\L \rightarrow \infty $ in \eq{e2/4.2}. We are
interested in its existence for the functions $f$ not necessarily
rapidly decreasing at $p\rightarrow \infty $.  In particular, when
constructing and studying subtraction operators below we will
encounter expressions like the following one:
\be[e2/4.3]
      \langle \cF \ast \cP^\a\Phi_{d} \rangle,
\ee
where $\cP^\a(p)$ is a polynomial of order $|\a|$
(\ssect{Isolated.Deltas}).

To study this expression, rewrite it as:
\be[e2/4.4]
     \lim_{\L \rightarrow \infty}
     \int^\L_{d} { d\l \over \l}
     \, \langle \cF, \phi_\l\, \cP^\a \rangle.
\ee
(For simplicity we have chosen the function $\Phi^\L(p)=1-\Phi_{d}(p\,
d/\L )$ to play the role of the upper cutoff.)  The integrand in
\eq{e2/4.4} can be estimated using \eq{e1/9.1}. Taking into account
that
\be[e2/4.5]
     \cS_{p\in \supp\eta_\l} [\cP^\a,d]
     = \const \, \l^{|\a|},
\ee
one finds:
\be[e2/4.6]
     |\langle \cF, \phi_\l\, \cP^\a \rangle|
     \leq  \l^{|\a|-\o} \L (\l ).
\ee
It follows that for $|\a|<\o $ the integral \eq{e2/4.4} converges
at the upper bound, and the following final estimate is true for
\eq{e2/4.3}:
\be[e2/4.7]
     |\langle \cF\ast \cP^\a\Phi_{d} \rangle|
     \leq   d^{|\a|-\o}\L (d).
\ee
So, we have shown that the functional $\cF$ bounded by the inequality
\eq{e1/9.1} with double-sided $\L$-functions can be extended onto the
space of smooth functions possessing zero of order $\o +1$ at $p=0$
and, perhaps, growing at $p\rightarrow
\infty $ as a polynomial of order not higher than $\o-1$.

\SUBSECTION{Special subtraction operators $\tilde{\protect\r}$.}
\label{Subtraction.r-tilde}%2/4.3

The subtraction operator $\r$ constructed in \ssect{Subtraction.r}
could be applied to any functional $\cF$ from the class under
consideration.  However, the estimate \eq{e1/9.23} with {\it
single-sided\/} $\L $-functions would be obtained for $\r\. \cF$ though
for $\cF$ the inequality \eq{e1/9.1} with {\it double-sided\/}
$\L$-functions holds.  This is essentially due to the presence of
cutoffs in {\it all\/} terms of the Taylor expansion in
\ssect{Subtraction.r}, while, as it follows from the results
of the preceeding subsection, for existence of the ``subtracted"
expression only the cutoff in the last term is necessary.

Indeed, define $\tr$ in the following way (cf.\
\ssect{Subtraction.r}):
\be[e2/4.8]
       \tilde\r \. \cF
     \bydef 
       \tilde\r_{0} \. \cF
     + \tilde{\D}_{(G)},
\ee
where
\be[e2/4.9-]
     \langle \tilde\r_0\.\cF,\vf\rangle
     \bydef
     \langle \cF\ast\tilde\r_0^{+}\vf\rangle,
\kern13mm\ee
\be[e2/4.9]
     \tilde{\D}_{(G)}
     \bydef
     \sum_{|\a|=\o}
     z_\a \d^\a,
\kern6.5mm\ee
\be[e2/4.10]
     \tilde\r_{0}^{+}\.\vf \bydef (1-\T^{\o ,\mu})\vf
\kern3mm\ee
(here $\ast$ is defined in \eq{e2/4.2}), and the operator of Taylor
expansion with a cutoff in the highest terms is defined as:
\be[e2/4.11]
     \T^{\o ,\mu} \.\, \vf (p)
     \bydef
     \T^{\o -1}\.\, \vf (p)
     + \Phi^{\mu}(p)\, \sum_{|\a|=\o} \cP^\a(p)
     \, \langle\d^\a,\vf \rangle.
\ee
Such subtraction operators will be called {\it special subtraction
operators\/} to distinguish them from those used in 
\ssect{Subtraction.r}.

Note that the definition \eqs{e2/4.8}-{e2/4.10} is self-consistent in
the sense that the finite renormalization $\tilde{\D}$ describes that
and only that arbitrariness which is involved in the definition of
$\tilde\r_{0}$ (the arbitrariness in the choice of the cutoff $\Phi^{\mu
}$).

A straightforward modification of the reasoning of
\ssect{Subtraction.d-Inequalities} leads to the inequality \eq{e1/9.23}
with {\it double-sided} $\L $-functions implied in $\cS[\vf ,d]$.

\SUBSECTION {Minimality condition for special subtraction operators.}
\label{Subtraction.Minimality}%2/4.5

Lastly, let us prove an important characteristic property (which can
be called {\it minimality condition}) for the special subtraction
operator:
\be[e2/4.16]
     \langle\tilde\r \. \cF , \cP^\a\rangle=0,
\ee
where $\cP^\a$ is a homogeneous polynomial of order $|\a| < \o $.

Indeed, taking into account that
\be[e2/4.17]
     \cS [\Phi^\l\cP^\a,\l]
     = \const \,\l^{|\a|}
\ee
(cf.\ \eq{e1/9.18}) one obtains:
\be[e2/4.18]
     |\langle \tilde\r \. \cF, \Phi^\l\cP^\a \rangle|
     \leq  \l^{|\a|-\o}\L (\l ).
\ee
Taking the limit $\l \rightarrow \infty$ one arrives at \eq{e2/4.16}.

We are fully equipped to turn to products of singular functions with
more complicated patterns of singularities.

\newpage\thispagestyle{myheadings}\markright{}
\PART {\rop s for Products of Singular Functions.}

Our aim in the following sections is to define and study two variants
of the \rop\ on hierarchies of graphs described above. First we define
a generic \rop\ (\sect{R-operation}) and prove the corresponding
localized version of the Bogoliubov-Parasiuk theorem (\sect{GenericR}).
The purpose of this exercise is to demonstrate in detail how our
formalism works using the familiar result as an illustration. From the
point of view of the theory of UV renormalization in coordinate
representation, our proof exhibits the essentially trivial
power-counting nature of the Bogoliubov-Parasiuk theorem and thus,
hopefully, helps (along with the derivation presented in
\cite{bog-shir} and the original letter of Bogoliubov \cite{bog52}) to
demystify the \rop. 

From the point of view of applications to the theory of asymptotic
expansions (which is our main aim), the very possibility to present
the proof in such a compact and explicit form (to be compared with the
reasoning of \cite{hepp69} or \cite{zav79}) has important and
far-reaching heuristic implications.  Indeed, to justify a
construction similar to \rop\ (e.g.\ the
\asop\ discussed in \sects{ExpansionObject}-{MainProof}) 
for all practical purposes it is normally sufficient ($a$) to exhibit
the underlying recursive distribution-theoretic pattern; ($b$) to
classify singularities and their intersections (in the case of the
\rop\ there is just one elementary pattern of overlapping
singularities corresponding to the recursive/inductive step of the
reasoning: singular linear subspaces intersecting at a point); ($c$)
determine the relevant indices of singularity etc.\ for each type of
intersections by power counting supplemented by analysis of ``model"
examples exhibiting salient features of how singularities interact at
each type of intersections.  After that a complete ``rigorous"
justification is a matter of a rather straightforward translation of
power-counting arguments into the language of $\cS$-inequalities.%

In \sect{SpecialR} we turn to what we call {\it special \rop} (usually
marked by tilde: \trop\ or, in formulae, $\tR$). The
\trop\ is a version of \rop\ designed to act on singular functions
possessing homogeneity properties in such a way as to preserve such
properties to a maximal degree. The main use of the \trop\ is in the
theory of \asop\ (see \ssect{Extension.As&R}).

In \sect{Variations} we investigate how variations of subtraction
operators affect the \rop\ as a whole. Interpreted from the point of
view of UV renormalization in coordinate representation, this result
represents the so-called renormalization-group transformation of an
individual diagram.  Again, our straightforward and rather simple
derivation of this result adds, we hope, to the understanding of the
fundamental structure of renormalization group in perturbative QFT.
That, however, is only an illustration. The main application of this
result is to the theory of \asop\ where it is used to exhibit the
structure of the latter (\ssect{Existence.Theorem}, remark $(iv)$).

\SECTION {Structure and definition of the  \rop.}
\label{R-operation}

We start in \ssect{R-op.Problem} by formulating the problem of
construction of the \rop\ on the graphs from the universum $\cG$.  We
require that the \rop\ satisfy a set of conditions enumerated
in \ssect{R-op.Axioms}. Those conditions naturally arise in the
problem of subtraction of UV divergences (\ssect{R-op.Comments},
remark $(ii)$), as well as in the problem of asymptotic expansions
(\ssects{As-op.Locality}-{As-op.Recursive}); here they are
regarded as axioms.  In \ssect{R-op.Structure} we derive---using the
decompositions of unit constructed in \sect{Decompositions}---the
structural properties of the \rop\ which follow directly from the
axioms.  This allows us in \ssect{R-op.Recursive} to present
a recursive definition of the \rop\ on a graph in terms of
\rop\ on its subgraphs. The definition fixes the structure of the
\rop\ up to subtraction operators which subtract divergences at
isolated points.

In \ssects{Consistency}-{Consistency.Property-iii} we investigate
consistency of the definition of \ssect{R-op.Recursive}.  For the sake
of uniformity we do that within the formalism used for the definition.
The corresponding proofs are quite dull and the results contain no
surprises. We believe, however, that such proofs have to be presented
explicitly at least once---even if only to show that they are not
worth worrying about. Another way to establish the consistency is via
the representation of \rop\ using regularization and local
conterterms, as described in \ssect{Append.Regularization}. Such a
representation, however, requires explicit knowledge of subtraction
operators, which is here not necessary and, therefore, mathematically
unaesthetic.

Finally, in \ssect{Consistency.Co-subgraphs} we discuss inclusion of
co-subgraphs into the general framework of the \rop, which completes
the definition of the \rop\ on the members of the double hierarchy of
the universum of graphs.

\SUBSECTION{The problem.}
\label{R-op.Problem}%{ss6.1}

Consider a graph $G$ and its coefficient function \eq{1/1.7},
\eq{1/1.6}.  We assume that all $F_g(p)$ are smooth for $p\not=0$ and
possess a singularity at $p\rightarrow 0$
(example ($iii$), \ssect{Graphs.Example}). A precise description of
such singularities will be given in \ssect{GenericR.Factors}.

The function $G(p)$ is smooth at all $p$ except for the points where
the argument of at least one of the factors $F_g$ in \eq{1/1.7}
turns to zero, i.e.\ it is smooth in the region $P^\0_G\backslash
\cup^\0_{g\in G}\pi_g$.  Therefore, $G(p)$ defines a distribution in
this region. The problem is to construct a distribution over the
entire $P_G$ starting from $G(p)$. It is a problem of extending a
linear functional (distribution) $G$ defined over the space of test
functions
\be[e1/6.2]
   \cD (P_G\backslash \cup_{g\in G}\pi_g),
\ee
to a functional over the space $\cD (P_G)$ which includes
\eq{e1/6.2} as a subspace.  The recipe of extension is called,
following \cite{bog52}, \cite{bog-shir}, the \rop\ (denoted as $\R$ in
formulae), while the distribution on $P_G$ which results from the
action of the \rop\ on $G(p)$ is denoted as $\R\. G(p)$ and called the
{\it renormalized graph}.  The action of the \rop\ is called {\it
renormalization}.  In pure mathematics the term ``regularization" is
used in such situations
\cite{gel-shi} but it denotes a different thing in quantum field
theory (cf.\ \ssect{Append.Regularization}). We prefer to follow the
terminology of quantum field theory. On the other hand, we treat the
\rop\ as a general mathematical construct and do not address questions
like whether or not a concrete version of the \rop\ yields unitary
$S$-matrix when applied to a specific set of Feynman diagrams, or
whether or not a given \rop\ preserves gauge invariance. Such
questions are beyond the scope of the present work.

\SUBSECTION{Axioms for \rop.}
\label{R-op.Axioms}%6.2

Let us enumerate the requirements which the \rop\ should satisfy, and
which are here accepted as axioms.  Their origin should be discussed
within the framework of the specific problems where a need for the
\rop\ emerges (concerning theory of UV renormalization in coordinate
representation see \ssect{SpecialR.Example}; concerning the theory of
asymptotic expansions, see \ssects{As-op.Locality}-{As-op.Recursive}.

The \rop\ should be defined not only for a graph but also for
each of its subgraphs and, more generally, for any graph from the
universum of graphs so that the conditions
below are to be satisfied for any $G$:

 ($i$)
$\R\. G$  is a distribution on $P_G$, i.e.\
$\R\. G\in \cD '(P_G)$;

 ($ii$) {\it The extension condition}.  $\R\. G$ is an extension of
$G$, i.e.\ for any $\vf$ from \eq{e1/6.2}:
\be
     \langle\R\. G, \vf \rangle
     = \langle G,\vf \rangle,
\kern15mm\ee
which means that $\R$ modifies $G$ only at singular points of the
latter;

 ($iii$) {\it The locality condition}.  Let
$\g \in S[G]$ be any non-empty $s$-subgraph of the graph $G$.
Then for any $\vf \in \cD (P_G\backslash \cup_{g\in
\Gg}\pi_g)$ one should have
\be
     \langle\R\. G, \vf \rangle
     = \langle\R\. \g , [\Gg \, \vf ]\rangle.
\ee
This condition means that, for any singular point, the factors that
are regular at that point are, so to say, transparent for the
\rop.  (Note that $\R\. \g$ on the r.h.s.---the full
notation is $\R\. \g^\0_{(G)}$, according to the conventions in
\ssect{Graphs.Graphs}---is the distribution induced on $P_G$ by the
corresponding distribution on $P_\g$---cf.\
\ssect{Subgraphs.Subgraphs}.)

The locality condition connects the \rop\ on a graph with the
\rop\ on all its $s$-subgraphs.

 ($iv$) {\it The factorization condition}.  If $G$ is a factorizable
graph (\ssect{Complete.Factorization}) 
i.e.\ $G=G_1\times G_2$, then $\R\. G=\R\. G_1
\times \R\. G_2$.  

\SUBSECTION{Remarks.}
\label{R-op.Comments}%6.3

$\quad$($i$) The four conditions given above fix the structure of
\rop\ up to local subtraction operators (see below
\ssect{R-op.Recursive}). All other conditions which may be imposed on
the \rop\ in applications should be satisfied by the choice of the
subtraction operators. As a rule, supplementary requirements have to
deal with conservation of various types of symmetries---kinematic,
global, gauge etc.  A requirement of a different type is the absence
of oversubtractions which will be discussed in
\ssect{GenericR.Oversubtraction}.

($ii$) If one considers applications to the theory of UV
renormalization in coordinate space then the following should be
noted.  In the Minkowsky space, the propagator in coordinate
representation $\D(x)$ is singular not only at the point $x=0$ but on
the light cone $x^2=0$ as well.  So our formalism in its present form
directly covers only coefficient functions of Euclidean Feynman
diagrams (extension to the non-Euclidean case is, nevertheless,
possible \cite{workinprogress}).  In \cite{lang-liesn} the conditions
on the \rop\ in the Euclidean space were found which allow one to
reconstruct the corresponding objects in the Minkowski space.  Those
conditions can in fact be regarded as a concretization of the
conditions ($i$)--($iv$) which take into account the details of
specific structure of Feynman diagrams in coordinate representation.

\SUBSECTION{Structure of R from axioms.}
\label{R-op.Structure}%6.4

The conditions of \ssect{R-op.Axioms} allow one to express the
\rop\ for a  graph $G$ in terms of the \rop\ for its
$s$-subgraphs.  For  factorizable  graphs such expression is given
directly by the factorization condition ($iv$).

First, let $G$ contain more than one maximal $s$-subgraph, i.e.\
$G\notin S[\cG]$.  Using the decomposition of unit $\eta
=\{\eta^\0_\G\}^\0_{\G \in S_{\max}[G]}$ introduced in
\ssect{Decomp.Maximal} one obtains:
\be[e1/6.3]
     \langle \R \. G,\vf \rangle
     = \sum_{\G \in S_{\max}[G]}
        \langle {\R\. G}, \eta^\0_\G  \vf \rangle.
\kern10mm\ee
The locality condition ($iii$) of \ssect{R-op.Axioms} can be applied to
the r.h.s.  One obtains:
\be[e1/6.4]
     \langle \R\. G, \vf \rangle
     = \sum_{\G \in S_{\max}[G]}
     \langle{ \R \. \G},
         [ \eta^\0_\G  \GG  \vf ]
     \rangle,
\ee
where the expression in square brackets represents a function which is
smooth everywhere.

Finally, let $G$ be a non-factorizable $s$-graph.  In this case the
singular plane coincides with the zero point of the space $P_G$.
The restriction of the distribution $\R\. G$ from ${\cal D}(P_G)$ to
$\cD (P_G\backslash \{0\})$ is denoted as $\R'\.  G$.  Take a
$\vf \in \cD (P_G\backslash \{0\})$.  Using the decomposition of
unit $\theta =\{\theta_\G\}_{\G
\triangleleft G}$ (\ssect{Decomp.Submaximal}) and the locality
condition, by analogy to the preceding case one obtains:
\be[e1/6.5]
     \langle\R'\. G,\vf \rangle
     =  \sum_{\G \triangleleft G}
        \langle \R\. \G ,
             [\theta_\G \GG  \vf ]
        \rangle.
\kern10mm\ee
Note that if we formally assume that the \rop\ for the empty graph is
a unit operation then the extension condition $(ii)$ might be
considered as a consequence of the locality condition $(iii)$ for $\g
=\emptyset$.

\SUBSECTION{Recursive definition of the \rop.}
\label{R-op.Recursive}%6.5

\hbox {\Eqs{e1/6.4}-{e1/6.5} form the basis of the following recursive
definition of the \rop:}

 $(a)$ For the empty graph $G=\emptyset $ let
\be[e1/6.6]
     {\R\. \emptyset \bydef \emptyset} \equiv 1.
\kern29mm\ee
(Recall our conventions concerning empty graphs,
\ssect{Graphs.Simplified}.)

 $(b)$ If $G=G_1\times G_2$ then
\be[e1/6.7]
     \R\. G \bydef \R \. G_1 \times \R\. G_2.
\kern15mm\ee

 $(c)$ If $G\notin S[\cG]$ then
\be[e1/6.8]
     \R\. G
     \bydef \sum_{\G \in S_{\max}[G]} (\R\. \G )
       [\eta^\0_\G \GG ],
\ee
where $\eta^\0_\G$ are defined in
\ssect{Decomp.Maximal}.

 $(d)$ If $G\in S_{0}[\cG]$ then first the restriction $\R'\. G$
of $\R\. G$ from $\cD (P_G)$ onto $\cD (P_G\backslash \{0\})$ is
defined by:
\be[e1/6.9]
     \R'\. G
     \bydef \sum_{\G \triangleleft G} 
          (\R\. \G )
          [\theta_\G \GG ].
\kern10mm\ee
In the special case $S[G]=\{G,\emptyset\}$ one has:
\be[e1/6.10]
     \R'\. G(p) \equiv G(p).
\ee

Before completing the definition note that the problem of
constructing $\R\. G$ from $\R'\. G$ is that of extension of a
functional from the subspace $\cD (P_G\backslash \{0\})$ onto the
space $\cD (P_G)$. On the other hand since $\R\. G$ has only one
singularity localized at an isolated point (the origin) of the space
$P_G$, one can say that the construction of $\R\. G$ from $\R'\. G$
consists in subtracting the singularity:

$(e)$ Let us introduce a formal operator $\r_{(G)}$ that carries out
such a subtraction for a class of distributions from $\cD
'(P_G\backslash \{0\})$ and yields a distribution from $\cD '(P_G)$.
Then one can write:
\be[e1/6.11]
     \R\. G =\r \. \R'\. G.
\ee
We call $\r$ {\it $($local$)$ subtraction operator\/} (the convention
of \ssect{Graphs.Graphs} allows us to omit the subscript $(G)$).%
\footnote{
Note that our use of the term ``subtraction operator' differs from the
BPHZ convention \cite{bog-shir}. If $M$ is the subtraction operator in
the sense of BPHZ then $\r=1-M$.
\relax}
The subtraction operators have been
studied in detail in \sect{Subtractions}.

 {\bf Remark.} Representations \eq{e1/6.4}, \eq{e1/6.5} hold for
factorizable graphs as well. Therefore, when studying the structure of
the \rop\ defined by \eqs{e1/6.7}-{e1/6.11}, we may ignore \eq{e1/6.7}
and use the general definitions \eq{e1/6.9}, \eq{e1/6.11} (or
\eq{e1/6.8}) also for factorizable graphs.  The factorization property
can be recovered by an appropriate choice of the subtraction
operators $\r$ for factorizable graphs. This allows one not to
distinguish the case of factorizable graphs when proving inequalities
for the \rop.

\SUBSECTION{Consistency of the definition of the \rop.}
\label{Consistency}

One has to verify that the operation $\R$ defined above satisfies the
axioms of \ssect{R-op.Axioms}. This, strictly speaking, is necessary
despite the fact that the definition was derived from the axioms
because we did not check consistency of the latter. Also, one has to
prove that $\R$ thus defined is independent of the choice of
decompositions of unit $\{\eta^\0_\G\}$ and $\{\theta_{\G }\}$ used in
\eq{e1/6.8}, \eq{e1/6.9}.

For the sake of uniformity and in order to demonstrate different
aspects of the formalism, we will outline a proof that involves only
the notions used in the definitions. An alternative reasoning is based
on the representation of the \rop\ in terms of a regularization and
counterterms (\ssect{Append.Regularization}).

The factorization property $(iv)$ holds by definition \eq{e1/6.7} and
uniqueness of the decomposition of a graph into non-factorizable
subgraphs (\ssect{Complete.Factorization}).  The property $(i)$ can be
ensured by an appropriate choice of the subtraction operator; this is
postponed till \sect{GenericR}. Other properties can be proved by
induction. That is, considering the \rop\ on a graph $G$, we assume
that consistency of the \rop\ has been proved for all $s$-subgraphs of
$G$ (except for $G$ itself). Note that the starting point of induction
is ensured by the definition of the \rop\ on the empty subgraph---see
part $(a)$ of the definition in \ssect{R-op.Recursive}.

We start by verifying  the property $(ii)$ and  in
\ssect{Consistency.Independence}  prove independence of  the choice of
decompositions of unit $\{\theta \}$ and $\{\eta \}$.  In
\ssect{Consistency.Property-iii} the
locality condition is verified.

For $G\notin S[\cG]$, take $\vf $ from \eq{e1/6.2} and use
the definition \eq{e1/6.8} in the form of \eq{e1/6.4}. The expression
$\tilde{\vf}=[\eta^\0_\G \GG \vf ]$
considered as a function of the proper variables of $\G $ is a test
function allowing application of $(ii)$ to $\G $.  This allows one to
replace $\R\. \G \rightarrow \G$, carry out summation over $\G $, and
obtain the property $(ii)$ for $G$.  The case $G\in S[\cG]$ is
treated analogously; one should only note that for $\vf$ from
\eq{e1/6.2} $\R\. G$ may be replaced by $\R'\. G$.

\SUBSECTION{Independence of the choice of $\{\theta \}$ and $\{\eta \}$.}
\label{Consistency.Independence}

For example, let us demonstrate that the r.h.s.\  of \eq{e1/6.9} is not
changed by replacement of $\{\theta \}$ by another decomposition of  unit
$\{\tilde{\theta}\}$  satisfying  the same general requirements. In fact
for  any $\vf \in \cD (P_G\backslash \{0\})$,  substitution  of
$1= \sum \tilde{\theta}_{H}$ into the r.h.s.\ of \eq{e1/6.9} results in the
expression
\be[e1/7.1]
     \sum_{\G \triangleleft G}
     \sum_{H\triangleleft G} \,
     \langle \R \. \G ,
             [ \tilde{\theta}_{H} \theta_\G
               \GG
               \vf
             ]
     \rangle.
\ee
The support of the function in the square brackets is such that
\be
    \supp [\tilde{\theta}_{H} \theta_\G \GG \vf] \cap\pi_g
    = \emptyset ,
\ee
provided $g\notin \G $ or $g\notin H$
i.e.\ for $g\notin \g =\G \cap H $ (note that $\g $ thus defined is an
$s$-subgraph).  Then one can use the locality condition for the pair
$\g <\G $, which is true by the inductive assumption. Instead of
\eq{e1/7.1} one obtains
\be
     \sum_{\G \triangleleft G}
     \sum_{H\triangleleft G} \,
     \langle\R\. \g ,
           [ \tilde{\theta}_{H} \theta_\G
             \Gg \, \vf
           ]
     \rangle,
\ee
where the equality $\GG \,\Gamma\backslash\g =\Gg $
has been used.  The summand is invariant under simultaneous exchange
$\theta \leftrightarrow \tilde{\theta}$, $\G \leftrightarrow H$.
After the exchange is done, one can invert the reasoning and obtain
\eq{e1/6.9} with $\theta $ replaced by $\tilde{\theta}$, as required.

Independence of the choice of $\{\eta \}$ is proved analogously.

\SUBSECTION{Proof of locality $(iii)$.}
\label{Consistency.Property-iii}%7.3

We will verify the locality condition $(iii)$ only for the case of 
non-factorizable graphs. Note that we can use the results of
\ssect{Consistency.Independence} and assume that the \rop\ is
independent of the choice of decompositions of unit for the graph $G$.
We will present the reasoning only for the case of an $s$-graph $G$.
Graphs with several maximal $s$-subgraphs can be considered in a
similar manner.

Assume that $\g \in S[G]$ and $\g $ is not a submaximal subgraph (the
case $\g \triangleleft G$ is trivial). Choose $\vf \in \cD
(P_G\backslash \cup_{g\in \Gg}\pi_{g(G)})$.  In other
words $\supp\vf $ intersects only singular planes $\pi_g$,
$g\in \g $. For such $\vf $ one may replace
$\R\. G$ by $\R'\. G$ and apply the definition \eq{e1/6.9} in the form
of \eq{e1/6.5}. Using the fact that the definition is independent of
the choice of $\{\theta \}$ and choosing $\{\theta \}$ so that
$\supp\theta_\G\cap \supp\vf \not=\emptyset $ only for $\G > \g
$, one obtains:
\be[e1/7.6]
     \langle\R\. G,\vf \rangle
     = \sum_{\g <\G \triangleleft G}
       \langle \R\. \G,
           [ \theta_\G \GG  \vf ]
       \rangle
\ee
The support of the test function in square brackets lies within the
support of $\vf $ and therefore intersects only $\pi_g$,
$g\in \g $. So, one may use the locality
condition for the subgraph $\g $ in the graph $\G $ which holds by
inductive assumption. Instead of the r.h.s.\ of \eq{e1/7.6}, one
obtains:
\be
     \sum_{\g <\G \triangleleft G}
     \langle\R\. \g,
         [\theta_\G \Gg \, \vf ]
     \rangle.
\ee
Now one can restore the summation over all $\G $. And since only
$\theta_\G$ depend on $\G $, one obtains the r.h.s.\ of the locality
condition $(iii)$, which completes the proof of self-consistency.

\SUBSECTION{\rop\ on co-subgraphs.}
\label{Consistency.Co-subgraphs}

As was indicated in \ssect{Complete.Co-subgraphs}, there is a second
way to generate new graphs from a given graph besides taking complete
singular subgraphs---namely, one can consider the {\it
co-subgraphs}.  

We have seen that one can define the \rop\ on the hierarchy consisting
of a graph $G$ and all its subgraphs $\g $, and the operations
$\R_{(G)}$ and $\R_{(\g )}$ are closely related.  However, one can
also define an \rop\ on each $[\Gg]_\g$ (we denote such operation as
$\R_{(\Gg )})$.  In the special case when $G$ factorizes into a
product of $\g $ and $\Gg $ the three operations $\R_{(\Gg )}, \R_{(\g
)}$ and $\R_{(G)}$ are related by the factorization condition ($iv$),
\ssect{R-op.Axioms}. However, in the most general case we can 
assume that there is no a priori relationship between $\R_{(\Gg )}$ and
$\R_{(G)}$.

\SECTION{Generic \rop.}
\label{GenericR}

Describing the general structure of the \rop\ in
\ssect{R-op.Recursive} we made no assumptions about singularities of
the factors $F_g(p)$, so that the subtraction operators could not
have been specified. A class of such operators for removing power
singularities was introduced in \sect{Subtractions}. In the present
section we complete the reasoning. In \ssect{GenericR.Factors} the
conditions on $F_g$ (in fact, the weakest ones) are formulated,
which ensure correctness of the \rop\ based on generic subtraction
operators.  After some definitions in \ssect{GenericR.Index}, in
\ssect{GenericR.Theorem} the theorem of existence and properties of the
generic \rop\ is presented and discussed. The theorem that we prove
is a localized and somewhat generalized version of the
Bogoliubov-Parasiuk theorem.  In
\ssects{GenericR.Proof.Plan}-{GenericR.Proof.PowerCounting} the
proof of the theorem is presented.  Finally, in order to better
understand the \rop, in \ssect{GenericR.Oversubtraction} we discuss
effects of the oversubtractions of
\ssect{Subtraction.Oversubtraction}.

Since here we only concentrate on local integrability,
all $\L$-functions are single-sided throughout this section (for
definitions see \ssect{Isolated.Lambda}).

\SUBSECTION{Properties of factors.}
\label{GenericR.Factors}%10.1

First of all, as in \ssect{R-op.Problem} we demand that the factors
$F_g$ outside $p=0$ be smooth:
\be[e1/10.1]
     F_g(p)\in C^\infty(P_g\backslash \{0\}).
\ee
Assume that at $p\rightarrow 0$ all $F_g$ possess power
singularities of the type of $|p|^{-d_g}$ (perhaps with soft
corrections), and that taking derivatives in $p$ modifies the
singularity in a ``natural" way. To be more accurate, we assume that
for any multiindex $\a$ (including $\a =0$) the following
inequality is valid:
\be[e1/10.2]
     |D^\a F_g(p)|
     \leq {\L (|p|)
              \over
           |p|^{d_g+|\a|}},
     \qquad p\not=0.
\ee
Note that in general the $\L $-functions in \eq{e1/10.2} depend on
$\a $.  Euclidean propagators (massive and massless, in coordinate
or momentum representation) satisfy the above condition.

We assume for simplicity that all $d_g$ in \eq{e1/10.2} are integer.
However the structure of proofs and the results (except for some minor
details) do not depend on this assumption.

\SUBSECTION {Divergence index.}
\label{GenericR.Index}%10.2

Now assume that for a graph $G$ the coefficient function is
constructed according to \eq{1/1.1}, \eq{1/1.7}.  For any subset
$H\subset G$ define:
\be[e1/10.3]
     d_{H} \bydef \sum_{g\in H}d_g, 
\kern12mm\qquad d_{\emptyset}\bydef 0,
\ee
and for $s$-subgraphs $\G \in S[G]$ (\ssect{Complete.s-subgraphs}) 
define also the {\it index of divergence}:
\be[e1/10.4]
     \o_\G \bydef d_\G - \dim P_\G,
     \qquad \o^\0_{\emptyset}\bydef 0,
\ee
where, in accordance with \ssect{Complete.Planes},
\be[e1/10.5]
     \dim P_\G=\hbox{codim} \;\pi^\0_{\G (G)},
     \qquad \dim P_{\emptyset}=0.
\ee
Note that the value of the divergence index is an intrinsic property
of $\G $, i.e.\ it does not depend on the larger graph $G>\G $, unlike
e.g.\ $\dim \pi^\0_\G$ which does.

We will call the \rop\ constructed following the algorithm of
\ssect{R-op.Recursive} and using the subtraction operators defined in
\ssect{Subtraction.r}, {\it generic} $R${\it -operation}. Its properties are
formulated in the following theorem.

\SUBSECTION{Theorem (existence and properties of generic \rop).}
\label{GenericR.Theorem}%10.3

\begingroup
\it
Assume that all $F_g$ satisfy the conditions \eq{e1/10.1} and
\eq{e1/10.2}. Then for the generic \rop\ on any $s$-graph $G$
$($\ssect{Complete.Ordering}$)$ the following is true$:$

 $(a)$ $\displaystyle\R \. G \in \cD '(P_G)$ and $\displaystyle \R
\. G$ satisfies the conditions of\/ \ssect{R-op.Axioms};

 $(b)$ For any $\displaystyle \vf \in \cD (P_G)$ the following
inequality holds:
\be[e1/10.6]
     |\langle\R \. G,\vf \rangle|
     \leq d^{-\o_G} \cS [\vf,d],
\ee
where
\be[e1/10.8]
     d = \rad\supp\vf,
\ee
and the $\L$-functions implied in $\cS $ are single-sided.

\endgroup

 {\bf Remarks.} $(i)$ The statement of the theorem implies that
analogous results, in particular the estimate \eq{e1/10.6}, are true
for any subgraph $\G< G$ as well.

$(ii)$ The estimate \eq{e1/10.6} holds also for all $\vf$ such
that $\rad\supp\vf \leq d$ (see the remark in
\ssect{Isolated.d-Inequalities}).

$(iii)$
For completeness' sake, it may be interesting to note that the upper limit
of summation in $\cS $ on the r.h.s.\  of \eq{e1/10.6} is given by the
following fairly obvious expression \cite{fvt-vvv}:
\be[e1/10.7]
     N_G
     = \max \{0,\max_{\G \in S[G]^\0,\G \not=\emptyset}
       \o_\G+1\}.
\ee
This number is exactly the {\it order\/} of the distribution $\R\.  G$
(for a definition see e.g.\ \cite{reed-sim}).

 $(iv)$ In general the $\L $-functions that are implicit in $\cS $ in
\eq{e1/10.6} are non-trivial (i.e.\ differing from constants) even in
the case when all $\L =\const$ in the estimates \eq{e1/10.2} for
individual factors (cf.\ the remarks at the end of
\ssect{Subtraction.d-Inequalities}).  This, however, may not be the
case if {\it all} positive divergence indices $\o_\G$, $\G \in S[G]$,
are {\it non-integer}.

 $(v)$ If at $p\rightarrow \infty $ all $F_g(p)$ together with their
derivatives have polynomial growth rates, then $\R\. G$ can be
naturally extended to the Schwartz space of rapidly decreasing
functions $\cS (P_G)$. In that case one can obtain inequalities
expressing continuity of $\R\. G$ in the topology of $\cS (P_G)$
in terms of the seminorms defining that topology, e.g.\
\be[e1/10.9]
     \|\vf\|_{k,l} =
     \sum_{|\a|\leq k,|\b|\leq l}{\ }
       \sup_{p\in P}
       | p^\a D^{\b}\vf (p) |.
\ee
However such inequalities are useless for our purposes because they
contain no information about the local structure of $\R\. G$ near
$p=0$.

 $(vi)$ Finally, note that the well-known result by Bogoliubov and
Parasiuk \cite{bog-par} and Hepp \cite{hepp69} on finiteness of
renormalized Feynman diagrams is an integral version of Theorem 1: in
fact in the Bogoliubov-Parasiuk-Hepp theorem the finiteness of the
expression $\langle\R\. G, \vf \rangle$ is proved for some special
cases of $\vf $ like $\vf (p)=\exp (ip q)$.  The new estimate
\eq{e1/10.6} contains more information about the distribution $\R\. G$
(the form of dependence on $d$).  Due to this
fact---paradoxically---there exists a possibility to present a rather
short and straightforward proof of the theorem, which makes an
effective use of the recursive structure of the \rop.

\SUBSECTION{Plan of the proof.}
\label{GenericR.Proof.Plan}

The crucial part of our definition of the \rop\ for a graph $G$---%
provided that the \rop\ is defined for all subgraphs $\G< G$---includes
two steps. First, the functional $\R'\. G\in \cD '(P_G\backslash
\{0\})$ is constructed; then the subtraction operator is applied to
it. The inequality \eq{e1/10.6} has exactly the form
\eq{e1/9.23} and will follow immediately from the results of
 \sect{Subtractions} if the corresponding inequality of the form
\eq{e1/9.1} is established for $\R'\. G$.

The exact formulation of the desired inequality for $\R'\. G$ and
the necessary inductive assumptions are presented in
\ssect{GenericR.Proof.EstimateR'}. The
inequality itself is proved in
\ssects{GenericR.Proof.UsingInduction}-{GenericR.Proof.PowerCounting}.

\SUBSECTION{Estimate for $\protect\R'\. G$.}
\label{GenericR.Proof.EstimateR'}%11.1

We assume that the graph $G$ is non-empty. (In the case $G=\emptyset $
the theorem is valid by definition---see our conventions for the empty
subgraph in \ssect{R-op.Recursive} and
\ssect{GenericR.Index}. This ensures correctness of the starting point
of our inductive proof.) Assume that for each $\G< G$
the theorem has been proved and, in particular, for any $\vf \in \cD
(P_\G)$ the following estimate holds:
\be[e1/11.1]
     |\langle \R\. \G , \vf \rangle|
     \leq d ^{-\o^\0_\G}\cS [\vf ,d],
\ee
where
\be
     d = \rad\supp\vf.
\ee
It has to be proved that for any $\eta \in \cD (P_G\backslash
\{0\})$ and any $\vf \in \cD (P_G)$ the following estimate
holds:
\be[e1/11.2]
     |\langle \R'\. G, \eta_\l \vf \rangle|
     \leq
     \l^{-\o^\0_G}
     \cS _{p\in \supp\eta_\l} [\vf ,\l ]
\ee
where
\be
     \eta_\l(p)\equiv \eta (p/\l ).
\ee
Note that a simple power counting (see below
\ssect{GenericR.Proof.PowerCounting}) yields:
\be[e1/11.3]
     |G(\l  p^\0_{0})|
     \leq
     \l^{-\o^\0_G- \dim P_G} \L (\l )
\ee
for $p^\0_{0}$ not belonging to any of the singular planes of $G$.
Comparing \eq{e1/11.3} and \eq{e1/11.2} one may say that the
subtraction of divergences from subgraphs preserves the power
character of the singularity of $G$ at zero (which is in itself
important since it allows application of the subtraction operators)
and, moreover, that the index of this power singularity ($\o^\0_G$) is
not changed. (Note that there exists another version of the \rop\ for
which the latter statement is not true---see below
\ssect{GenericR.Oversubtraction})

\SUBSECTION{Using inductive assumptions.}
\label{GenericR.Proof.UsingInduction}%11.2

To derive the inequality \eq{e1/11.2} one first takes into account the
definition of $\R'\. G$ \eq{e1/6.9}. One obtains:
\be[e1/11.4]
     |\langle \R'\. G, \eta_\l \vf \rangle|
     \leq
     \sum_{\G \triangleleft G}
     |\langle
        \R\. \G_{(G)},
        [ \theta_\G \,\eta_\l\,
          \GG \,
          \vf ]
     \rangle|.
\ee
To estimate individual terms on the r.h.s.\ one could use
\eq{e1/11.1}.  However, in \eq{e1/11.1} $\R\. \G$ are considered as
distributions over $P_\G$ while in \eq{e1/11.4} one encounters the
corresponding distributions induced on $P_G$ (cf.\
\ssect{Subgraphs.Subgraphs}).  This can be taken into account as
follows.

The functional $\R\. \G_{(G)}$ is defined on the space
$P^\0_G=P^\0_{\G }\oplus\pi^\0_\G$.  To obtain an estimate
analogous to
\eq{e1/11.1} for the distribution $\R\.
\G_{(G)}\in \cD '(P_G)$ one should only take into account the
additional dimensions in $P_G$ as compared to $P_\G$.  It is easy
to obtain that:
\be[e1/11.5]
     |\langle \R\. \G_{(G)} \vf' \rangle|
     \leq
     {d'}^{-\o^\0_\G + \dim \pi^\0_\G}
     \cS  [\vf', d'],
\ee
for any $\vf'\in \cD (P_G)$ such that $\rad\supp\vf'\leq d'$
(cf.\ the remark $(ii)$ in \ssect{GenericR.Theorem}).  Take
$\vf'=\theta_\G \eta_\l \GG \vf, \quad d =
\rad\supp \theta_\G \eta_\l$ and use \eq{e1/11.4} to estimate
\eq{e1/11.5}. Taking into account that $d'=\l \, \const$, one obtains:
\be[e1/11.6]
     |\langle \R'\. G, \eta_\l \, \vf \rangle| \leq \sum_{\G
\triangleleft G} \l^{-\o^\0_\G + \dim \pi^\0_\G} \, \cS [ \theta_\G \,
\eta_\l \, \GG \, \vf ,\l ].
\ee
The problem is thus reduced to studying the dependence of seminorms on
the r.h.s.\ of \eq{e1/11.6} on $\l $ and $\vf $.

\SUBSECTION{Power counting.}
\label{GenericR.Proof.PowerCounting}%11.3

Let us first take into account that it is sufficient to evaluate the
seminorms in \eq{e1/11.6} over the set $p\in\supp(\theta_\G \eta_{\l
})$.  Then one estimates the seminorms via the seminorms of individual
factors (property $(c)$,
\ssect{Isolated.Seminorms2}; heuristically: the leading singularity of
a product is a product of singularities of the factors):
\be[e1/11.7]
     \cS  [ \theta_\G \eta_\l \GG \vf ,\l ]
     \leq
     \cS  [ \theta_\G \eta_\l,\l ]\,
     \cS _{p\in\supp\theta_\G \eta_\l}\,
     [ \GG ,\l ]\,
     \cS _{p\in\supp\eta_\l}[\vf ,\l ].
\ee
It has been taken into account here that the last seminorm on the
r.h.s.\ of \eq{e1/11.7} may be evaluated over a wider set
$\supp\eta_\l \supset \supp (\eta_\l \theta_\G)$ (property
$(a)$, \ssect{Isolated.Seminorms2}) so that this factor
acquires exactly the form needed in the final result \eq{e1/11.2}.

The first factor on the r.h.s.\  of \eq{e1/11.7} contains no interesting
dependences except for $\l $, but it can be easily verified that it
is actually independent of $\l $:
\be[e1/11.8]
     \cS [ \theta_\G \eta_\l,\l ] = \const.
\ee
Here the fact that $\theta_\G(p) \eta_\l(p) =\theta_\G(p/\l )
\eta (p/\l )$ (see \ssect{Decomp.Submaximal}) and property $(b)$,
\ssect{Isolated.Seminorms2} have been used.

It remains to evaluate the second seminorm on the r.h.s.\ of
\eq{e1/11.7}.  Let $H=\GG$. It is easy to see that
$\supp\theta_\G\, \eta_\l=\l K$ where $K$ is a compact set
independent of $\l $ and non-intersecting singularity planes of any of
the factors in $H$. Let us prove that
\be[e1/11.9]
     \cS _{p\in \l  K}[H(p),\l ]
     \leq   \L (\l ) / \l^{d_{H}}, \qquad\l >0,
\ee
for any $H\subset G$ and any compact $K$ such that
$K\cap\pi_g=\emptyset $ for each $g\in G$ ($d_{H}$ is defined in
\eq{e1/10.3}).

The inequality \eq{e1/11.9} is rather obvious, it reflects the fact
that $D^\a H(p)$ behaves as $|p|^{-d_{H}-|\a|}$ when $p$
tends to zero along non-singular directions.

First, one can use the property $(d)$ of seminorms
(\ssect{Isolated.Seminorms2}) and obtain:
\be[e1/11.10]
     \cS _{p\in \l  K}[H,\l ]
     \leq  \prod_{g\in H}
     \cS _{p\in \l  K}[g(p),\l ].
\ee
Now it is easy to understand that the assumed properties of the
individual factors ensure that each factor on the r.h.s.\ can be
estimated by
\be[e1/11.14]
     \L (\l ) \l^{-d_g}.
\ee
Indeed, to perform a transition to variables $p_g\in P_g$ for each
factor one uses the chain rule of differentiation of a complex function:
\be[e1/11.11]
     \cS _{p\in\l  K}[g(p),\l ]
     \leq  \cS _{p_g\in \l  l_g(K)}[F_g(p_g),\l ].
\ee
Denote:
\be
     r_g \bydef \inf_{p_g\in l_g(K)}|p_g|> 0, 
\ee\be
     R_g \bydef \sup_{p_g\in l_g(K)}|p_g|\geq r_g.
\ee
Then the r.h.s.\ of \eq{e1/11.11} is estimated by the expression
\be[e1/11.13]
      \cS _{\l  r_g
      \leq |p_g|\leq \l  R_g} [ F_g(p_g),\l ].
\ee
Eq.\eq{e1/11.14} now follows from the assumption \eq{e1/10.2} and
property $(h)$ of $\L $-functions, \ssect{Isolated.Lambda}.

Combining all the results \eqs{e1/11.10}-{e1/11.14} and using the
properties of $\L $-functions (\ssect{Isolated.Lambda}) one obtains
\eq{e1/11.9} as required.

For $H=G$, $m=0$, $K=\{p_{0}\}$, taking into account the relation
\eq{e1/10.4} one obtains \eq{e1/11.3} as a special case.

Finally, recalling the definitions in \ssect{GenericR.Index}, one
obtains:
\be[e1/11.15]
     -\o^\0_\G+\dim \pi^\0_\G
     -d_\GG = -\o^\0_G
\ee
This completes the proof of the theorem of \ssect{GenericR.Theorem}.

\SUBSECTION{Oversubtractions.}
\label{GenericR.Oversubtraction}%12.1

Consider now an $s$-graph $G$. Let us construct the oversubtracted
\rop\ using the definitions of \ssect{Subtraction.Oversubtraction}, 
and study the corresponding estimate for $\R\. G$.

Assume that the \rop\ with oversubtractions is constructed for all the
subgraphs $\G< G$. Assume by induction that for each subgraph $\G< G$
an estimate of the form \eq{e1/12.6} has been proved:
\be[e1/12.8]
     |\langle \R\. \G ,\vf \rangle|
     \leq 
     d^{-\bar{\o}^\0_\G}\cS [\vf,d].
\ee
Here (cf.\ \ssect{Subtraction.d-Inequalities})
$\bar{\o}^\0_\G\geq \o^\0_\G$ is the
effective singularity index.  The difference between
$\bar{\o}^\0_\G$ and $\o^\0_\G$ as
defined by \eq{e1/10.4} is due to oversubtractions either in subgraphs
$\g< \G $, or in $\G $.

Let us construct the functional $\R'\. G$ and study its effective
singularity at zero. Reasoning as in
\ssect{GenericR.Proof.EstimateR'}, one obtains the following
inequality instead of \eq{e1/11.6} which was assumed in
\ssect{GenericR.Proof.EstimateR'}, due to the difference between
\eq{e1/12.8} and \eq{e1/11.1}:
\be[e1/12.9]
     |\langle  \R'\. G,\eta_\l \vf \rangle|
     \leq
     \sum_{\G \triangleleft G}
        \l^{-\bar{\o}^\0_\G
            +\dim\pi^\0_\G}
        \cS [ \theta_\G \eta_\l \GG \vf ,\l ].
\kern8mm\ee
Using the results of \ssect{GenericR.Proof.PowerCounting}
one obtains the following inequality instead of \eq{e1/11.2}:
\be[e1/12.10]
     |\langle \R'\. G,\eta  \vf \rangle|
     \leq
     \l^{ -\bar{\o}^\0_\G-d_\GG
          +\dim \pi^\0_\G}
     \sum_{\G \triangleleft G}
        \cS _{p\in \supp\eta_\l}[\vf ,\l ].
\ee
Now change the order of summation over $\G $ and $k$ and take into
account \eq{e1/11.15}and the property $(i)$ of single-sided
$\L$-functions (\ssect{Isolated.Lambda}). One gets:
\be[e1/12.11]
     |\langle \R'\. G,\eta_\l \vf \rangle|
     \leq
     \l^{-\o^{\textbox{eff}}_G}
     \cS _{p\in\supp\eta_\l}[\vf ,\l ],
\ee
where the effective divergence index is
\be[e1/12.12]
     \o^{\textbox{eff}}_G
     = \o^\0_G
     + \max_{\G \triangleleft G}
       (\bar{\o}^\0_\G-\o^\0_\G).
\ee
This instructive result means that each oversubtraction in a subgraph
adds to the effective singularity of the graph.

Now fix $\O_G \geq \o^{\textbox{eff}}_G$ and apply $\r^{\O_G}$ to
$\R'\. G$.  Then taking into account \eq{e1/12.6} one can
obtain for $\R\. G$ an estimate
\be[e1/12.14]
     |\langle \R\. G ,\vf \rangle|
     \leq 
     d^{-\bar{\o}^\0_G}\cS [\vf,d].
\ee
in which
\be[e1/12.13]
     \bar{\o}^\0_G
     =
     \left\{
     \vcenter{
        \hbox{ $\O_G,\quad\kern0.5mm\hbox{if} \quad \O_G\geq 0$,}
        \hbox{ $\o^{\textbox{eff}}_G, \quad \hbox{if} \quad \O_G<0$.}
     }
\right.
\ee
Note that if oversubtractions were done in at least one of the
subgraphs $\G \in S[G]$ then the effective singularity index $\bar{\o
}^\0_G$ for $\R\. G$ can in general become greater than
$\o^\0_G$.  In other words, the effect of
oversubtraction propagates to higher-level subgraphs.  This
phenomenon is, of course, well-known in the BPHZ framework
\cite{zim70}.

It is possible to determine the upper limit of summation implied in
$\cS$ \cite{fvt-vvv}. The corresponding expression, however, is
cumbersome and unilluminating.

To close the discussion of the oversubtracted \rop\ note the
following. The upper bound of summation $N_G$ which is implicit in
$\cS $ in inequalities like \eq{e1/12.14} is precisely the
order of the distribution $\R\. G$ \cite{vsvlad}, \cite{reed-sim}.
Using this notion one can say that the \rop\ without oversubtractions
yields, for any subgraph $\G \in S[G]$, the distribution $\R\. \G$ of
the minimal order as compared to \rop s with oversubtractions. This is
a characteristic property of the generic \rop\ without
oversubtractions.

There exists a more subtle classification of distributions with
respect to their order (see \cite{reed-sim}, Chapter V, exercise
32). In terms of that classification the \rop\ without
oversubtractions, unlike the case of oversubtractions, yields a
distribution not just of order $N_G$ but ``of order $N^{-}_G$".

\SECTION{Special operation $\protect\tR$.}
\label{SpecialR}

The generic $R$-operation, however important it may be from a purely
theoretical viewpoint, is of little immediate use in applications.
Specific problems impose specific restrictions that an $R$-operation
should satisfy.  Thus, in the theory of $As$-operation one has to
deal with graphs whose factors possess certain homogeneity properties;
thus, expanding a scalar propagator $(p^2+m^2)^{-1}$ in powers of
$m^2$ one obtains a sum of pure powers of $p^2$. The leading behaviour
of such factors at infinity is described by the same power exponent as
their singularity at zero (precise definitions are given in
\sect{Isolated.Lambda}).  The special $R$-operation (also \trop\ and
in formulae $\tR$) that we now proceed to discuss, should
preserve those homogeneity properties to a maximal degree.

The structure of the operation $\tR$ and the form of estimates
for it are the same as in the case of the generic $R$-operation,
except that generic subtraction operators are replaced by ``special''
ones, $\tr$, and single-sided $\L $-functions are replaced
everywhere by double-sided ones. The ``special" analogue of theorem
of \ssect{GenericR.Theorem} is presented in \ssect{SpecialR.Theorem}. Note
that from \ssect{SpecialR.Theorem} it follows directly that the
$\tilde{R}$-operation satisfies the minimality condition (see
\ssect{Subtraction.Minimality}) which is very important in application.
The proof of the theorem practically coincides with that of theorem
\ssect{GenericR.Theorem}; the only subtlety arises in the treatment of
factorizable graphs as discussed in \ssect{SpecialR.Proof}.  Finally,
in \ssect{SpecialR.Scaling} we discuss how the $\tilde{R}$-operation
affects scaling properties of the graphs on which it acts, which has
close relation to the minimality condition
\ssect{Subtraction.Minimality}.

\SUBSECTION{Homogeneity properties of factors.}
\label{SpecialR.Homogeneity}

Assume that the functions $F_g(p)$ satisfy the condition
\eq{e1/10.1}---smoothness outside the point $p = 0$---and are bounded
by inequalities similar to \eq{e1/10.2}:
\be[e2/5.1]
     |D^\a F_g(p)| \leq {\L (|p|) \over|p|^{d_g+|\a|}},
\qquad p\in P_g,\quad p \not= 0,
\ee
where $\L$-functions are double-sided. These conditions are satisfied
e.g.\ by almost-homogeneous singular functions, i.e.\ by the functions
such that:
\be[e2/5.2]
     F_g (\l \cdot p) 
     = \l^{-d_g} \, \sum^{L_g}_{k=0} F_{g,k}(p)
                                      \, \ln^k \l, 
     \qquad \l > 0,
     \quad p \not= 0.
\ee
Define $d_{H}$ and $\o^\0_\G$ by \eq{e1/10.3} and
\eq{e1/10.4}, respectively.

For products of such singular functions one can repeat the reasoning
of \ssect{GenericR.Proof.PowerCounting} and obtain an estimate
analogous to \eq{e1/11.9} but with double-sided $\L $-functions:
\be[e2/5.3]
     \cS _{p\in t\, K} [H(p),t] \leq {\L (t)\over t^{d_{H}}},\qquad
t > 0,
\ee
where
\be
     H\subset G \quad \hbox{ and } \quad K \cap \pi_g = \emptyset
\quad \hbox{\ for all } \quad g \in H.
\ee
For the products of singular functions of the described type an
$R$-operation can be constructed following the algorithm of
\ssect{R-op.Recursive} but using the special subtraction operators
$\bf\tilde r$ 
defined in \ssect{Subtraction.r-tilde}.  Such an $R$-operation
will be called {\it special $R$-operation\/} or {\it $\tilde{R}$%
-operation\/} and denoted in formulae as $\bf\tilde R$.  Its properties are
summarized in the following theorem.

\SUBSECTION{Theorem (existence and properties of
            $\tR$).}
\label{SpecialR.Theorem}

\begingroup
\it
Assume that all $F_g \in C^\infty(P_g\backslash \{0\})$ and
they satisfy the axioms \eq{e2/5.1}.  Then for any $s$-graph $G:$

 $(a)$ The operation $\tR$ defines the functional $\tR\. G \in
\cD'(P_G)$ which satisfies the conditions of \ssect{R-op.Axioms}.

 $(b)$ For any $\vf\in\cD (P_G)$ the following inequality holds:
\be[e2/5.4]
     |\langle \tR\. G, \vf \rangle| 
     \leq
     d^{-\o^\0_G} \cS [\vf, d],
\ee
where $d = \rad\supp\vf$ and all $\L $-functions are double-sided.

 $(c)$ $\tR\. G$ satisfies the following minimality condition $($cf.\
\eq{e2/4.16}$):$ if $\cP^{a}(p)$ is a homogeneous polynomial of
order $|a| < \o^\0_G$ then
\be[e2/5.7]
     \langle \tR \. G, \cP^{a} \rangle = 0.
\ee

\endgroup

{\bf Remarks.} The remarks $(i)$--$(iii)$, $(v)$ from
\ssect{GenericR.Theorem} remain valid here.

$ (iv)$ All the $\L $-functions in the estimate \eq{e2/5.4}---unlike
\eq{e2/10.6}---are double-sided. This means that the functional
$\tR \. G$ possesses, roughly speaking, the same power behaviour both
at infinity and at zero. In this respect note the importance not only
of the bound \eq{e2/5.1} which is stronger than \eq{e1/10.2} but also
of using the special subtraction operators $\tR $ (cf.\ the beginning
of \ssect{Subtraction.r-tilde}).

 $(vi)$ The identity \eq{e2/5.7} is crucial in the application of $\tR
$ to the study of Euclidean \asop\ (see
\sects{As-expansion}-{MainProof}).

\SUBSECTION{Proof.}
\label{SpecialR.Proof}%5.3

The proof here is very similar to that of theorem of
\ssect{GenericR.Theorem} on  generic $R$-operation and falls into three
parts.  First, one derives an estimate for $\cF = \tR '\. G$ similar to
\eq{e1/11.1} in exactly the same way as it was done in
\ssect{GenericR.Proof.UsingInduction} from an inductive assumption
about properties of $\tR \. \G$ for $\G < G$ since all the properties
of $\L $-functions needed there are also valid for double-sided $\L
$-functions. Second, one applies the operator $\tR $ whose properties
have been studied in \ssect{Subtraction.r-tilde}, to $\tR '\. G$.  The
only non-trivial step is to consider the case of factorizable graphs.

It was noted in \ssect{R-op.Recursive} that for factorizable graphs,
$\R\. G$ can be defined not by \eq{e1/6.7} but by first constructing
$\R'\. G$ as for non-factorizable $s$-graphs and then applying to it
the subtraction operator $\r$ while the factorization condition $(iv)$
of \ssect{R-op.Axioms} is ensured by an appropriate choice of the
finite renormalization $\D_G$.

Let us demonstrate that an analogous result holds also for the operation
$\tR $. The complication here is that the finite
renormalization $\tilde{\D}_G$ contains a lesser number of
arbitrary constants.

First, let us show that
\be[e2/5.8]
     \langle (\tR \. G_1 \times \tR \. G_2), \cP^{a} \rangle = 0
\ee
not only for polynomials
\be[e2/5.9]
     \cP^{a}(p) = \cP^{a_1}(p_1) \times \cP^{a_2}(p_2)
\ee
such that $|a_1| < \o_1$, $|a_2| < \o_2$, but also for all
polynomials with $|a_1| + |a_2| < \o^\0_G = \o_1 +
\o_2$ (recall that the homogeneous polynomials $\cP^{a}$ defined in
\ssect{Isolated.Deltas} can always be
chosen factorized).  To this end choose the cutoff function used in
the definition of the operation $\ast$, eq.\eq{e2/4.2}, in the
factorizable form of \ssect{Decomp.Radial}. Then one obtains:
\be[e2/5.10]
     \langle( \tR \. G_1 \times \tR \. G_2) \ast \cP^{a} \rangle
     = \lim_{\l \rightarrow \infty} 
     ( \langle 
         \tR \. G_1, \Phi^\l_1 \,\cP^{a_1} 
       \rangle \times 
       \langle \tR \. G_2, \Phi^\l_2 \, \cP^{a_2} 
       \rangle 
     ).
\ee
Making use of \eq{e2/5.4} for $\tR \. G_1$ and $\tR \. G_2$ results in
the inequality:
\be[e2/5.11]
     |\langle \tR \. G_\a, \Phi^\l_\a \, \cP^{a_{\a
}} \rangle| \leq \l^{|a_\a|-\o_\a} \, \L (\l ),\qquad
\a = 1,2.
\ee
So the expression under the limit in \eq{e2/5.10} is bounded by:
\be[e2/5.12]
     \l^{|a_1|+|a_2|-\o_1-\o_2} \, \L (\l ) =
\l^{|a|-\o^\0_G} \, \L (\l ),
\ee
which tends to zero when $\l \rightarrow \infty $ for $|a| <
\o^\0_G$, as required.

It is obvious that an appropriate choice of the finite renormalization
$\D_G$ \eq{e2/4.9} with summation over $\a$ extended to $|\a|<\o$,
ensures the equality:
\be[e2/5.13]
     \tr _{0}\. \tR '\. G + \D_G = \tR \. G_1 \times \tR \. G_1.
\ee
Now apply both sides of \eq{e2/5.13} to a polynomial $\cP ^{a}$, $|a|
< \o^\0_G$, using the opera\-tion~$\ast $. On the
r.h.s., one obtains zero. Then use the minimality condition
\eq{e2/4.16} for the l.h.s. From the resulting equality $\langle
\D_G, \cP ^{a}
\rangle = 0$ it follows that $z_{a} = 0$ for all $|a| <
\o^\0_G$ in the renormalization $\D_G$.  Therefore,
the subtraction operator which ensures factorizability of the
operation $\tR $ for a factorizable graph belongs to the class of
special subtraction operators.

The possibility to define the operation $\tR $ for the case of
factorizable graphs by the general equations \eq{e1/6.11}, \eq{e1/6.9}
allows one not to distinguish this case when deriving estimates.

This completes the proof.

\SUBSECTION{$\tR$ preserves scaling properties.}
\label{SpecialR.Scaling}

A property of the special operation $\tR $ which is exteremely useful
in applications, is that it does not change the form of scaling of the
initial expression $G(p)$ under the scaling of the argument $p
\rightarrow \l \, p$.  It turns out that the operation $\tR $---%
unlike the generic $R$-operation---violates the homogeneity properties
of the initial functional in a minimal degree, namely, by
introducing logarithmic corrections. The inductive proof of this fact
is straightforward, once the definitions are presented. However,
convenient explicit expressions for the scaling of $\tR \. G(p)$ are
more difficult to obtain, and this is postponed till
\ssect{SpecialR.VariationTheorem}, where we discuss the general
problem of how an \rop\ changes if the subtractions operators are
changed (because the non-trivial part of deriving scaling for
$\tR\.G(p)$ is equivalent to studying effects of rescaling the
cutoff parameter $\mu$ in the definition of subtraction operators
$\tr$, \eqs{e2/4.8}-{e2/4.11}). 

First we define the dilatation operator:
\be[e2/6.1]
     \hbox{Dil}_\l\. \cF (p) \bydef \cF (\l p),\qquad \l >
0.
\ee
Consider a functional $\cF $ satisfying \eq{e1/9.1} with double-sided
$\L$-functions.  Assume that the functional $\cF $ is {\it
almost-homogeneous}, i.e.\ (cf.\ \eq{e2/5.2}):
\be[e2/6.2]
     \hbox{Dil}_\l\. \cF  = \l^{-d_{\cF}} 
     \sum_n \:\log^n \l \,\cF _{[n]},
\ee
where $d_\cF $ is integer. Note that $\cF _{[0]}=\cF $.
The functionals $\cF _{[n]}$, $n>0$, will be called {\it
associated with} $\cF $. To simplify formulae it is convenient not
to show the limits of summation explicitly but rather to assume that
$\cF _{[n]}= 0$ for $n<0$ and for $n>L_{\cF}$ where $L_{{\cal
F}}$ is a non-negative integer.

Using the bound \eq{e1/9.1} one can see, first, that in \eq{e2/6.2}
\be[e2/6.3]
     d_{\cF} = \dim P + \o .
\ee

One finds from the definition \eq{e2/4.8} that $\tr \. \cF $ is
also almost-homogeneous in the sense of \eq{e2/6.2} and:
\be[e2/6.4]
     (\tr \. \cF )_{[n]} = \tr \. \cF _{[n]} + \D_{[n]}({\cal
F}) ,
\ee
where the action of $\tr $ on the associated functionals is defined
as follows:
\be[e2/6.5]
     \tr \. \cF _{[n]} 
     \bydef 
     \tr_{0}\. \cF _{[n]} 
     +
     \sum_{|\a |=\o_{\cF}} \d^\a \, {1\over n} \, 
           \langle
               \cF _{[n-1]}, \cP ^\a\, \phi_{\mu} 
           \rangle
\ee
(this definition makes sense for $n \not= 0$ ; however, curiously
enough, for $n = 0$ the r.h.s.\ contains the uncertainty 0/0 which
corresponds to the arbitrary constants in \eq{e2/4.9}); also:
\be[e2/6.6]
     \D_{[n]}(\cF ) 
   = \d_{n,L_\cF +1} \, 
     \sum_{|\a|=\o_{\cF}} \d^\a\, {1 \over n} \, 
     \langle \cF_{[n-1]}, \cP^\a\, \phi_{\mu} \rangle
\ee
for $n \not= 0$, and $\D_{[0]}(\cF ) = 0$, which agrees with
\eq{e2/6.6} because the first factor in \eq{e2/6.6} is the Kronecker symbol
and $L_\cF  \geq 0$.

It should be stressed that the counterterm $\D_{[n]}(\cF )$
contains only derivatives of order $\o $, the divergence index of
$\cF $.

One can see from \eqs{e2/6.4}-{e2/6.6} that the maximal $n$ for which
$(\tR \. \cF )_{[n]}\not= 0$ (or, equivalently, the maximal power
of $\log \l $ in $\hbox{Dil}_\l\. \tR \. \cF )$ is equal to:
\be[e2/6.7]
     L_{\tR \. \cF} = L_\cF  + 1 .
\ee
That is, $\tR $ introduces an additional logarithmic correction to the
dominant singularity. Nevertheless, one sees that the special
subtraction operators preserve the almost-homogeneity property.

Now, assuming that all $F_g(p)$ in a graph $G$ are
almost-homogeneous, it is straightforward to verify from the
definition of $\tR $ (\ssect{R-op.Recursive}) that $\tR \. G$ is also
almost-homogeneous.

Concerning explicit expressions for the scaling of $\tR \. G$, note
the following. The special subtraction operators defined in
\eqs{e2/4.8}-{e2/4.11} introduce an additional parameter $\mu$
(strictly speaking, to each subgraph $\g $ there corresponds a
parameter $\mu_{(\g )}$, but without loss of generality one can assume
all $\mu_{(\g )}$ to be proportional to one $\mu$). Now, it is easy to
understand that studying scaling of $\tR \. G$ with respect to $p^\0_G$
is equivalent to studying scaling of $\tR \. G$ with respect to this
parameter $\mu $. On the other hand, variations in $\mu $ are a
special case of variations in the form of the cutoffs 
$\Phi^{\mu}(p^\0_G) \equiv \Phi(p^\0_G /\mu )$ 
(cf.\ \eq{e2/4.11}) which are equivalent to
variations in the arbitrary constants $z_\a$.  Therefore, the
problem reduces to the more general one of studying explicit structure
of variations of the $R$-operation under changes of the arbitrary
finite counterterms. This fundamental problem is addressed in the
following section.

\SECTION {\rop s and counterterms.}
\label{Variations}

As we have seen, the definitions of both the generic $R$-operation and
the special \trop\ contain arbitrary parameters (the constants $z_{\a
}$ in \eq{e1/9.13} and \eq{e2/4.9}). However, since our definitions of
the $R$-operations are recursive, it is not immediately clear what is
the structure of variations with respect to those parameters of the
renormalized graph as a whole. To uncover that structure is an
important part of the theory of $R$-operations, as well as the
aim of the present section.

There are at least two situations where this problem arises. One is
the theory of UV renormalization in coordinate representation, and the
corresponding result is known as {\it renormalization group
transformation} (the renormalization group and its applications are
discussed e.g.\ in \cite{bog-shir}). Note that the so-called {\it
Zimmermann identities\/} in the theory of composite operators (see
e.g.\ \cite{zim70}, \cite{zav79}) are, essentially, renormalization
group transformations for the case of $R$-operation with
oversubtractions. Another situation is the theory of \asop, where
knowledge of the structure of variations of the operation $\tR$ is
used to extricate the non-analytical dependences in the expansion
parameter.

Since there is no essential difference between the generic and special
cases of the $R$-operation, we will not consider them separately.  In
the technical \ssect{SpecialR.Deltas} we introduce formal definitions
of the \rop\ on products involving $\d$-functions. The theorem
describing variations of \rop\ is formulated in
\ssect{SpecialR.VariationTheorem}. Its simple and straightforward
inductive proof is presented in \ssect{SpecialR.Variation.Proof}.

In \ssect{Append.Regularization} we represent \rop\ in terms of
regularization and infinite counterterms to make connection of our
definition of the \rop\ with the more conventional ones. Such
representations are of considerable practical and heuristic value
(cf.\ the reasoning in
\cite{bog-shir}, \cite{fvt91}, as well as the fact that calculations
are normally done using such representations---cf.\ the
state-of-the-art calculations mentioned in the Introduction).  They
exhibit the fact that \rop s modify the products they act on
only at singular points, demonstrate independence of the choice of
decompositions of unit (cf.\ \ssect{Consistency.Independence}) etc.

On the other hand, the use of a regularization requires an additional
analytical investigation of a limiting procedure ({\it removal of
regularization}). We take this opportunity to introduce a so-called
{\it natural regularization\/} based on the cut-offs involved in the
definitions of subtraction operators (see \eq{e1/9.2} and
\eq{e1/9.11}). This natural regularization does not require any
additional analysis of limiting procedures.

Finally, in \ssect{SpecialR.Example} we briefly explain how the
results of the preceeding subsections are translated into the language
of the theory of UV renormalization in coordinate representation. This
should help one to connect our abstract constructions with more
familiar results. The main application, however, will be to the theory
of \asop\ (\ssect{Existence.Theorem}, remark $(iv)$).

\SUBSECTION{$R$-operation on products with $\d$-functions.}
\label{SpecialR.Deltas}

To facilitate further manipulations with the $R$-operation it is
convenient to make a specific choice of the $\d $-functions $\d^\a_\g$
and the corresponding dual set of homogeneous polynomials $\cP ^\a_\g$
introduced in
\ssect{Isolated.Deltas}.
Our final results will not depend on this choice.

Define:
\be[e3/2.6]
     \d^\a_\g(p^\0_\g) \bydef \left[ {(-)^\a \over
\a!} \right] \cD ^\a_\g \d (p^\0_\g),
\ee
where $\a $ is a standard multiindex and $\cD ^\a_\g$ are partial
derivatives with respect to $p_\g$.  The set of polynomials that is
dual to \eq{e3/2.6} is
\be[e3/2.7]
     \cP ^\a_\g (p^\0_\g) \bydef p^\a_\g .
\kern26mm\ee
As was mentioned in \ssect{Consistency.Co-subgraphs}, in the theory of
$R$-operation one encounters objects obtained from products $G(p^\0_G)$
by replacing a subgraph $\g$ by a $\d $-function $\d^\a_\g$.
Let us define an $R$-like operation $\R_{(G)f}$ ($f$ stands for
``factorizable") on such products.

Consider the elementary identity:
\be[e3/2.9]
     \d^{\b}_\g(p^\0_\g) \,\vf (p^\0_\g) 
     \equiv
     \mathop{\sum_{\b'} \sum_{\b ''}}_{\b '+\b ''=\b}
       \d^{\b'}_\g(p^\0_\g) (\b ''!)^{-1} 
       [\cD ^{\b''}_\g \vf(p_\g)]_{p_\g=0} ,
\ee
where $\vf $ is an arbitrary function depending on $p_\g$.

Replace $\vf (p_\g)$ by $\Gg (p^\0_G) = \Gg (p_\g,p^\0_\Gg)$ in
\eq{e3/2.9}.  The r.h.s.\ will contain the expressions
\be[r.h.s.expr??]
     [\cD ^{\b''}_\g \Gg (p_\g,p^\0_{\Gg})]_{p_\g=0},
\ee
that are exactly the co-subgraphs defined in
\ssect{Consistency.Co-subgraphs}, and the operation $\R$ was defined
on co-subgraph in \ssect{Consistency.Co-subgraphs}.  (Recall that no
relation was assumed between the operation $\R$ on graphs and
co-subgraphs---see \ssect{Consistency.Co-subgraphs}; such relation may
exist but we will never use it.) Then one can define the operation
$\R_{(G)f}$ on the product $\d^{\b}_\g\, G
\backslash \g $ by applying $\R$ termwise to the expressions on the
r.h.s.\ of \eq{e3/2.9}
\be[e3/2.8]
     \R_{(G)f} \.  
     \left[ 
         \d^{\b}_\g(p_\g) \Gg(p^\0_G) 
     \right] \kern6.5cm
\ee\be\kern2cm
     \bydef 
         \mathop{\sum_{\b'}\sum_{\b ''}}_{\b '+\b''=\b} 
         \d^{\b'}_\g(p_\g) (\b''!)^{-1} 
         \R_{(\Gg)} \. 
         \left[ \cD ^{\b''} \Gg (p^\0_G) \right]_{p_{\g}=0} .
\ee
For $\g =\emptyset $, eq.\eq{e3/2.8} formally becomes:
\be[e3/2.10]
     \R_{f} \. G \equiv \R \. G.
\ee
\Eq{e3/2.8}, essentially, says that $\R_{f}$ commutes with the
$\d$-functions provided the other factors in the product are properly
projected onto the subspace singled out by the $\d $-function.

The operation $\R'_{f}$ is defined as a projection of \eq{e3/2.8} to
$P_G\backslash \{0\}$.

\SUBSECTION {Theorem (structure of variations of the $R$-operation).}
\label{SpecialR.VariationTheorem}

\begingroup
\it
Let $\R_{b}$ and $\R_{a}$ be two $R$-operations defined on the
universum of graphs and differing in the choice of subtraction
operators. Then the following relation is true:
\be[e3/2.11]
     \R_{a}\.  G(p^\0_G) 
     = 
     \sum_{\g \leq G} \sum_\b Z^\b_\g \, 
     \R_{b,f} \.  [ \d^\b_\g(p_\g) \, \Gg (p^\0_G) ] ,
\ee
where the summation over\/ $\g $ includes $\g =\emptyset$ $($in which
case we define\/ $\o_{\emptyset} \bydef -\infty $ and\/
$\d_{\emptyset}\bydef 1)$, while the summation over $\b $ is
restricted by the conditions $|\b | = \o^\0_\G$ for the case of the
special operations, $|\b | \leq \o^\0_\G$ for the case of generic
$R$-operation, and $|\b | \leq\o^{\textbox{eff}}_\g$ for
$R$-operations with oversubtractions.  Further, $Z^{\b}_\g$ are finite
constants such that
\be[e3/2.12]
     Z^{0}_{\emptyset} = 1.
\ee
$Z^{\b}_\g$ depend on the two $R$-operations $($and, of course,
$\g)$ but are independent of G. Moreover, if $\g $ factorizes as $\g =
\g_1\times\g_2$ $($so that $p_\g = (p_{\g_1}, p_{\g_2})$ and
$\b = (\b_1, \b_2) )$ then $Z$ also factorizes: $Z^{\b
}_\g = Z^{\b_1}_{\g_1} \times Z^{\b_2}_{\g_2}$.

\endgroup

\SUBSECTION{Proof.}
\label{SpecialR.Variation.Proof}

\hbox{\Eq{e3/2.11} can be proved by induction over the graphs of the
universum, as follows.}

For the graphs $G$ which have no $s$-subgraphs one has $\R \equiv \r$
and \eq{e3/2.11} takes the form
\be[e3/2.13]
     \r_{a}\.  G = \r_{b}\.  G + \sum_{\b} Z^{\b}_G \,
\d^{\b}_G ,
\ee
which follows from the definition of $\r$, \ssect{Subtraction.r}.
This establishes correctness of the starting point for the induction.

Further, assume that \eq{e3/2.11} holds for any $\G < G$ and $[\cD
^{\b}\Gg ]_{p_\g=0}$ where $\G $ and $\g $ are
subgraphs of $G$. (A formula similar to \eq{e3/2.11} but with
different coefficients is then valid for the incomplete operation
$\R'$, except that summation will run over $\g < \G $ etc.)  By
definition:
\be[e3/2.14]
     \R'_{a}\.  G 
     = 
     \sum_{\G \triangleleft G} \theta_\G (\R_{a}\.\G ) (\GG ).
\ee
Substitute \eq{e3/2.11} (taken for $\G $ instead of $G$) into the
r.h.s.\ of \eq{e3/2.14} and change the order of summations over $\g $
and $\G $:
\be[e3/2.15]
     \R'_{a}\.  G 
     = \sum_{\g < G} \sum_{\b} Z^{\b}_\g
       \sum_{\G \triangleleft G, \G \geq \g} 
       \theta_\G \, \GG \, \R_{b,f} \.  
       [ \d^{\b}_\g \, \G\backslash\g ] .
\ee
The functions $\theta_\G$ are arbitrary within the limits specified
in \ssect{Decomp.Submaximal} and a simple geometrical fact is that
$\theta_\G$ can be chosen so that $[\cD ^{\b}_\g\theta_{\G
}]_{p_\g=0} = 0$ (\ssect{Decomp.Submaximal}).  Taking this into
account and using \eq{e3/2.9} and the definition \eq{e3/2.8} one
obtains:
\be
     \R'_{a(G)}\.G = 
     \sum_{\g < G} \,
     \mathop{\sum_{\b '}\sum_{\b''}}
      _{\b '+\b ''=\b^{\vphantom{i}} }  \,
     \mathop{\sum_{\b'''}\sum_{\b^{iv}}}
      _{\b '''+\b^{iv}=\b'} 
     Z^{\b '+\b''}_\g 
     ( \b ''! \b^{iv}! )^{-1} \d^{\b '''}_\g
\ee\be[e3/2.16]
     \kern3cm\times \sum_{\g\leq\G \triangleleft G} 
     [ \theta_\G]_{p_\g=0}\, [ \cD ^{\b^{iv}} \GG ]_{p_\g=0}\,
     \R_{b(\Gg )}\.  [\cD ^{\b''}\G\backslash\g ]_{p_\g=0} .
\ee
The sum in the last line re-assembles into
\be
     \R'_{b(\Gg)}\.  [ \cD ^{\b^{iv}+\b ''}\Gg ]_{p_\g=0}
\ee
(one can check this using \eq{e3/2.14} specified to $[\cD ^{\b}G
\backslash \g ]_{p_\g=0}$ instead of $G$ and $b$ instead of $a$,
which is valid by the inductive assumption; one can also see that the
functions $[\theta_\G]_{p_\g=0}$ constitute a decomposition of
unit with the necessary properties). Renaming the summation indices
and using the notation \eq{e3/2.8} for $\R'_{b}$ one rewrites the
above as follows:
\be[e3/2.17]
     \R'_{a}\.  G = \sum_{\g < G} \sum_{\b} Z^{\b}_\g \,
\R'_{b,f} \. [\d^{\b}_\g \, \Gg ].
\ee
(Note that the constants $Z^\b_\g$ here are the same as in
\eq{e3/2.15}.)
To transform this into $\R_{a}$ one only has to apply the
corresponding subtraction operator $\r_{a}$. It should be noted here
that the terms in the sum on the r.h.s.\ of \eq{e3/2.17} satisfy all
the criteria of \ssect{Subtraction.Problem} for applicability of
subtraction operators. On the r.h.s., one will have:
\be[e3/2.18]
     \r_{a}\.  \R'_{b,f} \.  [ \d^{\b}_\g \, \Gg ]
     = \R_{b,f}\.  [ \d^{\b}_\g \, \Gg ] 
     + \sum \hat{Z}^{\b,\b '}_{G,\g} \, \d^{\b '}_G.
\ee
Using this and rearranging the terms in the sum, one arrives at
\eq{e3/2.11}.  The limits of summation over $\b $ indicated after
\eq{e3/2.11} easily follow from \eq{e3/2.18}.

Note that the above reasoning remains valid for factorizable graphs,
so that one can assume that \eq{e3/2.11} has been proved for
factorizable subgraphs.  Then factorizability of $Z_\g$ for
factorizable subgraphs $\g $ can be seen as follows:
\be
     \R_{a}\. \g = \R_{a}\. \g_1 \times \R_{a}\. \g_2
\kern7.92cm\ee\be
     = \Big( \R_{b}\. \g_1 + \ldots + \sum_{\b_1}
Z^{\b_1}_{\g_1} \d^{\b_2}_{\g_2} \Big) \, \Big( \R_{b}\.
\g_2 + \ldots + \sum_{\b_2}Z^{\b_2}_{\g_2}
\d^{\b_1}_{\g_1} \Big)
\ee\be
     = \R_{b}\. \g + \ldots + \sum_{\b_1}\sum_{\b_2} \Big(
Z^{\b_1}_{\g_1} Z^{\b_2}_{\g_2} \Big)
\d^{\b_1}_{\g_1} \d^{\b_2}_{\g_2},\kern3.04cm
\ee
where the dots denote the terms that contain non-trivial factors
besides $\d $-functions. Comparing the obtained expression with
\eq{e3/2.11} for $G=\g $, one arrives at the desired result, which
completes the proof of the theorem.

Considering $\R_{a}$ as an arbitrary varying, and $\R_{b}$ as a fixed
basic $R$-operation, one can interpret eq.\eq{e3/2.11} by saying
that the arbitrariness of the $R$-operation is fully described by the
finite constants $Z$ in the representation \eq{e3/2.11}.

\SUBSECTION {Natural regularization and infinite counter\-terms.}
\label{Append.Regularization}

Let us recast our definition of \rop s into a form involving a
regularization and infinite counterterms. The definition of
subtraction operators \eq{e1/9.10} can be rewritten as
\be[eA.1]
     \r^\0_{(\G )}\. \cF  
     = \lim_{\e_\G\rightarrow +0}
\Big\{ \Phi_{\e_\G} \, \cF  + \D_{\e_\G} \Big\},
\ee
where
\be[eA.2]
     \D_{\e_\G} = \sum_{|a|\leq \o^\0_\G}
z_{a,\e_\G} \, \d^{a}_\G,
\ee
and $z_{a,\e_\G}$ are numeric expressions depending on ${\cal
F}$.  The braced expression is naturally interpreted as a regularized
functional $\cF $ plus counterterms.

The generalization is straightforward. For each non-factorizable
$s$-subgraph $\g$ define $\e_\g>0$ and the cutoff
$\Phi_{\e_\g}(p)$, $p\in P_\g$ (cf.\ \ssect{Decomp.Radial}).
Let $\{\e\}$ be the collection of all such $\e_\g$.
Define the limiting procedure of removing regularization as:
\be[eA.3]
     \lim_{\{\e \}\rightarrow 0} \bydef \prod_{\g \in S[G]}
\lim_{\e_\g\rightarrow +0},
\ee
the order of limits being fixed so that if $\G > \g $ then taking
$\e_\g \rightarrow 0$ should precede taking $\e_\G
\rightarrow 0$. Define the regularized graph as:
\be[eA.4]
     G^{\{\e \}} \bydef \prod_{\g \in S[G]} \Phi_{\e_\g}
     G \,\in \cC^\infty(P_G).
\ee
Regularized subgraphs are defined similarly.

It is straightforward to check by induction in subgraphs that
\be[eA.5]
     \R\. G = \lim_{\{\e \}\rightarrow 0}[\R\. G]^{\{\e
\}},
\ee
where
\be[eA.6]
     [\R\. G]^{\{\e \}} = G^{\{\e \}} + \sum_{\g_1\ldots
\g_n} \D^{\{\e \}}(\g_1)\ldots \D^{\{\e \}}(\g_n)\,
[ G\backslash (\g_1 \cup\ldots \cup\g_n) ] ^{\{\e \}},
\ee
where summation runs over all sets of non-factorizable and pair-wise
non-intersecting $s$-sub\-graphs, while
\be[eA.7]
     [G\backslash (\g_1 \cup \ldots \cup\g_n)]^{\{\e \}} 
     =
     \left( 
        \mathop{\prod_\g}_{\g \not\leq \g_1...\g \not\leq \g_n} \Phi_\g 
     \right)\,
     G\backslash (\g_1 \cup \ldots \cup\g_n),
\ee
and $\D^{\{\e \}}(\g )$ are counterterms of the form of
\eq{e1/9.13} with $\o = \o^\0_\G$ and $z_\a =
z_{\a\{\e \}}$, localized on the corresponding singular
planes $\pi_\g$.  (We will not need explicit expressions for the
counterterms.)

Eq.\eq{eA.6} has the desired form of a standard definition of the
$R$-operation.

Note that the apparent dependence on various decompositions of unit
used in the definition of the $R$-operation in \ssect{R-op.Recursive}
disappeared in \eq{eA.6}.

In the case of special $R$-operations, there is also an implicit
regularization at the upper limit of integration (recall
\eq{e2/4.2}). Then all one has to do is to replace
$\e_\G$ by the pair $(\e_\G, \L_\G)$, the
lower-limit cutoff $\Phi_{\e_\G}$ by a two-sided cutoff
$\Phi^{\L_\G}_{\e_\G}$, and instead of $\lim_{\e_{\G
}\rightarrow +0}$ in \eq{eA.1} one will have
$\lim_{\e_\G\rightarrow +0}
\lim_{\L_\G\rightarrow +\infty}$
(note that the two limits commute).  Otherwise, the above equations
remain the same.

It is worth stressing once more that {\it any\/} regularization
involves, perhaps implicitly, all the above limiting
procedures---because the very definition of an integral in infinite
limits or of an integral of a singular function involves at
intermediate steps cutoffs equivalent to those used above.  Therefore,
the apparent complexity of the above expressions is simply a
manifestation of the inherent complexity of the objects we are dealing
with.

\SUBSECTION{Example: theory of UV renormalization in
            coordinate representation.}
\label{SpecialR.Example}

To make the connection of \eq{e3/2.11} with the ordinary
renormalization group clearer, let us briefly explain what the result
\eq{e3/2.11} looks like in the case of Bogoliubov's $R$-operation in
coordinate representation.  The variable $p^\0_G$ now has the form
$p^\0_G = (x_1,\ldots x_n)$ where the space-time coordinates
$x_1,\ldots x_n$ correspond to vertices of the Feynman diagram
$G$. Each factor of $G$ has the form $D(x_i-x_{j})$ and each factor
is singular in the sense of \ssect{Graphs.Factors} 
and example ($iii$) of
\ssect{Graphs.Example}. Each complete singular
subgraph in the sense of our formalism is exactly a collection of
non-intersecting UV subgraphs (in the Bogoliubov-Parasiuk
definition), while both the $\d $-functions and the
constants $Z$ in \eq{e3/2.11} factorize into the corresponding
products for subgraphs.

Diagrammatically, replacing an UV subgraph $\g (x_{i_1}\ldots
x_{i_{m}})$ by a (derivative of) $\d $-function
\be
     \d (x_{i_1}\ldots x_{i_{m}}) = \d (x_{i_1}- x_{i_2}) \times
\ldots ,
\ee
as done on the r.h.s.\ of \eq{e3/2.11}, corresponds to ``shrinking $\g
$ to a point", i.e.\ effectively replacing $\g $ by a new vertex to
which there corresponds a coordinate variable. Thus one restores an
equation with a familiar structure (which we, as is customary,
represent in a somewhat schematic form):
\be[BP-R-op]
     \R_{a}\. G = \sum_{\g_1\ldots\g_N} \Big( \prod_i
z^{ab}_{\g_i} \Big) \R_{b}\. G/(\g_1\ldots\g_N),
\ee
where the operation / on the r.h.s.\ denotes shrinking the
corresponding subgraphs to a point. Summing \eq{BP-R-op} over all $G$
from a perturbation series and performing combinatorial exponentiation
in a standard fashion, one can rewrite such a renormalization-group
transformation in terms of finite renormalizations of fields and
parameters of the Lagrangian. We will consider issues of
exponentiation in a somewhat different and more general context when
the theory of UV renormalization will be studied within the framework
of momentum representation and the \asop, where an equation similar to
the above will be obtained \cite{gms}, \cite{rmp2}.

One can perform a similar excercise with the results of
\ssect{Append.Regularization} and obtain a formula similar to
\eq{BP-R-op} except that there will be no 
$\R$ on the r.h.s.\ (its role will be played by the regularization)
while the finite coefficients $z^{ab}$ will be replaced by
coefficients $Z^\e$ which diverge as the regularization is removed.
Such a representation of the \rop\ is used e.g.\ in the heuristic
reasoning of \cite{bog-shir}.

\newpage\thispagestyle{myheadings}\markright{}   \PART  {As-expansions
and Extension Principle.}

In this part of our work we turn to our central topic---asymptotic
expansions of products of singular functions in the sense of
distributions and principles of their construction. The mathematical
problem itself is new (its relevance to, and importance for, the
theory of multiloop Feynman diagrams was pointed out in \cite{fvt82},
\cite{fvt91}).  Its interesting feature is that it considers the 
distinctly applied problem of asymptotic expansions from a rather
abstract point of view of functional analysis. The heuristic potential
of such an approach was amply demonstrated by discoveries of the
calculational algorithms mentioned in the Introduction.

In this part we consider the problem of asymptotic expansions from a
general point of view. First we discuss definitions and general
properties of asymptotic expansions in the sense of distributions
(\sect{As-expansion}). The set of such properties turns out to be
rich enough to allow one to study the general structure of the \asop\
on graphs; in particular, we formulate the key locality condition
\cite{fvt91} which reveals the recursion structure of the expansion
problem and allows one to reduce the problem to the case of a
singularity at an isolated point (\sect{As-operation}).  This last
step of the solution is explained in \sect{ExtensionPrinciple} where
the so-called extension principle is presented which offers a recipe
to construct \asex s in a very general situation.  

The totality of the properties of \asex s together with the
extension principle constitute a universal and flexible constructive
context in which to consider specific problems.

\SECTION{General properties of $\protect\As$-expansions.}
\label{As-expansion}%1

In this section we discuss the definition and general properties of
asymptotic expansions of functions taking values in spaces of
functions or distributions. The content of this section is practically
independent of the discussion of \rop s in the preceeding sections.
Concerning general properties of asymptotic expansions of
numeric-valued functions see e.g.\ \cite{erdelyi},
\cite{fedoriuk}.

In \ssect{As-exp.Definition} the definition of the so-called {\it
As}-expansions (the special case of asymptotic expansions we are
dealing with) is presented which is essentially due to Erdelyi (see
\S1.2 in \cite{erdelyi}), in \ssect{As-exp.Generalizations} possible
generalizations are considered which are likely to be of interest in
future developments.  In \ssect{As-exp.Uniqueness} a very important
property of uniqueness of the \asex s is pointed out. Then
in \ssect{As-exp.Notations} a convenient system of notations is fixed,
and in \ssect{As-exp.Functional} we clarify the meaning of the phrase
that the remainder term of the \asex s vanishes at a given
rate for some important classes of functional spaces (we follow
\cite{vsvlad} in defining convergence in spaces of distributions and
test functions). In \ssect{As-exp.LinearMaps} and
\ssect{As-exp.DirectProducts} we study the behaviour of {\it
As}-expansions under linear and bilinear mappings of spaces in which
the expanded functions take values.  In \ssect{As-exp.LinearMapsInverse} it is
shown how the problem of extension of approximating functionals
emerges in a rather general context.  Finally, in
\ssect{As-exp.Globalization} we consider how
the \asex s obtained in different regions for the same
distribution are glued together into one \asex\ valid
everywhere in the union of all the regions.

\SUBSECTION{Expansion parameter.}
\label{As-exp.Parameter}%1.1

Our standard notation for the expansion parameter is $\k $.  We
always assume that $\k>0$ while the expansions are considered at
$\k \rightarrow + 0$.

Strictly speaking, different $\k $-dependent expressions that we  will
deal with are defined  on  different  intervals $(0, \k_{0}$).
However  we  are  only interested in studying the limit
$\k \rightarrow  0$. So, it is sufficient that  all  the objects and
operations on them make  sense  simultaneously  on  a  small  but
non-empty interval of values of $\k $. Such an interval will always
exist in all our reasonings.

The dependence of functions and distributions on $\k $ will be
denoted as $G(p,\k)$. If a function is considered as an element of
a linear space, the notation $G_\k$ is used.

\SUBSECTION{\asex s.}
\label{As-exp.Definition}%1.2

Let $L$ be a linear topological space. For a function $f: \k
\rightarrow  f_\k \in  L$  the expression
\be[e2/1.1]
     f_\k \ \equals^{L}\  o(\k^N)
\ee
means that $f_\k / \k^N \rightarrow 0$ as $\k
\rightarrow 0$ in the topology of {\it L}.  In
\ssect{As-exp.Functional} the meaning of this will be defined more
precisely for some classes of $L$.

Consider a function $\k \rightarrow G_\k \in L$.  We are
interested in asymptotic expansions of the form
\be[e2/1.2]
     G_\k
     \ \mathrel{\mathop{\0\simeq\0}^{L}_{\k \rightarrow 0}}\ %
     \sum_n G_{\k,n} ,
\ee
where the formal sum runs over integer $n$, while $G_{\k,n} = 0$
for $n < N_{0}$; $N_{0} > - \infty $ depends on $G_\k$. The
expression \eq{e2/1.2} means that for any $N$
\be[e2/1.3]
     G_\k - \sum_{n\leq N} G_{\k,n} 
     \ \equals^{L}\  o(\k^N) .
\ee
For the special case $N = N_{0} - 1$ one has
\be[e2/1.4]
     G_\k \equals^{L} o(\k^{N_{0}-1}) .
\ee
In the present work we deal only with expansions in
integer powers and logs of $\k $. This means that the dependence
of $G_{\k,n}$ on $\k $ is as follows:
\be[e2/1.5]
     G_{\k,n}
     = \k^n\,
     \sum^{\0 I_n<+\infty}_{k=0} 
     \ln^k \k \, G_{n,k}\, ,
\ee
where $G_{n,k} \in L$ are independent of $\k$.  The dependence on
$\k $ of the form \eq{e2/1.5} can be conveniently called {\it
power-and-log dependence} and expansions of the form \eq{e2/1.2},
\eq{e2/1.5} can be  called  {\it power-and-log  expansions}.

Expansions of the form \eqs{e2/1.2}-{e2/1.5} will be called {\it
As}-{\it expansions}.

\SUBSECTION{Generalizations.}
\label{As-exp.Generalizations}%1.3

A major part of our formalism can be easily applied to more general
cases, e.g.\ when non-integer powers of $\k $ are allowed in
expansions \eq{e2/1.2}.  It is sufficient to say that the summation in
\eq{e2/1.2} runs over real $n$ while for any real $N$ there exists
only a finite number of functions $G_{\k,n}\not= 0$ with $n < N$.

An analogous generalization is possible with respect to powers of $\ln
\k$ in \eq{e2/1.5}.

\SUBSECTION{Uniqueness.}
\label{As-exp.Uniqueness}%1.4

The continuity of linear operations in $L$ immediately results in a
straightforward generalization of the well-known property of
asymptotic expansions of numeric-valued functions (see \cite{erdelyi},
\cite{fedoriuk}): if the \asex\ exists then it is unique.
This fundamental property will be frequently used
(\ssects{As-exp.LinearMaps}-{As-exp.DirectProducts}) because it
ensures that asymptotic expansions commute with all kinds of
operations---algebraic and other. For example, if the expressions to
be expanded satisfy some condition of an algebraic type, then \asex s
will inherit such a property. In particular, the uniqueness property
has important implications for the second, combinatorial, part of the
expansion problem in pQFT: the algebraic properties of non-expanded
diagrams (e.g.\ due to gauge invariance) are inherited termwise by the
expanded expressions in a straightforward manner.  This eliminates a
major stumbling block of the conventional theory of asymptotic
expansions of Feynman diagrams \cite{factoriz}.%
\footnote{
For an example of how gauge properties of \asex s of Feynman
diagrams can be studied, see \cite{pivo-gauge}.
\relax}

\SUBSECTION{Some notations.}
\label{As-exp.Notations}%1.5

Let us introduce convenient notations for \asex s. If
there exists an \asex\ \eq{e2/1.2}, \eq{e2/1.5} then in
virtue of \ssect{As-exp.Uniqueness} the following operations are
well-defined:
\be[e2/1.6a]
     \as_n\.G_\k \bydef  G_{\k,n},
\kern15.4mm\ee\be[e2/1.6b]
     \As \bydef  \sum_n \as_n,
\kern3.4mm\ee\be[e2/1.6c]
     \As_N \bydef \sum_{n\leq N} \as_n,
\kern4mm\ee\be[e2/1.6d]
     \D_N \bydef  1 - \As_N  .
\ee
Now eqs.\eq{e2/1.2} and \eq{e2/1.3} can be rewritten as:
\be[e2/1.2'-1.3']
     G_\k 
     \mathrel{\mathop{\0\simeq}^{L}} 
     \As\. G_\k,
     \qquad
     \D_N\. G_\k \equals^{L} o(\k^N) .
\ee
If the \asex\ does not exist for $G_\k$ then the
operations \eqs{e2/1.6a}-{e2/1.6d} are undefined.

We will deal with three types of spaces $L$: the spaces $\cD'(\cO )$
of distributions over open regions $\cO \subset P$ in Euclidean space,
and the spaces of smooth functions $\cC^{\infty }(\cO )$ and $\cD (\cO
)$.  Our problems are such that the
\asex s for distributions are the object of construction
while the expansions for smooth functions are assumed to be known
(normally, the latter are Taylor expansions).  Therefore it is
convenient to use special notations $\T$ and $\t$ for smooth
functions, instead of $\As$ and $\as$.  For the remainder term we
always use the notation $\D_N$.

\SUBSECTION{\asex s in functional spaces.}
\label{As-exp.Functional}%1.6

Let us clarify the meaning of \eq{e2/1.2'-1.3'} which expresses the
asymptotic character of the expansion \eq{e2/1.2}, for various types
of $L$.

If $L$ is equal to:
$(i)$ $\cC^\infty(\cO )$,
$(ii)$ $\cD (\cO )$,
$(iii)$ $\cD '(\cO )$ then \eq{e2/1.2'-1.3'} means,
respectively, that:

 $(i)$
for any compact $K \subset  \cO $ and any integer $\o  \geq  0$
\be[e2/1.7]
     \|\D_N\. G(p,\k)\|^{\o}_{p\in K} 
     = o(\k^N)
\ee
(the seminorms $\|\cdot\|$ were introduced in
\ssect{Isolated.Seminorms2}; note that \eq{e2/1.2'-1.3'} makes sense
also when $G(p,\k)$ (and therefore $\D_N\.  G(p,\k)$)
belongs to the class $\cC^\infty(K)$ only for $\k < \k_K$
where $\k_K$ depends on $K$);

 $(ii)$
there exists a compact $K \subset  \cO $ such that for all
sufficiently small $\k $ and all $\o  \geq  0$
\be[e1.8]
    \supp \D_N\. G_\k \subset  K,
    \qquad
    \|\D_N\. G(p,\k)\|^{\o}_{p\in \cO}
    = o(\k^N) ;
\ee

 $(iii)$
for any $\vf  \in  \cD (\cO )$
\be[e2/1.9]
     \langle \D_N\. G_\k, \vf  \rangle
     = o(\k^N) .
\ee

Now, let us study general properties of \asex s.

\SUBSECTION{\asex s and linear mappings.}
\label{As-exp.LinearMaps}%1.7

Let $A$ be a continuous linear mapping of one topological space $L$
into another $\bar{L}$. If there exists the \asex\ for
$G_\k \in L$ then the \asex\ for $A\. G_\k
\in \bar{L}$ also exists and is defined by:
\be[e2/1.10]
     \As\. A\. G_\k = A\. \As\. G_\k .
\ee
In other words, operations $\As$, $\as$, $\D_N$ can be ``carried
through" $A$.  On the r.h.s.\ of \eq{e2/1.10}, $A$ is applied termwise
to the series $\As\. G_\k$.

{\bf Examples.}
$(i)$ Multiplication of $\vf_\k \in \cD (\cO )$
by a constant and 

$(ii)$ a pointwise multiplication by $\Phi \in
\cC^\infty(\cO )$; 

$(iii)$ the pointwise multiplication of $\Phi_\k \in C^{\infty
}(\cO )$ by $\vf \in \cD (\cO )$ (the mapping
$\cC^\infty(\cO ) \rightarrow \cD (\cO )$);

$(iv)$ the application of a distribution from $\cD '({\cal
O})$ which is independent of $\k $ to a $\k$-dependent test
function $\vf_\k$ (the mapping $\cD (\cO )
\rightarrow $ complex numbers); 

$(v)$ multiplication of $G_\k \in \cD '(\cO )$ by
$H \in \cC^\infty(\cO )$ etc.  

$(vi)$ A somewhat non-trivial example of a different kind is the
differentiation of distributions from $\cD '(\cO )$:
\be[e2/1.11]
     \As\. (\partial  G_\k) = \partial  (\As\. G_\k) .
\ee

$(vii)$ Another important example is {\it localization}, i.e.
restriction of a distribution from $\cO $ onto $\cO ' \subset
\cO $:
\be[e2/1.12]
\As\.(G_\k |_{\cO '}^\0)
=
(\As\. G_\k) |_{\cO '}^\0 .
\ee

\SUBSECTION{\asex s and linear mappings: the inverse problem.}
\label{As-exp.LinearMapsInverse}%1.8

Consider the following fundamental problem which is inverse to the one
studied in the preceding subsection: given the expansion $\As\. A\.
G_\k$, it is necessary to construct $\As\. G_\k$.
Without pursuing full rigour, let us discuss the general structure of
the problem.

Denote as $\cL $ and $\overline\cL $ the spaces that are
(topologically) conjugate to $L$ and $\overline{L} = A\.L$,
respectively.  Then $G_\k$ and $A\.G_\k$ can be considered as
linear functionals over $\cL $ and $\overline\cL $, respectively.  Let
$\overline{\vf} \in \overline{\cL}$.  Then the equation
\be
     \langle \overline{\vf}, (A\. G_\k) \rangle
     = \langle (\overline{\vf}\. A), G_\k \rangle
\ee
defines a vector $\overline{\vf}\. A \in \cL $ and demonstrates
that the mapping $A$ can be treated as a restriction of the functional
$G_\k$ from $\cL $ onto a subspace $\cL _{A} \subset
\cL $.  The expansion $\As\. A\. G_\k$ induces an expansion
for the restriction of $G_\k$ onto $\cL _{A}$ and then the
problem of construction of $\As\. G_\k$ is transformed into the
problem of extension of the \asex\ of a functional from a
subspace onto the entire space on which the functional is defined.
The latter problem will be studied in all generality in
\sect{ExtensionPrinciple}.

\SUBSECTION{Expansions of direct products.}
\label{As-exp.DirectProducts}%2/1.9

Consider a continuous bilinear mapping $L' \times L'' \rightarrow L$
where $L$, $L'$, $L''$ are linear topological spaces. We will denote
this mapping by $\times $.  Assume that $G'_\k \in L'$ and
$G''_\k \in L''$ both possess \asex s. Then
\be[e2/1.13]
     \As\. (G'_\k \times  G''_\k)
     =  (\As\. G'_\k) \times  (\As\. G''_\k) .
\ee
The r.h.s.\ should be understood as the result of termwise
multiplication of the two expansions with a subsequent reordering of
the terms in increasing powers of $\k $, so that
\be[e2/1.14]
     \as_n \. (G'_\k \times  G''_\k)
     =   \sum_{l+m=n} (\as_l \. G'_\k)
     \times  (\as_m \. G''_\k) .
\ee
The proof of this statement follows easily from a useful 
summation-by-parts formula:
\be[e2/1.15]
    \D_N\. (G'_\k \times  G''_\k)
    =   (\D_{N-M_{0}}\. G'_\k) \times  G''_\k
    +   \sum_{l\leq N-M_{0}}
        (\as \. G'_\k)
        \times  (\D_{N-m}\. G''_\k),
\ee
where $M_{0}$ is the lower bound of summation in the {\it
As}-expansion for $G''_\k$.

{\bf Examples.} $(i)$ $G_\k \in \cD '(\cO )$, $\vf_\k \in
\cD (\cO )$ while the bilinear mapping is just the application of the
distribution to the test function. Then
\be[e2/1.16]
     \langle G_\k, \vf_\k \rangle
     \mathrel{\mathop{\0\simeq\0}_{\k \rightarrow 0}}
     \langle \As\. G_\k, \T\. \vf_\k \rangle
\ee
($\T$ is defined in \ssect{As-exp.Notations}).

 $(ii)$ $\cO = \cO _1 \times \cO _2$, $G(p,\k) =
G_1(p_1,\k)\times G_2(p_2,\k)$ where $p = (p_1,p_2)$, $p_i
\in \cO _i$, $i=1,2$, $G_\k \in \cD '(\cO )$, $G_{i\k} \in \cD
'(\cO _i)$, $i=1,2$. Then
\be[e2/1.17]
     \As\. (G_{1\k}\times G_{2\k})
     = (\As\. G_{1\k})\times (\As\. G_{2\k}) .
\ee

$(iii)$ Under the conditions of \ssect{As-exp.LinearMaps}, assume that
the mapping $A$ depends on $\k$ too.  For instance, let $G(p,\k) \in
\cD '(P)$ and $A_\k$ be the translation $G(p,\k) \rightarrow G(p +
\k \times p^\0_{0},\k)$.  Then:
\be[e2/1.18]
     \As\. A_\k 
     = \sum^\infty_{n=0}
          {\k^n\over n!}
          \left(p^\0_{0} {\partial \over \partial p}\right)^n
          \times \As .
\ee
This is easily verified by reducing the problem to the example $(i)$
by replacing $p \rightarrow p - \k \, p^\0_{0}$. One can consider a
diffeomorphism of the class $\cC^\infty$ parametrically depending on
$\k $ instead of the translation; in this case the coordinate
functions defining the diffeomorphism should possess {\it
As}-expansions in $\k $.

\SUBSECTION{Transitions from local to global \asex s.}
\label{As-exp.Globalization}%1.10

Note also the following rule of ``gluing" \asex s which
follows from their uniqueness property (\ssect{As-exp.Uniqueness}):
given \asex s for restrictions of a distribution $G_\k$
onto $\cO _1$ and $\cO _2$, then they coincide within $\cO _1
\cap \cO _2$ and there exists the \asex\ for $G_\k$
on $\cO _1 \cup \cO _2$ which is a termwise ``gluing" of the two
expansions.

The properties of the \asex s described in this section
are sufficient for a rather detailed study of general structure of
the \asex s for products of singular functions depending
on $\k $.  This is the subject of the following section.

\SECTION{\asop\ on graphs and subgraphs.}
\label{As-operation}%2

Here we will use the results of \sect{As-expansion} to study the
general recursive structure of the operation $\As$ on the
$\k$-dependent graphs from the universum of 
\sects{Graphs}-{CompleteSubgraphs}.  The main purpose is to connect
the \asex\ of a graph with those of its subgraphs.  

In \ssect{As-op.Formal} the formal expansion for a graph is
constructed and the problem of construction of the expansion for the
graph in the sense of the distribution theory (\asop) is
formulated.  In \ssect{As-op.Locality} the locality condition is
derived and its implications are studied.  Finally, in
\ssect{As-op.Recursive} the definition of the \asop\ on graphs is
discussed.

\SUBSECTION{Formal expansions on graphs.}
\label{As-op.Formal}%2.1

Within the framework of the universum of graphs $\cG$
(\sects{Graphs}-{CompleteSubgraphs}) we will assume that for any $g
\in \cG$ the functions $F_g$ depend on the parameter $\k $ in
such a way that for any $\k,\ F_g(p,\k) \in \cC^\infty(P_g)$ and
there exists the \asex\ of the form \eq{e2/1.2},
\eq{e2/1.5}:
\be[e2/2.1]
     F_g(p,\k)
     \mathrel{\mathop{\0\simeq}
     ^{\cC^\infty(P_g\backslash \{0\})}_{\k \rightarrow 0}}
     \T_\k\. F_g(p,\k) .
\ee
It is assumed (and this is the essence of the problem) that individual
terms on the r.h.s.\ are singular at $p \rightarrow 0$ (cf.\ example
($i$), \ssect{Graphs.Example}). The explicit form of the expansion
\eq{e2/2.1} will not be needed in this section.

In order to avoid unnecessary complications, the mappings $l_{g(G)}$
will be assumed to be independent of $\k$.  This does not affect the
general structure of the \asop\ and this restriction will be removed
in subsequent sections.

Now take a graph $G \subset \cG$ and $g \in G$. From \eq{e2/2.1},
taking into account the results of \ssect{As-exp.LinearMaps} and
\ssect{As-exp.DirectProducts}, one concludes that there exists the {\it
As}-expansion:
\be[e2/2.2]
     g_{(G)}(p,\k)
     \mathrel{\mathop{\0\simeq\0}
     ^{\cC^\infty(P_G\backslash \{\pi_g\})}_{\mathstrut\k \rightarrow 0}}
     \qquad \T_\k\. g_{(G)}(p,\k) ,
\ee
where $\pi_g$ is the singular plane defined in the usual way
(\ssect{Complete.Planes}). Let us emphasize that in the context of the
expansion problem the singularity of the factor $g$ in the sense of
\ssect{1/1.1} is to be understood as the singularity of the {\it
expansion} \eq{e2/2.2}.  It is important to distinguish such
singularities from those which the factors may in general possess at
$\k \not= 0$ i.e.\ {\it before} the expansion.

For any graph $G$ one obtains the \asex\ of the form:
\be[e2/2.3]
     G(p,\k) \mathrel{\mathop{\0\simeq\0}
     ^{\cC^\infty(\scriptcrcaccent{P}_G)}
     _{\k \rightarrow 0}} \T_\k\. G(p,\k)
     =  \prod_{g\in G} \T_\k\. g_{(G)}(p,\k),
\ee
where $\crcaccent{P}_G \bydef P_G \backslash \cup_{g\in G}
\pi_g$.

It is convenient to call the expansion \eq{e2/2.3} {\it formal} in
order to distinguish it from the expansions in the sense of the
distribution theory which are well-defined for any $p$ and will be
constructed later.

The structure of the singularities of the formal expansion \eq{e2/2.3}
is fully described using the notions of $s$-subgraphs, their singular
planes etc.\ introduced in \sect{CompleteSubgraphs}.  All those
notions can and will be freely used here without any changes.

At $\k \not= 0$ the graph $G(p,\k)$ defines a distribution from $\cD
'(P_G)$.  Consider the problem of constructing the $As$-expansion for
$G(p,\k)$ in the sense of distribution theory:
\be[e2/2.4]
     G(p,\k) 
     \mathrel{\mathop{\0\simeq\0}^{\cD '(P_G)}_{\k \rightarrow 0}} 
     \As\. G(p,\k) = \ ?
\ee
Let us study the conditions which the \asop\ on graphs should satisfy.

\SUBSECTION{The locality  condition.}
\label{As-op.Locality}%2/2.3

Take an arbitrary region $\cO \subset P_G$ and a subgraph $\G \in
S[G]$ such that the singularities of the expansion $\T_\k\.
\GG $ do not intersect $\cO $.  Then from \eq{e2/2.2} using
the results of \ssect{As-exp.Globalization} one obtains the formal
expansion:
\be[e2/2.5]
     \GG
     \mathrel{\mathop{\0\simeq\0}
     ^{\cC^\infty(\cO )}
     _{\k \rightarrow 0}}
     \T_\k\. \GG
     = \prod_{g\in \GG} \T_\k\. g .
\ee
Consider now the graph $G$ as a distribution in $\cO $. It can be
factorized as:
\be[e2/2.6]
     G|^\0_\cO
     =  (\GG)|^\0_\cO
        \, \G|^\0_\cO .
\ee
In virtue of \eq{e2/2.5} and \ssect{As-exp.LinearMapsInverse} it is
clear that if one wishes the \asex\ to exist for the
l.h.s.\ of \eq{e2/2.6} in the sense of the distribution theory then
the \asex\ should exist also for $\G$.  (Note that it
should be possible to evaluate $\As\. \G $ within its proper space
$P_\G$---cf.\ \ssect{As-exp.LinearMaps} and
\ssect{Subgraphs.Example}. Making use of the uniqueness property of the
\asex s \ssect{As-exp.Uniqueness} and the example $(vii)$
from \ssect{As-exp.LinearMaps}, one obtains the locality condition%
\footnote{
cf.\ the locality condition 
for \rop\ in \ssect{R-op.Axioms}.
\relax}:
\be[e2/2.7]
     \As\. G|^\0_\cO
     = (\T_\k\. \GG)
        \times (\As\. \G)|^\0_\cO .
\ee
For $\G = \emptyset $ it is convenient to define
\be[e2/2.8]
     \As\. \emptyset  \bydef  1
\ee
(cf.\ \ssect{R-op.Recursive}).  Then for $\G =
\emptyset $ one can take $\cO = \crcaccent{P}_G$ and the r.h.s.\ of
\eq{e2/2.7} is transformed into the r.h.s.\  of the formal  expansion
\eq{e2/2.3}  whereas the equality resulting from \eq{e2/2.7} becomes the
``initial  condition"  for  the explicit construction of the operation
$\As$.

Let us now study the implications of the locality condition
\eq{e2/2.7} (cf.\ the reasoning concerning the structure of the
$R$-operation in \ssect{R-op.Structure}). Take an arbitrary graph $G$.
Using a decomposition of unit $\{\eta^\0_\G\}$ isolating
the singularities of maximal $s$-subgraphs $\G \in S_{\rm max}[G]$ 
(\ssect{Decomp.Maximal}), reasoning in analogy to the preceding
subsection and taking into account the results of
\sect{As-expansion}, one obtains:
\be[e2/2.9]
     \As\. G =
     \sum_{\G \in S_{\rm max}[G]}
     \eta^{\0}_\G\,
     (\T_\k\. \GG)\, (\As\. \G) .
\ee

Now let $G$ be an $s$-graph (\ssect{Complete.Ordering}). Denote
\be[e2/2.10]
     \As'\. G \bydef  \As\. (G|^{\0}_{P_G\backslash \{0\}}) .
\ee
Analogously, using a decomposition of unit $\{\theta_\G\}$ from
\ssect{Decomp.Submaximal} which isolates the singularities of
submaximal subgraphs $\G \triangleleft G$, one obtains:
\be[e2/2.11]
     \As'\. G 
     = \sum_{\G \triangleleft G} \theta_\G
       \, (\T_\k\. \GG)
       \, (\As\. \G) .
\ee
Finally, using the results of \ssect{As-exp.LinearMapsInverse} for a
factorizable graph $G = G_1 \times G_2$ one obtains:
\be[e2/2.12]
     \As\. G = \As\. G_1 \times  \As\. G_2 .
\ee

\SUBSECTION{Recursive definition of the \asop .}
\label{As-op.Recursive}%2/2.4

Let us define the \asop\ on the graphs of the corresponding types by
the equations \eq{e2/2.8}, \eq{e2/2.9}, \eq{e2/2.11} and \eq{e2/2.12}.

This definition is recursive (cf.\ the definition of the $R$-operation
in \ssect{R-op.Recursive} and the discussion in
\ssect{R-op.Structure}): before constructing the \asop\ on a graph $G$
it is necessary to construct the \asop\ on all subgraphs $\g < G$.  Let
us emphasize that this part of the definition of the \asop\ is
completely independent of the character of the singularities of the
expansions \eq{e2/2.2} of individual factors at $p \rightarrow 0$.
Note also that the present case is even easier than the case of the
$R$-operation in that the uniqueness of the \asex s
(\ssect{As-exp.Uniqueness}) ensures independence of the \asop\ of the
choice of the decompositions $\{\eta^\0_\G\},
\{\theta_\G\}$ etc.

However, the definition is yet incomplete: if $G$ is a
non-factorizable $s$-graph and the \asop\ is constructed for all
$\g< G$ then eq.\eq{e2/2.11} defines the expansion only on
$P_G\backslash \{0\}$:
\be[e2/2.13]
     G_\k 
     \mathrel{\mathop{\0\simeq\0}
     ^{\cD '(P_G\backslash \{0\})}
     _{\k \rightarrow 0}}
     \As'\. G_\k .
\ee
In other words, the operation $\As'$ allows one to expand the
expression $\langle G_\k, \vf \rangle$ only for test functions
vanishing in some neighbourhood of the point $p = 0$. But we wish to
construct the \asex\ of the form \eq{e2/2.4} valid on {\it
all} $\vf \in \cD (P_G)$ and the means described in
\sect{As-expansion} are insufficient for this purpose.

The next section is devoted to the solution of this problem.

\SECTION{Extension principle and transition from $\protect\As'$
to $\protect\As$.}
\label{ExtensionPrinciple}%3

The problem of construction of the expansion $\As\. G$ \eq{e2/2.4}
from $\As'\. G$ \eq{e2/2.13} for an $s$-graph $G$ is a special case of
a very general problem of extension of an approximating functional
from a subspace onto the entire space so that the approximation
property be preserved (cf.\ \ssect{As-exp.LinearMapsInverse}).
Therefore, it is interesting to discuss it in as general terms as
possible.  Such a discussion is presented in
\ssects{Extension.Essence}-{Extension.Series}.  In particular in
\ssect{Extension.Essence} the extension principle is proved. The
extension principle is an abstract construction related in spirit to
the one used in the Hahn-Banach theorem on extension of functionals
bounded by seminorms.  In \ssect{Extension.Practical} this
construction is transformed into a form convenient for applications
and in \ssect{Extension.Series} a recipe for construction of infinite
asymptotic expansions is presented.

In \ssect{Extension.Example} the scheme of \ssect{Extension.Series} is
used to construct the \asex\ for a simple case of an
$s$-graph which possesses no non-trivial $s$-subgraphs.  However, the
reasoning and notations are chosen so as to allow immediate transition
to the general case without any change in formulae.  Lastly, in
\ssect{Extension.As&R} we outline the plan of justifying the formulae
derived in \ssect{Extension.Example} for the case of a general
$s$-graph $G$.

\SUBSECTION{The extension principle \protect\cite{fvt82}.}
\label{Extension.Essence}%3.1

Let $L$ be a linear space and $L^{\ast}$ be the space of linear
functionals on $L$ with the topology of pointwise convergence on any
vector $v \in L$. Assume that a functional $G_\k \in L^{\ast}$
depending on $\k $ is given and it is required to construct $G_{\k,N}
\in L^{\ast}$ approximating $G_\k$ on $L$ to $o(\k^N)$. More
precisely, for any $v \in L$ the following should hold:
\be[e2/3.1]
     \langle (G_\k - G_{\k,N}), v \rangle = o(\k^N) .
\ee
Assume also that a functional $G'_{\k,N}$ is given which approximates
$G_\k$ to $o(\k^N)$ on a subspace $L_N \subset L$. Then there
exists $G_{\k,N} \subset L^{\ast}$ coinciding with $G'_{\k,N}$ on
$L_N$, and it is unique up to $o(\k^N)$.

The construction of $G_{\k,N}$ is in fact elementary. Let $L^{\rm
T}_N$ be an algebraic complement to $L_N$ in $L$ so that any $v
\in L$ is uniquely decomposed into a sum of the form $v = v^\0_N +
v^{\rm T}_N$ where $v^\0_N \in L_N$ and $v^{\rm T}_N \in L^{\rm
T}_N$. Define
\be[e2/3.2]
     \langle G_{\k,N}, v \rangle
     = \langle G'_{\k,N}, v^\0_N \rangle
     + \langle G_\k, v^{\rm T}_N \rangle .
\ee
Now it is easy to verify \eq{e2/3.1} while the uniqueness within
$o(\k^N)$ follows from the fact that if $G^{(2)}_{\k,N}$ is another
functional possessing the same properties as $G_{\k,N}$ then
\be
     G_{\k,N} - G^{(2)}_{\k,N}
     = 
     (G_\k - G^{(2)}_{\k,N}) - (G_\k - G_{\k,N})
     \equals^{L} o(\k^N).
\ee
If one deals with topological spaces and continuous functionals then
to ensure continuity of $G_{\k,N}$ it is sufficient to require that
the complement $L^{\rm T}_N$ should be topological, i.e.\ that the
projections $v \rightarrow v^\0_N$ and $v \rightarrow v^{\rm T}_N$
should be continuous. If $L_N$ is closed while $\dim L^{\rm T}_N < +
\infty $, then this condition is automatically satisfied (see
sect.1.9.7 in \cite{edwards}). It is just the case we will mainly
need.

\SUBSECTION{Practical form of the extension principle.}
\label{Extension.Practical}%3.2

Now we can present the following recipe of constructing $G_{\k,N}$.
Let $\r\. G'_{\k,N}$ be an arbitrary extension of $G'_{\k,N}$ onto
$L$. Fix a basis $v^\a$ in $L^{\rm T}_N$ and the dual set of
functionals $\d^\a$ equal to zero on $L_N$:
\be[e2/3.3]
     \langle \d^\a, v^{\b} \rangle
     =  \d_{\a \b} \qquad (\hbox{the Kronecker symbol}).
\ee
Then $G_{\k,N}$ can differ from $\r\. G'_{\k,N}$ only by a linear
combination of $\d^\a$:
\be[e2/3.4]
     G_{\k,N}
     = \r\. G'_{\k,N}
     + \sum_\a C_{N,\a} \, \d^\a 
     + o(\k^N) ,
\ee
where $C_{N,\a}$ are numeric coefficients which may depend on
$\k$. Since the desired $G_{\k,N}$ exists and is unique, it is
possible to obtain $C_{N,\a}$ by evaluating both sides of
\eq{e2/3.4} on $v^\a$. Thus one obtains the following {\it
consistency conditions}:
\be[e2/3.5]
     C_{N,\a}
     = \langle (G_\k - \r\. G'_{\k,N}), v^\a \rangle
     + o(\k^N) .
\ee

 {\bf Remark.} We stress that $C_{N,\a}$ is defined only up to
the indicated accuracy. The recipe \eq{e2/3.4}, \eq{e2/3.5} is
convenient in that it allows one to choose the most natural $\r$ and
$v^\a$ independently.  One may even consider sequences
$v_\L^\a$, $\L \rightarrow \infty $ which do not
necessarily converge in $L$, it is only required that in the limit
$\L \rightarrow \infty $ the coefficients $C_\a$ converged
and changed only by finite quantities of order $o(\k^N)$.

\SUBSECTION{Expansions in the form of infinite series.}
\label{Extension.Series}%3.3

Let us now discuss how expansions in the form of a series can be
obtained. Consider a sequence of subspaces $L \supset L_1 \supset
\ldots \supset L_{N-1} \supset L_N$ and their complements $\emptyset
\subset \ldots \subset L^{\rm T}_{N-1} \subset L^{\rm T}_N$.  For
any $N$ the functional $G'_{\k,N}$ satisfying the conditions of
\ssect{Extension.Essence} with the
corresponding $L_N$ can be represented as:
\be[e2/3.6]
     G'_{\k,N} =\sum_{N_{0}\leq n\leq N} f_{\k,n} .
\ee
In fact we have assumed that $G'_{\k,N}$ is defined only on $L_N$,
so that $f_{\k,n}$ are defined only on $L_n \supset L_{n+1} \supset
\ldots $\@ Then to represent $G_{\k,N}$ also as a sum similar to
\eq{e2/3.6} one can use---instead of \eq{e2/3.4}---the following
construction which is equivalent to the above for each $N$.

Choose the basis $v^\a$ so that $v^\a \in L^{\rm
T}_{|\a |} \cap L^{\rm T}_{|\a |-1}$ where $|\a |$ is an
integer function of $\a$. (The condition \eq{e2/3.3} is assumed to
be satisfied.) It is easy to see that $G_{\k,N}$ can be represented
as:
\be[e2/3.7]
     G_{\k,N}
     =   G_{\k,N-1} + \r\. f_{\k,N} +
         \sum_{|\a |\leq N}C_{N,\a}\, \d^\a .
\ee
This representation should be compared with the one that follows from
\eq{e2/3.6}:
\be[e2/3.8]
     G'_{\k,N} = G'_{\k,N-1} + f_{\k,N} .
\ee
Note that for any $N$ the operator $\r$ extends $f_{\k,N}$ from
$L_N$ onto the the entire $L$.

Now eq.\eq{e2/3.5} should be replaced by
\be[e2/3.9]
     C_{N,\a}
     = \langle (G_\k - G_{\k,N-1}
     - \r\. f_{\k,N}), v^\a \rangle
     + o(\k^N) .
\ee
As a result one obtains the expansion
\be[e2/3.10]
     G_\k
     \ \mathrel{\mathop{\0\simeq\0}^{L^{\ast}}_{\k \rightarrow 0}}
     \ \sum_n \Big\{ 
           \r\. f_{\k,n}
         + \sum_{|\a |\leq n} C_{n,\a}\, \d^\a 
     \Big\},
\ee
which has the form of the \asex\ \eq{e2/1.2}.

The above construction, providing a general recipe for how the problem
of the asymptotic expansion should be approached, does not guarantee,
however, that the dependence on $\k $ of the expression in braces in
\eq{e2/3.10} is of the power-and-log type of \eq{e2/1.5}. To study the
exact form of the dependence on $\k $ one needs to make further
assumptions on the structure of $G_\k$.

\SUBSECTION{Transition from $\protect\As'$ to $\protect\As$: an example.}
\label{Extension.Example}%3.4

We wish to apply the results of
\ssects{Extension.Essence}-{Extension.Series} to the problem which
remained open in \sect{As-operation} (see also \ssect{Extension.As&R}).

Consider a non-factorizable $s$-graph $G$. In the framework of
\ssects{Extension.Essence}-{Extension.Series},
let $L = \cD (P_G)$, $L^{\ast} = \cD '(P_G)$. $G(p,\k)$ will play
the role of $G_\k$, and the expansion $\As'_N\. G(p,\k)$
satisfying \eq{e2/2.13} will play the role of $G'_{\k,N}$ which
appears in \ssect{Extension.Series}.  Here $f_{\k,n} = \as'_n\.
G_\k$.

Here we consider the special case when $G$ does not possess any
$s$-subgraphs.  For example, $G(p,\k) = (|p| + \k)^{-d_G}$, then
$\As' = \T_\k$ (the Taylor expansion in $\k $---cf.\
\ssect{As-exp.Notations})
and $\as'_n\. G(p,\k) \propto  \k^n/|p|^{d_G+n}$.
In this  case \eq{e2/2.2} is obviously satisfied. However, we will preserve
general notations  so  that the general scheme of reasoning and basic
formulae  be  applicable  without changes in the most general case.

One should start with determining the maximal subspace $L_N \subset
\cD (P_G)$ on which
\be[e2/3.11]
     G \mathrel{\mathop{\0\simeq\0}^{L_N}} 
     \As'_N\. G + o(\k^N) .
\ee
(It is obvious that $L_N \supset \cD (P_G\backslash \{0\})$.)  The
divergence index at zero for $\As'_N\. G$ (cf.\
\ssect{Subtraction.Problem}) is determined by the highest term
(proportional to $\k^N$) and is equal to $\o^\0_G + N$
where $\o^\0_G = d_G - \dim P_G$ and $d_G$ is the
homogeneity index of $G(p,\k)$ under the simultaneous scaling of $p$
and $\k $.  Therefore $\As'_N\. G$ allows a natural extension onto
the subspace $\cD _{\o^\0_G+N+1}(P_G) \subset \cD
(P_G)$ which consists of test functions $\vf (p)$ possessing
zero of order $\o^\0_G + N + 1$ at $p = 0$ (cf.\
\ssect{Subtraction.Problem}).  It is natural to expect that
\be[e2/3.12]
     L_N = \cD _{\o^\0_G+N+1}(P_G) .
\ee
In the example under consideration it is easy to see that this is
indeed the case.  In fact one should verify that for $\vf \in
L_N$, $\int dp \, \vf (p)\, [\D'_N\.G(p,\k)] = o(\k^N)$.  To
do this let us define a neighbourhood of $p = 0$: $\cO _\k = \{p
\in P_G \mid |p| \leq \const\, \k \}$.  The desired estimate for the
integral over $\cO _\k$ is easily obtained after replacement $p
\rightarrow \k \, p$, while the one for the integral over $P_G
\backslash \cO _\k$ is obtained by estimating the expression in
square brackets by an expression proportional to
$\k^{N+1}/|p|^{d_G+N+1}$.

Now the recipe of \ssect{Extension.Series} can be realized in the
following way: $\r$ should perform an extension of every $f_{\k,n}$
from $\cD _{\o^\0_G+N+1}(P_G)$ onto $\cD (P_G)$
i.e.\ it should act without oversubtractions as defined in
\ssect{Subtraction.r}.  As $v^\a$, the vectors $v_\L^{\a
} \bydef \Phi^\L\, \cP^\a$ can be taken where $\Phi^\L(p)$
is the cutoff function introduced in \ssect{Decomp.Radial}, ${\cal
P}^\a(p)$ are the homogeneous polynomials introduced in
\ssect{Isolated.Deltas}, and $\d^\a$ is the corresponding dual
set of $\d $-functions; note that in this case $|\a|$ is just the
power of the polynomial $\cP^\a$, which agrees with
\ssect{Extension.Series}.  It would not be difficult to write down the
analogues of \eq{e2/3.9}, \eq{e2/3.10} but it turns out that the form
of the expansion simplifies drastically (with far-reaching
implications as regards the expansions of Feynman diagrams) with a
special---although very natural---choice of the subtraction operator
$\r$.

Special subtraction operators $\tr$ were defined in
\sect{Subtractions} (see \ssect{Subtraction.r-tilde}).  The operator
$\tr$ is defined for any $f(p) \in \cD '(P\backslash \{0\})$
which are homogeneous or ``almost-homogeneous" i.e.\ differing from
homogeneous by logarithmic corrections under the scaling $p
\rightarrow \l \, p$ (see below).  For the example under consideration
it is convenient to have in view the homogeneous function $f(p) =
|p|^{-d_{f}}$. The characteristic property of $\tr$ is that
although it does violate the homogeneity properties of the original
functional $f(p)$, which is unavoidable, but in a minimal degree,
i.e.\ the leading power behaviour is not changed but only logarithmic
corrections are introduced.  For instance, for the function $f(p) =
|p|^{-d_{f}}$:
\be[e2/3.14]
     \tr\. f(\l \, p)
     =  \l^{-d_{f}}
        \, \Big[ \tr\. f(p)
      + \ln  \l \,
         \sum_{|\a |=\o_{f}}
             C_\a\, \d^\a \Big],
\ee
where $C_\a$ are constants. Distributions which behave like
\eq{e2/3.14}  under  the scaling of $p$, perhaps with a larger number of
logarithmic corrections, can be conveniently referred to as {\it
almost-homogeneous}. (Note that the power-and-log dependence on $\k$
described in \ssect{As-exp.Definition} is a special case of
almost-homogeneous dependence.) 

Another important property of $\tr$ which is closely related to
\eq{e2/3.14}  is that $\tr\. f$ vanishes on polynomials
$\cP^\a$ of order $|\a | < \o_{f}$. More precisely,
\be[e2/3.15]
     \lim_{\L \rightarrow \infty}{\ }
     \langle \tr\. f, \Phi^\L\, \cP^\a \rangle
     = 0, \qquad\hbox{if}\quad|\a | < \o_{f}
\ee
(cf.\  \ssect{Subtraction.Minimality}).

Let us complete our calculation using $\tr$. With the choice of
$v^\L_\a$ as in \ssect{Extension.Example}, one can take the
limit $\L \rightarrow \infty$ in the expressions for $C_{n,\a
}(\k)$ (cf.\ the remark in \ssect{Extension.Practical}).  Moreover, in
virtue of \eq{e2/3.15}, $C_{n,\a} \not= 0$ only for $|\a | =
\o^\0_G + n$ in the final formulae (so that it is
sufficient to write $C_\a$ instead of $C_{n,\a})$. As a
result one obtains the following concretization of the formulae
\eq{e2/3.10} and \eq{e2/3.9}:
\be[e2/3.16]
     G_\k
     \mathrel{\mathop{\0\simeq\0}
     ^{\cD '(P_G)}
     _{\k \rightarrow 0}} \As\. G_\k
     = \tr\. \As'\. G_\k
     + {\bf \tilde{E}}_\k\. G_\k
     = \sum_n
       \Big[ \tr\. \as'_n\. G_\k
           + \sum_{|\a |=\o_n} C_{G,\a}(\k)\,
           \d^\a 
       \Big],
\ee
\be[e2/3.17]
     C_{G,\a}(\k) 
     =
     \lim_{\L \rightarrow \infty}
     \langle
        ( G_\k
        - \As'_{N-1}\. G_\k
        - \tr\. \as'_N\. G_\k  ),
        \Phi^\L\, \cP^\a
     \rangle ,
\ee
\be[e2/3.18]
     \o_n \bydef  \o^\0_G + n,
     \qquad \o^\0_G \bydef  d_G - \dim  P_G .
\ee
It is easy to check that $C_{G,\a}(\k)$ do not depend on the
choice of $\Phi^\L$.  Moreover, replacing $p \rightarrow \k \, p$ in
the integral on the r.h.s.\ of \eq{e2/3.17} and taking into account
\eq{e2/3.14}, for the example under consideration one obtains that
$C_{G,\a}(\k)$ have a power-and-log dependence on $\k$ (cf.\
\eq{e2/1.5}):
\be[e2/3.19]
     C_{G,\a}(\k)
     = \k^n\, (\const'_\a
     + \const''_\a\, \ln  \k)
\ee
(the relation between $\a $ and $n$ is established by the equation
$|\a | = \o_n$ and \eq{e2/3.18}). Thus, the expansion
\eq{e2/3.16} is indeed the \asex\ in the sense of
\ssect{As-exp.Definition}.

To complete the discussion of the example we would like to emphasize
the systematic use in our calculation of (almost-) homogeneity
properties of $G(p,\k)$: in proving the equality \eq{e2/3.12} in
\ssect{Extension.Example}; in constructing $\tr$;
in establishing the fact of power-and-log dependence of $C_{G,\a
}(\k)$ on $\k$ \eq{e2/3.19}.  Note that the homogeneity in $p$ of the
individual terms in $\As'\. G(p,\k)$ is closely related to the
homogeneity of $G(p,\k)$ with respect to the pair of arguments $p$ and
$\k$.

{\it A posteriori} there should be no wonder that the power-and-log
dependence on $\k$ in the \asex\ stems from some
almost-homogeneity properties of the original functions.

\SUBSECTION{From $\protect\As'\.G$ to $\protect\As\.G$.
$\protect\As$ and $\tilde{\protect\R}$.}
\label{Extension.As&R}%3.8

Let us now discuss the transition from $\As'\. G$ to $\As\. G$ for a
general non-factorizable $s$-graph $G$. All our reasoning implies that
the equations \eq{e2/3.16}, \eq{e2/3.17} obtained for $G$ without
non-trivial (non-empty) $s$-subgraphs, are preserved in the general
case as well.  The plan of justifying these formulae is as follows:

 $(i)$ Comparing \eq{e2/3.16} with \eqs{e2/4.8}-{e2/4.10} one sees
that the \asop\ can be considered as a composition of $\As'$ and the
special subtraction operator $\tr$ in which the finite
arbitrariness is fixed uniquely via consistency conditions:
\be[e2/3.20]
     \As = \tr^{(\k)}\. \As' .
\ee
If this remains true in the general case then due to the complete
similarity of structures of the \asop\ and of the $R$-operation (cf.\
\ssect{As-op.Locality} and \ssect{R-op.Recursive}) the \asop\ can be
treated as a composition of the formal expansion \eq{e2/2.3} and the
special $R$-operation $\tR^{(\k)}$ with specially chosen finite
counterterms:
\be[e2/3.21]
     \As = \tR^{(\k)}\. \T_\k .
\ee
The only point that should be mentioned here is that one has to assume
certain homogeneity properties for the individual factors in order to
be able to repeat the reasoning similar to that in the above example.

 $(ii)$ It should be proved that the equations \eq{e2/3.11},
\eq{e2/3.12} (see also their discussion) are true in the general case.

The combination of the results $(i)$ and $(ii)$ ensures automatic
validity of the reasoning of \ssect{Extension.Example} and of
the equations \eq{e2/3.16} and \eq{e2/3.17} in the case of a general
$s$-graph $G$. This will complete the construction of the \asop.

\newpage\thispagestyle{myheadings}\markright{}
\PART{Construction and Properties of the \asop.}

\SECTION{Object of Expansion.}
\label{ExpansionObject}%7

In this section we specify the class of products of singular functions
(graphs, in terms of \ssect{Graphs.Graphs}) depending on a parameter
for which we will construct the asymptotic expansions of the form
described in \sect{As-expansion}.

In \ssect{Object.Factors} the conditions are imposed on the elementary
factors $F_g$, which allow one to obtain asymptotic expansions in the
sense of the distribution theory for their products and, on the other
hand, ensure that the expansions run in powers and logarithms of the
expansion parameter. In particular, Euclidean propagators of the
standard perturbation theory in momentum representation satisfy these
conditions if their masses are proportional to the expansion
parameter.  In \ssect{Object.KappaDependence} the $\k$-dependence of
products of such functions (graphs) is considered.  In
\ssect{Object.Notations} singular planes and $s$-subgraphs are
introduced for the purposes of description of the singularities of the
formal expansion of the graph.  In \ssect{Object.Properties} an
auxiliary estimate is obtained for the functional defined by the
graphs of the described type on test functions confined within a
region of radius of order of the expansion parameter.

\SUBSECTION{Properties of factors.}
\label{Object.Factors}%7.1

Consider a class of singular functions $F_g$ depending on the
coordinate $p \in P_g$ and the parameter $\k > 0$. We assume that
all $F_g$ satisfy the following conditions:

 $(i)$ {\it Smoothness} in $p$ at $p
\not= 0$ for any $\k > 0$.

 $(ii)$ {\it Almost-homogeneity with respect to the pair of arguments},
i.e.{\ }for any $\l > 0$:
\be[e2/7.1]
    F_g(\l p, \l \k)
    =  \l^{-d_g}
    \, \sum^{L_g}_{k=0}
       F^{(k)}_g(p,\k)\, \ln^k \l  ,
\ee
where the numbers $d_g$ will be called {\it indices of
homogeneity}.

 $(iii)$ $F_g$ are bounded by the inequalities (cf.\ \eq{e1/10.2}
and \eq{e2/5.1}):
\be[e2/7.2]
     |D^\a F_g(p,\k)|
     \leq
     \left( {\k \over |p|} \right)^{N_g}
     \,
     {\L (\k)\, \L (|p|)  \over |p|^{d_g+|\a |}},
\ee
for any $p \not= 0$, $\k > 0$.  In \eq{e2/7.2} and everywhere below
all the $\L $-functions are double-sided (see the definition in
\ssect{Isolated.Lambda}).

 $(iv)$ $F_g$ are expanded at $\k \rightarrow 0$ as:
\be[e2/7.3]
     F_g(p,\k)
     = \sum^N_{n=N_g}
       F_{g,n}(p,\k) + o(\k^N) , \qquad p\not=0
\ee
and we assume that the functions $F_{g,n}$ for all $n$ satisfy the
following conditions:

 $(v)$ {\it Smoothness} with respect to $p$ at $p \not= 0$, $\k > 0$
(cf.\ $(i)$).

 $(vi)$ {\it Almost-homogeneity} with respect to both $p$ and $\k $
with the same index $d_g$ as for $F_g$ (see $(ii)$).

 $(vii)$ The following estimate (cf.\ \eq{e2/7.2}) is satisfied for
$F_{g,n}(p,\k)$:
\be[e2/7.4]
     |D^\a F_{g,n}(p,\k)|
     \leq
     \left({\k \over |p|}\right)^n
     \, {\L (\k)\, \L (|p|)
         \over
         |p|^{d_g+|\a |}} .
\ee

 $(viii)$ The remainder term of the series \eq{e2/7.3} is bounded by
an expression analogous to the r.h.s.\ of \eq{e2/7.4} for the first
discarded term:
\be[e2/7.5]
     \Big| D^\a
        \Big[ F_g(p,\k)
          - \sum_{n\leq N} F_{g,n}(p,\k)
        \Big]
     \Big|
     \leq
     \left({\k \over |p|}\right)^{N+1}
     \, {\L (\k) \, \L (|p|)
         \over
         |p|^{d_g+|\a |}},
\ee
for all $p$ outside the region
\be[e2/7.6]
     \cO ^{\k}_g
     \bydef
     \{ p \in  P_g \mid |p| \leq  \k \, \const \} .
\ee

Under the above conditions the series \eq{e2/7.3} approximates
$F_{g,\k}$ in the sense of $\cC^\infty(P_g\backslash \{0\})$ which
in the notations of \ssect{As-exp.Definition} can be written as:
\be[e2/7.7]
     F_{g,\k}
     \mathrel{\mathop{{\vphantom{0_0}}\simeq}
     ^{\cC^\infty(P_g\backslash \{0\})}
     _{\k \rightarrow 0}}
     \T_\k\.F_{g,\k}
     = 
     \sum_n \t_n\.F_{g,\k};
     \qquad {\bf t}_n\.F_g \bydef  F_{g,n} .
\ee
Note that \eqs{e2/7.3}-{e2/7.5} contain more information than
\eq{e2/7.7}.

In general the expansion \eq{e2/7.3} is not unique: indeed, one could
take $F_g$ itself as the first term of its expansion and set all the
other terms to zero, the conditions $(v)$--$(viii)$ are automatically
satisfied due to $(i)$--$(iii)$.  So let us introduce the following
requirement:

 $(ix)$
All $F_{g,n}$ have a {\it power-and-log dependence} on $\k$, i.e.
\be[e2/7.8]
     F_{g,n}(p,\k)
     = \k^n
     \,
     \sum^{M_g}_{k=0} \ln^k \k \, F^{(k)}_{g,n}(p)
\ee
(cf.\ \eq{e2/1.5}). Then $F_{g,n}(p,\k)$ are almost-homogeneous with
respect to {\it each} of their arguments. In simplest applications
(e.g.\ when $F_g$ are Feynman propagators in momentum representation
while $\k$ is proportional to masses) it is assumed that
$\t_n\.F_g$ are the terms of the Taylor expansion in $\k $ for the
function $F_g$.  In that case $M_g = 0$ in \eq{e2/7.8}.

The functions $F_g(p,\k)$ can be either singular at $p = 0$ or
regular. Let us impose one more condition on singular factors:

 $(x)$ If $F_g(p,\k) \notin \cC^\infty(P_g)$ then $F_g$ is
almost-homogeneous with respect to $p$ (hence it is such with respect
to $\k$ as well). In this case $F_g$ coincides with the first (and
the only) term of its expansion. Note that this condition is not a
serious restriction since a wide class of singular functions can be
represented as a factor satisfying $(x)$ multiplied by a regular one.

To simplify formulae we assume that all the numbers $d_g$ and
$N_g$ are integer. This restriction is obviously satisfied in
problems of standard perturbative QFT, but it is by no means
crucial for applicability of our techniques, which can be used with
appropriate modifications beyond the standard
perturbative framework.

\SUBSECTION{Dependence of graphs on the expansion parameter.}
\label{Object.KappaDependence}%7.2

Define the graph $G_\k$ similarly to \eq{1/1.7}:
\be[e2/7.9]
     G(p,\k) \bydef  \prod_{g\in G} g(p,\k),
\kern7.4mm\ee
where
\be[e2/7.10]
     g^\0_{(G)}(p,\k) \bydef  F_g(l_{g(G)}(p,\k), \k),
\ee
and $l_{g(G)}$: $P_G \rightarrow P_g$ are affine mappings.  We
assume that $l_g$ have the following structure (cf.
\ssect{Graphs.MoreExample}):
\be[e2/7.11]
     l_{g(G)}(p,\k)
     \bydef  l'_{g(G)}(p) + q_g\, \k,
\kern6mm\ee
where $l'_g$ are linear mappings $P_G \rightarrow P_g$ (cf.\
\ssect{Graphs.Graphs}) while $q_g \in P_g$ are constant vectors.
As a result each $g(p,\k)$ represents the corresponding function
$F_g$ with the argument translated by the vector $\k \, q_g$ (cf.\
\ssect{As-exp.DirectProducts}).

We could add to \eq{e2/7.11} a constant vector $l''_g \in P_g$
(when considering Feynman integrals in momentum representation this
would correspond to the presence of an external momenta in which no
expansion is performed). In that case one would encounter several
maximal $s$-subgraphs in $G$ (concerning subgraphs see
\ssect{Object.Notations}).  However, the arguments of
\ssect{As-op.Recursive} allow one to reduce such case to the one
considered here.

\SUBSECTION{Some notations.}
\label{Object.Notations}%7.3

To study the expansions of $G_\k$ let us introduce some useful
notions.

The {\it singular plane} for $g \in G$ is defined as:
\be[e2/7.12]
     \pi_{g(G)}\ \bydef\ \{ p \in  P_G \mid l'_{g(G)}(p) = 0 \}
\ee
(cf.\ \eq{1/4.1}).  The expansion $\T_\k\.g$ is singular
on $\pi_g$. As regards the factors $g_\k$, in general they are
singular (if they are singular at all) on other planes (which approach
$\pi_g$ as $\k \rightarrow 0)$ and can be regular on $\pi_{g(G)}$.

Now, following the scheme of \sect{CompleteSubgraphs} we introduce
the notion of complete singular subgraphs ($s$-subgraphs) and the
ordering in the set of all $s$-subgraphs of the graph $G$ which is
denoted $S[G]$.  The singular planes for subgraphs
$\G \in S[G]$ are defined by \eq{1/4.2}.

For $g \in  G$ define the $\k $-{\it vicinity} of the plane $\pi_g$:
\be[e2/7.13]
     \cO ^{\k}_{g(G)}
     \bydef
     (l'_{g(G)})^{-1} (\cO ^{\k}_g) .
\kern20mm\ee
It is important that the singularity of $g(p,\k)$ is always located
within $\cO ^{\k}_{g(G)}$. For an $s$-subgraph $\G $ the $\k
$-vicinity of the plane $\pi^\0_\G$ is defined by:
\be[e2/7.14]
     \cO ^{\k}_{\G (G)}
     \bydef
     \cap_{g\in \G} \cO ^{\k}_{g(G)} .
\kern26.4mm\ee
For any $H \in  G$ define:
\be[e2/7.15]
     N_{H}
     \bydef
     \sum_{g\in H} N_g,
     \qquad  N_{\emptyset} \bydef  0 ,
\ee
where $N_g$ are defined in \eq{e2/7.2}, and
\be[e2/7.16]\kern0.8mm
     d_{H} \bydef  \sum_{g\in H} d_g,
     \qquad\kern2.4mm d_{\emptyset} \bydef  0 .
\ee
For the $s$-subgraphs $\G $ also define (cf.\ \ssect{GenericR.Index}):
\be[e2/7.17]\kern9.5mm
     \o^\0_\G
     \bydef
     d^\0_\G - \dim  P^\0_\G,
     \qquad \o_{\emptyset} \bydef  0 .
\ee

\SUBSECTION{Properties of $G(p,\k)$.}
\label{Object.Properties}%7.4

Let us study the properties of the graph $G$ that will be needed for
the construction of the asymptotic expansion.

First of all note that from the properties of the mappings $l_g$
assumed in \ssect{Object.KappaDependence} it follows that $G$ is an
$s$-graph and one can assume that $\pi^\0_G = \{0\}$
(cf.\ \ssect{Complete.Ordering}). Then from the property $(ii)$
(\ssect{Object.Factors}) of the functions $F_g$ one obtains that the
product $G(p,\k)$ (and any of its subproducts as well) is
almost-homogeneous with respect to both arguments:
\be[e2/7.18]
     G(\l p,\l \k)
     = \l^{-d_G}
       \, \sum^{L_G}_{k=0}
       G_k(p,\k)\, \ln^k \l,\qquad \l  > 0 .
\ee
Assume that $G(p,\k)$ is locally absolutely integrable in $p$.  Then
$G$ defines a continuous functional on $\cD (P_G)$, and from the
almost-homogeneity of $G(p,\k)$ it follows that
\be[e2/7.19]
     \langle G(p,\k), \vf (p) \rangle
     = (\k /\k_{0})^{-d_G+\dim P_G}
       \sum^{L_G}_{k=0} \ln^k (\k /\k_{0})
       \, \langle G_k(p,\k_{0}), \vf (p\, \k /\k_{0}) \rangle,
\ee
where $\k_{0}$ is a positive number. One immediately obtains a useful
auxiliary estimate:
\be[e2/7.20]
     |\langle G_\k, \vf  \rangle|
     \leq
     \| \vf \|^0
     \, \k^{-\o^\0_G}\, \L (\k)
\ee
for any $\vf \in \cD (\cO ^{\k}_G)$, where $\|\cdot\|$ is the
seminorm defined in \ssect{Isolated.Seminorms1}.

If $G$ contains singular factors then the product $G(p,\k)$ can be
non-integrable.  In such case decompose $G$ as:
\be[e2/7.21]
     G = G^{\rm sing}\times G^{\rm reg},
\ee
where $G^{\rm sing}$ includes only singular factors $F_g$ and
$G^{\rm reg}$ includes only regular ones. The condition $(iii)$ of
\ssect{Object.Factors} allows one to apply the operation $\tR$
to $G^{\rm sing}$.  Note that due to the term $\k q_g$ in
\eq{e2/7.11} the factors $F_g$ in the product $G^{\rm sing}$ are
singular on the planes which differ from the planes $\pi_g$ defined
in \eq{e2/7.12} by a $\k$-dependent shift. So, when constructing the
operation $\tR$ for $G^{\rm sing}$ one deals with a hierarchy of
$s$-subgraphs which is different from $S[G]$.

Due to the conditions $(ii)$, $(iii)$ and $(x)$ as well as the scaling
properties of ${\tR}$ one can verify that the functional
$G^{R}_\k$ defined as
\be[e2/7.22]
     \langle G^{\rm R}_\k, \vf  \rangle
     \bydef
     \langle
         \tR\.G^{\rm sing}_\k,
         [ G^{reg}_\k\, \vf ]
     \rangle
\ee
is almost-homogeneous in the sense of \eq{e2/7.19}. Then the estimate
for it analogous to \eq{e2/7.20} is
\be[e2/7.23]
     |\langle G^{\rm R}_\k, \vf  \rangle|
     \leq
     \k^{-\o^\0_G} \cS[\vf,\k],
\ee
for $\vf \in \cD (\cO ^{\k}_G)$.

In what follows we will omit the superscript R in the notation $G^{\rm
R}_\k$ since the construction of the \asex\ does not
depend on whether $G_\k$ requires an $R$-operation or not.
OOB
The necessity to consider locally non-integrable functionals $G_\k$
arises in applications when one studies asymptotic expansions of UV
divergent Feynman diagrams and e.g. in the theory of asymptotic
expansions of Feynman diagrams in non-Euclidean asymptotic regimes.

\SECTION{Formal expansion and its properties.}
\label{FormalExpansion}%8

Here we study the properties of the formal expansion of the product
$G(p,\k)$. In \ssect{Formal.Taylor} some useful formulae for the
formal expansion (summation by parts) are presented.  In
\ssect{Formal.Bounds} we obtain estimates for the expansion and its
remainder term.  A possible way to relax the conditions imposed on
the factors is discussed in
\ssect{Formal.Generalizations}.

\SUBSECTION{Formal expansion $\protect\T_\k$.}
\label{Formal.Taylor}%8.1

The expansion for the function $g(p,\k) = F_g(p + \k q_g, \k)$ in
the sense of $\cC^\infty(P_g\backslash \{0\})$ is obtained from the
expansion for $F_g$ with the use of the results of
\ssect{As-exp.LinearMaps} and \ssect{As-exp.DirectProducts}.
It can be seen that the conditions $(v)$--$(ix)$
of \ssect{Object.Factors}  are satisfied for the resulting expansion.

As was mentioned in \ssect{As-op.Formal}, by formal multiplication of
the expansions for the factors one obtains the expansion for the
product $G(p,\k)$ in the sense of $\cC^\infty(\crcaccent{P}_G)$
where $\crcaccent{P}_G = P_G \backslash \cup_{g\in G} \pi_g$:
\be[e2/8.1]
     G_\k
     \ \mathrel{\mathop{\0\simeq\0}
      ^{\cC^\infty(\scriptcrcaccent{P}_G)}}\ {}
     {\bf T\.G}_\k
     = \sum_n\t_n\.G_\k .
\ee
We will also use the following recurrent relations for the terms of
the expansion \eq{e2/8.1} and for its remainder term (cf.\ the 
summation-by-parts formulae \eq{e2/1.14}, \eq{e2/1.15}):
\be[e2/8.2]
     \t_n\.G_\k
     = \sum_{m}\t_{m}\.\G_\k
     \times \t_{n-m}\.\GG_\k ,
\ee
\be[e2/8.3]
     (1 - \T_N)\.G_\k
     = \G_\k
       \times (1 - \T_{N-N_\G})\.\GG_\k
     + \sum^{N-N_\GG}_{m=N_\G}
       (1 - \T_{m})\.\G_\k
       \times \t_{N-m}\.\GG_\k.
\ee
Note that \eq{e2/8.3} may contain products of singular terms that were
not present in \eq{e2/8.1} (e.g.\ \ $g_1\times \t_k\.g_2$ for
$g_1\in \G $ and $g_2 \in \GG $ which is present in $\G
\times (1 - \T_{N-N_\G})\.\GG $ and in $(1 - \T_{m})\.\G
\times\t_{N-m}\.\GG)$.  After all summations are performed on the
r.h.s.\ of \eq{e2/8.3} such spurious singularities cancel, but when
studying each term in \eq{e2/8.3} separately one should pay proper
attention to them.

\SUBSECTION{Simple bounds for products away from singularities.}
\label{Formal.Bounds}%8.2

Let $H \subset G$. We are interested in a bound similar to
\eq{e1/11.9} for the expansion of $H_\k$ and its remainder term.
Let $K \subset P_G$ be a compact set such that for all $g \in H$
\be[e2/8.4]
     K \cap \pi_g = \emptyset  .
\ee
In the same way as in \ssect{GenericR.Proof.PowerCounting}, using an
equality of the form of \eq{e2/8.2} and the inequalities \eq{e2/7.4}
one can by induction (since $\t_n\.H_\k$ contains terms with
different asymptotics) obtain the following estimate:
\be[e2/8.5]
     \|\t_n\.H(p,\k)\|^{\o}_{p\in t\, K}
     \leq
     \left({\k \over t}\right)^n
     \, {\L (\k)\, \L (t)\over t^{d_{H}+\o}},
     \qquad t > 0 .
\ee
To obtain a similar bound for the remainder term of the series
$\T_N\.H_\k$ one should use an induction with respect to \ 
$\ldots\subset H'' \subset H'\subset\ldots \subset H_\k$ 
and use a representation of the
form \eq{e2/8.3} and \eq{e2/8.5}. Then for $t > 0$ such that
\be[e2/8.6]
     t\, K \cap \cO ^{\k}_{g(G)}
     = \emptyset, \qquad g \in  H
\ee
(actually, $t \geq  \k \, \const_{H,K}$) one obtains:
\be[e2/8.7]
     \| H - \T_N\.H \| ^{\o}_{p\in t\, K}
     \leq  \left(  {\k \over t}  \right)^{N+1}
     \,
     {  \L (\k)\, \L (t)
        \over
        t^{d_{H}+\o}  } .
\ee
This estimate is a formalized (and stronger) version of the statement
that $\T_N\.H_\k$ approximates $H_\k$ within $o(\k^N)$ in
the sense of $\cC^\infty(P_G \backslash \cup_{g\in H} \pi_g)$;
cf.\ \eq{e2/8.1}.

\SUBSECTION{Generalizations.}
\label{Formal.Generalizations}

A remark is in order.  Actually, our constructions will never use
directly the properties $(i)$--$(x)$ from \ssect{Object.Factors} but
only the estimates from
\ssect{Object.Properties},
\ssect{Formal.Bounds} and also some factorization properties
like $G = H\times (G\backslash H)$
within  certain regions of $P_G$. Therefore, the
conditions of  \ssect{Object.Factors}  may  be  changed
if  the mentioned estimates are
preserved. All that is required is that there exist a vicinity
$\cO ^{\k}$ of the point $l = 0$ outside which the estimates of the
form  of \eq{e2/8.7} hold and that the product $G(p,\k)$ be  locally
integrable (or  allow application of the operation $\tilde{R}$). This
ensures a straightforward  modification of our techniques to the case 
when the initial expression involves $\tR$.

\SECTION{Existence and properties of the As-operation.}
\label{Existence}%9

In this section we present explicit formulae for the {\it
As}-expansion which yield the asymptotic expansion of a functional
$G_\k$ in the sense of distribution theory
(\ssect{Existence.Construction}) and then we formulate a theorem on
the \asop\ (\ssect{Existence.Theorem}). The proof of the theorem is
relegated to a separate section.

\SUBSECTION{Construction of the As-operation.}
\label{Existence.Construction}%9.1

Our aim is to obtain the expansion of the functional $G_\k$ in the
sense of distribution theory (the \asex):
\be[e2/9.1]
     G_\k
     \mathrel{\mathop{\0\simeq\0}
       ^{\cD '(P_G)}_{\k \rightarrow 0}}
    \As\.G_\k = \sum_n \as_n\.G_\k .
\ee
The formal expansion $\T_N\.G_\k$ is insufficient since for large
enough $N$ the terms of the expansion become non-integrable on the
planes $\pi^\0_\G$, $\G \in S[G]$. We will construct
the desired expansion starting from \eq{e2/9.1} following the ideas of
\sects{As-expansion}-{ExtensionPrinciple}.

As explained in \ssect{Extension.As&R}, the \asex\ is
expected to have the form:
\be[e2/9.2]
     \As = \tR^{(\k)}\.\T_\k,
\ee
where $\T_\k$ is the operation of the formal expansion \eq{e2/8.1},
and $\tR^{(\k)}$ is the special operation $\tR$ (\sect{SpecialR}) with
finite counterterms specially fixed.  So, now it is sufficient to
obtain the explicit formulae for the counterterms of $\tR^{(\k)}$ and
prove that the \asop\ thus defined yields the desired \asex\ in the
sense of distributions for any $s$-graph $G$.

Assume that the form \eq{e2/9.2} for the \asop\ has been proved for
all subgraphs $\G < G$. Define the {\it incomplete As}-operation by
\eq{e2/2.11} (cf.\ the definition of the incomplete $R$-operation
$\R'$ in \ssect{R-op.Recursive}).  The operation $\As'$ yields the
\asex\ for $G_\k$ in the sense of distributions valid within the open
region $P_G\backslash \{0\}$. It is easy to see that the situation
here is fully analogous to the example in
\ssect{Extension.Example}. Therefore we define:
\be[e2/9.3]
     \As_N\.G_\k
     \bydef
     \sum_{n\leq N}
     \left\{
          \tr\.\as'_n\.G_\k
        + {\bf \tilde{E}}_n(\k)
     \right\} ,
\ee
where $\tr$ is the special subtraction operator \eq{e2/4.8}
while the finite renormalization is defined as
\be[e2/9.4]
     {\bf \tilde{E}}_n(\k)
     \bydef
     \sum_{|a|=\o^\0_G+n}
     \tilde{E}_{a}(\k) \, \d^{a}(p)
\ee
and the coefficients $\tilde{E}_{a}(\k)$ are as follows:
\be[e2/9.5]
     \tilde{E}_{a}(\k)
     \bydef
     \langle
         ( \D'_{|a|-\o^\0_G-1}\.G_\k
         - \tr\.\as'_{|a|-\o^\0_G}\.G_\k
         )\ast
         \cP^{a} 
     \rangle .
\ee
In \eq{e2/9.5}, $\cP^{a}(p)$ are polynomials of order $|a|$
(\ssect{Isolated.Deltas}) and the operation $\ast $ is defined by
\eq{e2/4.2}. All the terms of the expansion thus defined are  correct
distributions on $P_G$.

A straightforward calculation shows that for the expansion defined in
\eqs{e2/9.3}-{e2/9.5} the following holds:
\be[e2/9.6]
     \langle \D_N\.G_\k, \vf \rangle 
     =
     \langle
         \D'_N\.G_\k
         \ast
         \Big( 1 - \T^{\o^\0_G+N} \Big) \.\vf
     \rangle .
\ee
Now, to prove that the \asop\ thus defined yields the desired {\it
As}-expansion for the graph $G$, it is sufficient to verify that
\be[e2/9.7]
     \langle \D'_N\.G, \vf  \rangle = o(\k^N)
\ee
for any $\vf \in \cD _{\o^\0_G+N+1}(P_G)$. The
proofs of the propositions analogous to those of
\sect{ExtensionPrinciple} (e.g.\ the proof of the fact that
$\As'_N\.G_\k$ approximates $G_\k$ within $o(\k^N)$ on the
space of test functions possessing zero of the corresponding order
etc.) will in fact be contained in the proof of
\eq{e2/9.7}.

\SUBSECTION{Theorem (existence and properties of the $As$-operation).}
\label{Existence.Theorem}%9.2

\begingroup
\it

Let all $F_g$ satisfy the conditions $(i)$--$(x)$ of\/
\ssect{Object.Factors} and $G(p,\k)$ be defined by \eq{e2/7.9}. Then
for the $s$-graph $G$ satisfying \eq{e2/7.23} there exists a unique
expansion \eq{e2/9.1} $($its explicit expression is given by
\eq{e2/9.3}--\eq{e2/9.5}$)$ such that$:$

 $(a)$ The expansion \eq{e2/9.1} starts at the terms of order $N_G -
A_G$ where
\be[e2/9.8]
     A_G 
     = \max_{{\vphantom{0^0}}\G \in S[G]} 
       \{ \o^\0_\G + N^\0_\G \} ,
\ee
i.e.\  for any $n < N_G - A_G$
\be[e2/9.9]
     \As_n\.G \equiv  0 .
\ee
$($The explicit expression for $N_\G$ is given in \eq{e1/10.7}.$)$

 $(b)$ For any $n \geq N_G - A_G$, $\As_n\.G \in \cD '(P_G)$
and $\As_n\.G$ satisfies the locality condition of
\ssect{As-op.Locality}.

 $(c)$ All distributions $\as_n\.G$ are power-and-log functions of
$\k $ of order $n;$ the following estimate holds for any $\vf
\in \cD (P_G):$
\be[e2/9.10]
     |\langle\as_n\.G_\k, \vf \rangle|
     \leq
     \left( \frac{\k}{d} \right)^n \, 
     d^{-\o^\0_G} \cS[\vf,d] .
\ee

 $(d)$ $\As_N\.G_\k$ approximates $G_\k$ in the sense of $\cD
'(P_G)$ to $o(\k^N);$ moreover, for any $\vf \in \cD (P_G)$
\be[e2/9.11]
     |\langle \D_N\.G_\k, \vf \rangle|
     \leq
     \left( \frac{\k}{d} \right)^{N+1} \, 
     \L (\k) \, d^{-\o^\0_G} \, 
     \cS[\vf,d],
\ee
for $d \geq  \k \, \const$  or
\be[e2/9.12]
     |\langle \D_N\.G_\k,\vf \rangle|
     \leq
     \k^{-\o^\0_G} \cS[\vf,\k ]
\ee
for $d \leq \k \, \const$ $($in which case 
$\vf \in \cD (\cO ^{\k}_G))$.  In the above relations
\be[e2/9.14]
     d = \rad\supp\vf  ,
\ee
and the $\L$-functions implied in all $\cS$ are double-sided.

 $(e)$ The As-expansion has the following minimality property $($cf.\
\eq{e2/4.16}$):$
\begin{eqnarray}
\label{e2/9.15}
     \langle \as_n\.G \ast \cP^{a} \rangle = 0,
     &\qquad& 
     |a| < \o^\0_G + n ;
\\
\label{e2/9.16}
     \langle \D_N\.G \ast \cP^{a} \rangle = 0,
     &\qquad& 
     |a| \leq  \o^\0_G + N .
\end{eqnarray}

\endgroup

 {\bf Remarks.}
 $(i)$ The estimate \eq{e2/9.10} has the form characteristic of the
operation $\tR$ (cf.\ \eq{e2/5.4}) and is an auxiliary one. The
inequality \eq{e2/9.12} is a consequence of \eq{e2/9.11}. In fact
(cf.\ the remark in \ssect{Isolated.d-Inequalities}), for any $\vf
\in \cD (\cO ^{\k}_G)$ the inequality \eq{e2/9.11} must remain valid
if one chooses a larger $d$, namely $d = \k \,\const \geq \rad\supp
\vf $, which immediately results in \eq{e2/9.12}.

 $(ii)$ The inequality \eq{e2/9.12} does not contradict to the
$o(\k^N)$ behaviour of the remainder term despite the fact that the
r.h.s.\ of \eq{e2/9.12} can include $\k $ in powers lesser than $N$.
The point is that for any fixed $\vf \in \cD (P_G)$, as $\k
\rightarrow 0$, sooner or later $\k $ becomes less than $d$, and then
one must use \eq{e2/9.11}.

 $(iii)$ If the conditions $(ix)$--$(x)$ of \ssect{Object.Factors} are
not fulfilled then the dependence of $\as_n\.G_\k$ on $\k $ will
not be power-and-log (the statement $(c)$ of the theorem).
Nevertheless the estimates \eqs{e2/9.10}-{e2/9.12} will remain true.

 $(iv)$ Using the results of \ssect{SpecialR.VariationTheorem}, one
can explicitly extract all the non-trivial dependences on the
expansion parameter $\k$ in the \asop:
\be
     \As\.G_\k = \sum_{\g\leq G}
     \tR_f\.[({\bf \tilde{E}}_{f,\k}\.\g)\,(\Tk\.\Gg)] \kern4cm
\ee\be[e2/9.16+]
     \equiv
     \tR\.\Tk\.G  + 
     \sum_{\emptyset\leq\g\leq G}
     \tR_f\.[({\bf \tilde{E}}_{f,\k}\.\g)\,(\Tk\.\Gg)],
\ee
where ${\bf \tilde{E}}_{f,\k} = \sum_n {\bf \tilde{E}}_{f,n} $ and
${\bf \tilde{E}}_{f,n} $ is defined similarly to \eqs{e2/9.3}-{e2/9.5}
with all $\tr$, $\tR$ etc.\ replaced by $\tr_f$ etc., the
latter being defined by the formal relation $\tR_f=\tr_f\.\tR'_f$
(recall that $\tR_f$ and $\tR'_f$ as defined in
\ssect{SpecialR.Deltas} differ from $\tR$ and $\tR'$ by the fact that
the ``$f$''-operations are defined on products containing
$\d$-functions in a ``natural'' way). 

\def\TempSect{\ssect{Existence.Theorem}}

\SECTION{Proof of the theorem of \protect\TempSect.}
\label{MainProof}%10

As was the case with the $R$-operation (cf.\
\ssect{GenericR.Proof.PowerCounting}) the proof consists in a
meticulous but quite straightforward power counting.  Complications
here are due to the three facts: first, one has to deal with series
instead of a single product; second, the expansion involves a product
of expansions of several factors so that studying remainder of the
expansion requires the combinatorial trick of summation-by-parts;
third, one has to keep track of two parameters in estimates---$\k$ and
$d$.

The most natural way to prove the theorem is to use induction with
respect to $\G \in S[G]$ and $N$. We will assume that the theorem is
valid for all subgraphs $\G< G$. A correct starting point for the
induction in $\G$ is ensured by the fact that for the empty subgraph,
by the definition \eq{e2/2.8} one has:
\be[e2/10.1]
     \As\.\emptyset_\k
     \equiv
     \As_{0}\.\emptyset_\k
     \equiv  \emptyset_\k
     \equiv  1 .
\ee
Furthermore, we will assume that for the graph $G$ itself, the theorem
is valid for all $n \leq N - 1$.  To ensure a correct starting point
for the induction in $N$ it is sufficient to obtain an estimate
similar to \eq{e2/9.11} for the graph $G$ in the order $N_G - A_G
- 1$.%
\footnote{
Recall that $N_G - A_G$ is the order at which the expansion
$\As\.G_\k$ starts; the 1-dimensional example ($p\in[0,+\infty]$):
$\displaystyle
     \frac{1}{(p+m)^2} = \frac{\pi}{2m}\d(p)+O(\log m),
$
demonstrates that this order can be less than $N_G$, i.e.\ $A_G
\geq 0$.
\relax}
This is postponed till \ssect{Proof.StartingPoint}.

Under the above assumptions we will prove the validity of the
statements of the theorem in the order $N$ for the graph $G$.  First,
in \ssect{Proof.as'-estimate} we will obtain an estimate for the
functional $\as'_N\.G$.  In \ssect{Proof.DeltaEstimate} we will
obtain a bound for the functional $\D'_N\.G_\k$ defined on the
space $\cD _{\o^\0_G+N+1}(P_G)$.  In
\ssect{Proof.EstimatesProof}, using the representation
\eq{e2/9.6}, we will prove \eq{e2/9.11} for
$\D_N\.G_\k$.
Then we will obtain an estimate for the functional
$\tr_{0}\.\as'_N\.G$ and, finally, we will demonstrate that the
finite renormalization ${\bf \tilde{E}}_N(\k) + \tilde{\D}$ in
the definition \eq{e2/9.3} is of order $\k^N\, \L (\k)$.
This  completes  the  proof of  the estimates of the theorem.

In \ssect{Proof.Factorizable} the case of a factorizable graph $G$ is
considered. In \ssect{Proof.KappaDependence}
the $\k $-dependence of the individual  terms
of  the  expansion  generated  by  the \asop\ is studied, and
it is  demonstrated  that  the  series  runs  in powers and logarithms of
$\k $.

\SUBSECTION{An estimate for $\protect\as'_N\.G$.}
\label{Proof.as'-estimate}%10.1

First one uses the sector decomposition of unit $\{\theta_\G\}$ and
applies the recursion \eq{e2/8.2} to define $\as'_N\.G$ in each
sector. One obtains:
\be[e2/10.2]
     \as'_N\.G
     =
     \sum_{\G \triangleleft G}  \ \ \sum_{n<N-N_\GG}
     (\as_n\.\G) \, [ \theta_\G\, \t_{N-n}\.\GG ] .
\ee
The estimates of the form of \eq{e2/9.10} and \eq{e2/8.5} hold for
$\as_n\.\G$ and $\t_k\.\GG $, respectively---cf.\ {}
\eq{e2/5.4} and \eq{e2/5.3}.  One can follow the pattern of reasoning
used in studying the operation $\tilde{R}$ (\sect{SpecialR}, see also
\ssects{GenericR.Proof.Plan}-{GenericR.Proof.PowerCounting}).
Taking into account the  overall  factor
$\k^N\, \L (\k)$,  one  obtains  the following  estimate
for  the  functional  $\as'_N\.G$   defined   on   the   space
$\cD _{\o^\0_G+N+1}(P_G)$:
\be[e2/10.3]
     |\langle \as'_N\.G_\k \ast \vf \rangle|
     \leq
     \left( \frac{\k}{d} \right)^N \L (\k) 
     \sum_{k\geq \o^\0_G+N+1}
       \|\vf \|^k \, 
       d^{k-\o^\0_G} \, 
       \L (d).
\ee
(Here and below the upper bound of summation is omitted because it is
of no practical interest.)

\SUBSECTION{An estimate for $\D'_N\.G_\k$.}
\label{Proof.DeltaEstimate}%10.2

Consider the functional $\D'_N\.G_\k = G_\k -
\As'_N\.G_\k$ defined on the space $\cD
_{\o^\0_G+N+1}(P_G)$ which is the domain of definition
of the most singular term in $\D'_N\.G_\k$, namely,
$\as'_N\.G_\k$ (cf.\ \eq{e2/10.3}). Choose a cutoff function
$\Phi_\k$ (\ssect{Decomp.Radial}) so that
\be[e2/10.4]
     \supp (\theta_\G\, \Phi_\k)
     \cap
     \cO ^{\k}_{g(G)}
     = \emptyset, 
     \qquad g \in  \G  \triangleleft  G .
\ee
(The $\k $-vicinities $\cO ^{\k}_{g(G)}$ are defined in
\eq{e2/7.13}.) This is always possible due to the scale invariance of
the decomposition $\{\theta_\G\}$ (cf.\ \ssect{Decomp.Submaximal}
and the fact that if the conditions \eq{e2/10.4} are satisfied for
some $\k > 0$ then they are satisfied for any $\k > 0$).

Represent the functional $\D'_N\.G_\k$ as:
\be[e2/10.5]
     \langle (\D'_N\.G_\k) \ast  \vf \rangle
     =
     \langle
         (\D'_N\.G_\k)
         \ast \Phi^{\k}\, \vf
     \rangle 
     +
     \langle
        (\D'_N\.G_\k),
        \Phi_\k\, \vf
     \rangle
\ee
for any $\vf \in \cD _{\o^\0_G+N+1}(P_G)$. Let us
split the first term on the r.h.s.\ of \eq{e2/10.5} as:
\be[e2/10.6]
     \langle
       (\D'_N\.G_\k) \ast \Phi^{\k}\, \vf
     \rangle 
     =
     \langle
       (\D'_{N-1})\.G_\k \ast \Phi^{\k}\, \vf
     \rangle 
     -
     \langle
       (\as'_N\.G_\k), \Phi^{\k}\, \vf
     \rangle.
\kern4mm\ee
The functional $\D'_{N-1}\.G_\k$ coincides with $\D_{N-1}\.G_\k$
on $\vf \in \cD _{\o^\0_G+N+1}(P_G)$ so that one
can use the estimate \eq{e2/9.12} (which is satisfied by the inductive
assumption) for the first term in
\eq{e2/10.6}. Taking into account the  property $(e)$  of  the seminorms
(\ssect{Isolated.Seminorms2}) one obtains:
\be[e2/10.7]
     |\langle
         (\D'_{N-1}\.G_\k) \ast \Phi^{\k}\, \vf
     \rangle|
     \leq
     \sum_{k\geq \o^\0_G+N+1}
     \|\vf \|^k \, \k^{k-\o^\0_G} \,
     \L_k (\k) .
\ee
For the second term on the r.h.s.\ of \eq{e2/10.6}, from \eq{e2/10.3}
one has a similar estimate:
\be[e2/10.8]
     |\langle  
         (\as'_N\.G_\k)  \ast  \Phi^{\k}\, \vf
     \rangle|
     \leq
     \sum_{k\geq \o^\0_G+N+1}
     \|\vf \|^k \, \k^{k-\o^\0_G} \,
     \L_k (\k) .
\ee
The sum of \eq{e2/10.7} and \eq{e2/10.8} gives the desired estimate
for the first term in \eq{e2/10.5}.

Now consider the second term on the r.h.s.\ of \eq{e2/10.5}.  Let us
first obtain an estimate for the expression
\be[e2/10.9]
     \langle
         (\D'_N\.G_\k), \eta_\l\, \vf
     \rangle
\ee
(cf.\ \eq{e1/9.1}).  Using the sector decomposition of unit
$\{\theta_\G\}$ and representing the remainder term in each sector
using \eq{e2/8.3} (cf.\ \eq{e2/1.15}), one obtains:
\be
     \D'_N\.G
     =
     \sum_{\G \triangleleft G}
     \Big\{
        \D_{N-N_\GG}\.\G \,
        [\theta_\G\, \GG ]
\kern3cm\ee\be[e2/10.10]\kern2cm
     +   \sum_{n\leq N-N_\GG}
         (\as_n\.\G) \,
         [\theta_\G\, (1 - \T_{N-n})\.\GG ]
     \Big\}.
\ee
Note that for $p \in \supp \Phi_\k$ this representation is correct
since due to \eq{e2/10.4} neither the singularities of $\GG
$ nor those of $\t_k\.\GG $ fall within the sector $\supp
\theta_\G$, so that one deals with products of a distribution by
a smooth function (the latter is put into the square brackets in
\eq{e2/10.10}).

Consider the $n$-th term in the sector $\G $. By the inductive
assumption, one has (cf.\ \eq{e2/9.11}):
\be[e2/10.11]
     |\langle
         \as_n\.\G_\k, \vf
     \rangle|
     \leq
     \left(\frac{\k}{d}\right)^n \, 
     \L (\k)
     d^{-\o^\0_\G} \,
     \cS[\vf,d],
\ee
for any $\vf \in \cD (P_\G)$. Furthermore, from \eq{e2/8.7} one
has the following estimate for the remainder term of the formal
expansion in the square brackets:
\be
     \| (1 - \T_{N-n})\.\GG  \|^{\o}_{p\in t\, K}
\kern9cm\ee\be[e2/10.12]
\leq
     \left( {\k\over t} \right)^{N-n+1} \,
     { \L (\k) \, \L (d)
            \over
        t^{d_G-d_\G+\o}},
     \quad \hbox{if\ \ \ } t\, K \cap \cO ^{\k}_{g(G)}
     = \emptyset,\quad g \in  \GG.
\ee
Following the reasoning of \ssect{GenericR.Proof.UsingInduction},
\ssect{GenericR.Proof.PowerCounting}  one  obtains  an
estimate similar to \eq{e1/9.1}:
\be[e2/10.13]
     |\langle
        \as_n\.\G_\k,
        [ \theta_\G\, \eta_\l \,
          (1 - \T_{N-n})\.\GG \, \vf  ]
     \rangle|
     \leq
     {\k}^{N+1} \,
     \L (\k) \,\l^{-\o^\0_G}
     \cS_{p\in \supp\eta_\l}[\vf,\l ].
\ee
The r.h.s.\ of \eq{e2/10.13} has the same form for all $n$ and $\G $,
and a similar inequality can be obtained for the term $\D \.\G \,
[\theta_\G\, \GG ]$ in \eq{e2/10.10}.  Therefore,
carrying out summation one obtains the following estimate:
\be[e2/10.14]
     |\langle
        \D_N\.G_\k, \eta_\l\, \vf
     \rangle|
     \leq
     \left( \frac{\k}{\l}\right)^{N+1} \,
     \L (\k) \,
     \l^{-\o^\0_G} \,
     \cS_{p\in \supp\eta_\l}[\vf,\l ],
\ee
where $\supp\eta_\l \subset \supp \Phi_\k$, i.e.\ $\l \geq
\k\,\const(\eta)$.  Now, similarly to \ssect{Subtraction.Extension},
one can obtain the following estimate for the second term in
\eq{e2/10.5} from \eq{e2/10.14}:
\be[e2/10.15]
     |\langle
       \D'_N\.G_\k, \Phi_\k\, \vf
     \rangle|
     \leq
     \left(\frac{\k}{d}\right)^{N+1} \,
     \L (\k) \, d^{-\o^\0_G} \,
     \cS[\vf,d] .
\ee
Adding \eq{e2/10.7}, \eq{e2/10.8} and \eq{e2/10.15}, one finally obtains the
following estimate for $\D'_N\.G_\k$:
\be[e2/10.16]
     |\langle
       \D'_N\.G_\k  \ast  \vf
     \rangle|
     \leq
     \left( \frac{\k}{d} \right)^{N+1} \, 
     \L (\k)\,  d^{-\o^\0_G}\,
     \cS[\vf,d] ,
\ee
where $\vf \in \cD _{\o^\0_G+N+1}(P_G)$ and
$\rad\supp\vf \leq d \geq \k \,\const$.

\SUBSECTION{Starting point for the induction in $N$.}
\label{Proof.StartingPoint}%10.3

Let us prove the estimate for $\D_N\.G$ which ensures a correct
starting point for the induction with respect to $N$. As was noted at
the beginning of this section, the lowest $N$ for which such estimate
should be proved is $N_G - A_G - 1$, and in this case
$\D_N\.G_\k = G_\k$, so that in fact one should obtain an
estimate for $G_\k$. However, let us first determine $A_G$.

Assume that for each subgraph $\G < G$ the \asex\ starts
at the order $N_\G - A_\G$. Then the expansion $\As'\.G$ starts
at the order $N_G - \max^{\0}_{\G < G} A_\G$, but if
$\o^\0_G + N_G - \max^{\0}_{\G < G} A_\G >
0$ then a non-trivial renormalization ${\bf \tilde{E}}_n$
(eq.\eq{e2/9.4}) can appear at even lower order, namely, at the order
$(-\o^\0_G)$. So, in general the expansion $\As\.G$
starts at the order
\be[e2/10.17]
     N_G - A_G 
     =
     \max_{\G < G} \{ A_\G \} .
\ee
Solving this recursion, one obtains the expression \eq{e2/9.8} for
$A_G$.

We can now turn to proving the estimate \eq{e2/9.11} for
$\D_{N_G-A_G-1}\.G_\k \equiv G_\k$.  The reasoning will
closely follow \ssect{Proof.DeltaEstimate}.

One starts by performing the splitting as in \eq{e2/10.5}.  From
\eq{e2/7.23} one estimates the first term on the r.h.s.\ of
\eq{e2/10.5} as:
\be[e2/10.18]
     |\langle
        \D_{N_G-A_G-1}\.G_\k, \Phi^{\k}\, \vf
     \rangle|
     \leq
     \k^{-\o^\0_G} \cS[\vf {\bf,\k}],
     \qquad\hbox{for any}\quad
     \vf  \in  \cD (P_G).
\ee
To estimate the second term on the r.h.s.\ of \eq{e2/10.5}, one notes
that the sum over $n$ in \eq{e2/10.10} becomes zero while the
remaining term in the braces is:
\be[e2/10.19]
     \Big(
        \D_{N_\G-A_\G-1}\.\G
     \Big) \,
     [ \theta_\G \, \GG_\k ]
     \equiv
     \G_\k \, [\theta_\G \, \GG_\k] .
\ee
One can use \eq{e2/9.11} which is satisfied for $\G_\k$ in
\eq{e2/10.19}  by  the  inductive assumption, and for $\GG $
one can use \eq{e2/8.5}. Then  one  obtains  the  following estimate for an
expression similar to \eq{e2/10.9}:
\be[e2/10.20]
     |\langle
        \D_{N_G-A_G-1}\.G_\k,
        \eta_\l\, \vf
     \rangle|
     \leq
     \left( \frac{\k}{d} \right)^{\nu} \, 
     \L (\k) \, \l^{-\o^\0_G } \,
     \cS[\vf {\bf,\l}],
\ee
where $\nu = N_G - \max^\0_{\G < G} A_\G$ (one can
assume that $\supp\eta_\l \subset \supp \Phi_\k)$.  From
\eq{e2/10.20} one can obtain an estimate for the second term in the
representation
\eq{e2/10.5} which coincides with \eq{e2/10.15} with $N = \nu - 1$.

Combining the estimates \eq{e2/10.18} and \eq{e2/10.15} for the  two  terms
on  the r.h.s.\  of \eq{e2/10.5}, one arrives at the following inequality:
\be[e2/10.21]
     |\langle
         G_\k,\vf
     \rangle|
     \leq
     \k^{-\o^\0_G} \, \cS[\vf,\k]
   + (\k /d)^{\nu} \, \L (\k) \, 
     d^{-\o^\0_G} \, \cS[\vf,d] ,
\ee
which is valid for $d \geq \k \, \const$. Since $d \geq \k \, \const$,
one can multiply each term in the braces by an appropriate
non-negative power of $[d / (\k \, \const)] \geq 1$ and then extract
$\k $ in the maximal power which is equal to $\min_{k\geq 0} \{ k -
\o^\0_G, \nu \} = N_G - A_G$.  Then one obtains the
final estimate:
\be[e2/10.22]
     |\langle
        \D_{N_G-A_G-1}\.G_\k, \vf
     \rangle|
     \equiv
     |\langle
        G_\k, \vf
     \rangle|
     \leq
     \left( \frac{\k}{d} \right)^{N_G-A_G} \, 
     \L (\k) \, d^{-\o^\0_G} \, \cS[\vf,d],
\ee
which is valid for $d \geq \k \,\const$ and coincides with
\eq{e2/9.11} for $N = N_G - A_G - 1$.

\SUBSECTION{Proof of the estimates \protect\eq{e2/9.10} and
\protect\eq{e2/9.11}.}
\label{Proof.EstimatesProof}%10.4

First, if $\o^\0_G + N < 0$ then the functionals
$\as'_N\.G$ and $\D'_N\.G$ are well-defined on the entire space
$\cD (P_G)$ and the estimates \eq{e2/10.3} and \eq{e2/10.16} for
them are of the desired form.

Assume now that $\o^\0_G + N \geq 0$. For $\D_N\.G$
one uses the representation \eq{e2/9.6}.  Then from \eq{e2/10.14} and
\eq{e2/10.16} one can obtain for $\D_N\.G$ the desired inequality
\eq{e2/9.11} in the same way as in \ssect{Subtraction.r-tilde}.

Let us turn now to the functional $\as_N\.G$. (Note that all the
constructions that follow are valid in the case $\as'_N\.G \equiv 0$
as well.  This ensures a correct treatment of the case when the full
asymptotic expansion starts at a lower order in $\k $ then
$\as'\.G$---cf.\ the beginning of this section and the calculation of
$A_G$ in \ssect{Proof.StartingPoint}.)  One represents the
definition of $\as_N\.G$ as:
\be[e2/10.23]
     \as^{\vphantom{\prime}}_N\.G
     = 
     \tr_{0}\.\as'_N\.G
   + \tilde{\D}
   + {\bf \tilde{E}}_N(\k)
\ee
(cf.\ \eq{e2/9.3} and the definition of the operator $\tr$
\eq{e2/4.8}). Using the results of 
\ssect{Subtraction.r-tilde} one can obtain an inequality just
of the form of \eq{e2/9.10} from the estimate \eq{e2/10.3} for the
first term in \eq{e2/10.23}. To convince oneself that
\eq{e2/9.10} holds for the functional $\as_N\.G$, one has to prove
that the finite renormalization $\tilde{\D} + {\bf \tilde{E}}_N$ in
\eq{e2/10.23} is of the order $\k^N\, \L (\k)$. To do
this, represent the definition \eq{e2/9.5} as:
\be[e2/10.24]
     z_{a} + \tilde{E}_{a}(\k)
     =
     \langle
       (\D_{N-1}\.G_\k),
       (\Phi^{\mu}\, \cP^{a})
     \rangle
     +
     \langle
       (\D'_N\.G_\k)  \ast  (\Phi_{\mu}\, \cP^{a})
     \rangle,
\ee
where $\Phi^{\mu}(p)$ is the cut-off function which has been used in the
definition of the operator $\tr_{0}$ \eq{e2/4.10} while
$\cP^{a}$ are polynomials of the order $|a| \leq
\o^\0_G + N$.  Note that the finite renormalization
constants $z_{a}$ cancel on the r.h.s.\ of \eq{e2/10.24}.

The first term in \eq{e2/10.24} is bounded by
\be[e2/10.25]
     \k^N\, \L (\k) .
\ee
For the second term one can obtain the following estimate (cf.\
\ssect{Subtraction.UniformProblem} and \eq{e2/4.7}):
\be[e2/10.26]
     \k^{N+1} \, \L (\k) \, \mu^{-1} \, \L (\mu) .
\ee
As a result one concludes that ${\bf \tilde{E}}_N + \tilde{\D}$ is
of the order $\k^N\, \L (\k)$ and, therefore,
$\as_N\.G$ satisfies \eq{e2/9.10}.
This completes the proof of the estimates \eq{e2/9.10}, \eq{e2/9.11}
for the expansion $\As\.G$ to order~$N$.

\SUBSECTION{The case of a factorizable graph $G$.}
\label{Proof.Factorizable}%10.5

When constructing the \asex\ it is convenient---as in the
study of the $R$-operation (\sect{GenericR} and
\sect{R-operation})---not to distinguish this case and to prove the
estimates in the same way as for non-factorizable graphs.  To ensure
correctness of this, one should prove equivalence of the definition
\eq{e2/2.12} to \eq{e2/2.11} with \eq{e2/3.16}.

Note that the operator $\tr^{(\k)}$ in \eq{e2/3.20} for a general
$s$-graph $G$ is in fact uniquely fixed by the minimality conditions
\eq{e2/9.15}, \eq{e2/9.16}. Therefore, to demonstrate equivalence of
the definitions \eq{e2/2.12} and \eq{e2/2.11}, \eq{e2/3.16} for a
factorizable $s$-graph $G = G_1 \times G_2$ it is sufficient to prove
that for any polynomial $\cP^{a}(p)$, $|a| \leq \o^\0_G + N$, the
following identity holds:
\be[e2/10.27]
     \langle \Big(
          G_1 \times  G_2 -
          \sum_{n+m\leq N}(\as_n\.G_1)
          \times
          (\as_{m}\.G_2)
        \Big)
        \ast \cP^{a}
     \rangle = 0
\ee
(cf. \eq{e2/5.8} in \ssect{SpecialR.Proof}).

Let us transform the bracketed expression in \eq{e2/10.27}, using
\eq{e2/1.15}:
\be   
     G_1 \times G_2 
   - \sum_{n+m\leq N}(\as_n\.G_1) \times (\as_{m}\.G_2)
\kern2cm\ee\be[e2/10.28]\kern2cm
    = G_1 \times (\D_{N-N_1+A_1}\.G_2) 
    + \sum_n(\D_n\.G_1)
      \times  (\as_{N-n}\.G_2),
\ee
where $N_1 - A_1$ is the order at which the expansion $\As\.G_1$
starts (\ssect{Proof.StartingPoint}).  Now one can use the
estimates \eq{e2/10.22} and \eq{e2/9.11} to estimate the first term on
the r.h.s.\ of \eq{e2/10.28}, while for each term in the sum one uses
\eq{e2/9.11} and \eq{e2/9.10}. Then one derives \eq{e2/10.27} in the
same way as in \ssect{SpecialR.Proof}.

Thus, factorizable $s$-graphs can be studied on an equal footing with
non-factorizable ones.

\SUBSECTION{Form of dependence on $\k$.}
\label{Proof.KappaDependence}%10.6

Let us demonstrate that the expansion \eq{e2/9.1} is indeed the {\it
As}-expansion in the sense of \ssect{As-exp.Definition}, i.e.\ it is a
series in powers and logarithms of the expansion parameter. To do
this, assume that for any $\G < G$ each term of the expansion
$\as_n\.\G (p,\k)$ is almost-homogeneous with respect to each of its
arguments separately (therefore, it is such with respect to both
arguments).  Assume also that the remainder term of the expansion,
$\D_{N-1}\.G(p,\k)$, is almost-homogeneous with respect to both
arguments.

Construct the functional $\as'_N\.G$ using \eq{e2/10.2}. In the same
way as in \ssect{SpecialR.Scaling}, one can see that it is
almost-homogeneous with respect to $p$. Then the functional
$\tr\.\as'_N\.G$ is also almost-homogeneous with respect to
$p$. The finite renormalization ${\bf \tilde{E}}_N$ which is a sum
of derivatives of $\d $-functions is homogeneous with respect to $p$,
moreover, the index of homogeneity here is the same as for
$\tr\.\as'_N\.G$.  Therefore, the functional $\as_N\.G$ is
almost-homogeneous with respect to $p$.

Consider now the remainder term $\D'_N\.G$.  Using the recurrent
relation \eq{e2/10.10} one sees that it is almost-homogeneous with
respect to both arguments.  The transition from $\D'_N\.G$ to
$\D_N\.G$ via \eq{e2/9.6} does not spoil the almost-homogeneity with
respect to both arguments (cf.\ an analogous property of the special
subtraction operator $\tr$ in
\ssect{SpecialR.Scaling}).
Then  the  functional  $\as_N\.G$ which is just the difference of
$\D_N\.G$ and $\D_{N-1}\.G$ is also  almost-homogeneous  with
respect to both arguments.

Comparing the properties of almost-homogeneity of $\as_N\.G$ with
respect to both arguments and with respect to $p$ only one concludes
that $\as_N\.G$ is almost-homogeneous with respect to each of its
arguments separately.  In particular, $\as_N\.G$ is a power-and-log
function of $\k $:
\be[e2/10.29]
     \as_N\.G(p,\k)
     = 
     \k^N \, \sum^{L_{G,N}}_{k=0} \ln^k \k \, 
     (\as_N\.G)^{(k)}(p) ,
\ee
where ($\as_N\.G)^{(k)}(p)$ are almost-homogeneous with respect to
$p$.

This completes the proof of the theorem of \ssect{Existence.Theorem}.

\newpage\thispagestyle{myheadings}\markright{}

\newpage
\centerline{\large\sc Figure captions.}

Fig.~1. (a)  Diagrammatic representation of the product
\eq{Eq.Example1}. (b) Geometry of singularities in the $(x,y)$-space.
For example, the horizontal $x$-axis corresponds to the singularity
$y=0$, etc.

Fig.~2. (a) The diagrammatic image of \eq{Eq.Example1.Subgraph} which
coincides with a subdiagram of Fig.~1a. (b) The diagrammatic
representation of the co-subgraph obtained by replacing the subgraph
(a) in Fig.~1a by $\d(y)$ as described in
\ssect{Complete.Co-subgraphs}. 

Fig.~3. (a) The two-loop diagram corresponding to \eq{Eq.Example3}.
The fat lines correspond to the propagators with the heavy mass $M$.
(b) The geometry of singularities of the expansion in $m$ of the
integrand of Fig.~3a. The blob is the region where the test functions
considered after \eq{oiuy} differ from zero.

Fig.~4. (a) The diagrammatic representation of the $i$-graph
\eq{serapW}. (b) Its coordinate representation version. The
corresponding analytical expression is a product of three
propagators---$(x^{-2})^3$. There is only one singular point, $x=0$,
to which all three propagators contribute simultaneously which
corresponds to the fact that the graph has only one UV subgraph
according to the Bogoliubov-Parasiuk definition.

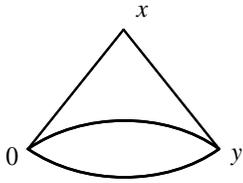
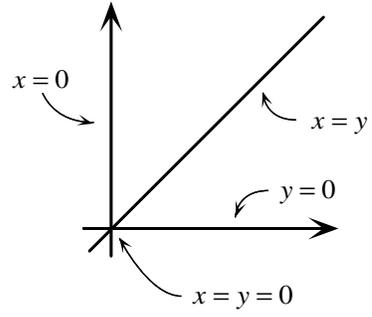
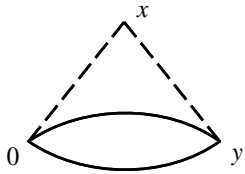
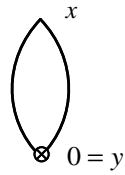
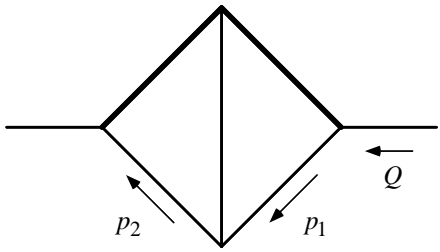
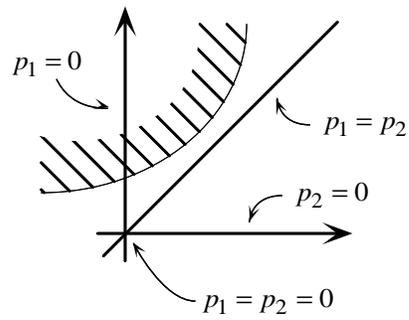
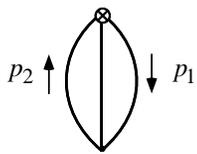
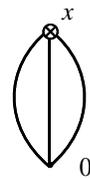
\end{document}